\documentclass[preprint]{elsarticle}

\journal{Journal of Computational Physics}

\usepackage{graphicx}
\usepackage{epstopdf, epsfig}
\usepackage{tcolorbox}
\usepackage{todonotes}
\usepackage{amsmath}
\usepackage{amssymb}
\usepackage{ulem}

\graphicspath{
{figures/}
}

\DeclareMathOperator\erfc{erfc}
\newcommand{\uvec}{\mathbf{u}}
\newcommand{\nint}{\mathbf{n}_\Gamma}

\newcommand{\deriv}[2]{\frac{\partial #1}{\partial #2} }
\newcommand{\jump}[1]{\left[ #1 \right]_\Gamma}
\newcommand{\dirac}{\delta_\Gamma}

\usepackage{times}
\usepackage{fancyhdr}
\usepackage{fancybox}
\usepackage{arydshln}
\usepackage{natbib}
\usepackage[labelfont=bf,font={small}]{caption}
\usepackage{subcaption}
\usepackage{tikz}

\usepackage{placeins}

\begin{document}

\begin{frontmatter}

\title{Comparison of interface capturing methods for the simulation of two-phase flow in a unified low-Mach framework}
\author[em2c]{Victor Boniou \corref{mycorrespondingauthor}}
\cortext[mycorrespondingauthor]{Corresponding author}
\ead{victor.boniou@centralesupelec.fr}

\author[em2c]{Thomas Schmitt}
\author[em2c,fedM]{Aymeric Vi\'e}

\address[em2c]{Laboratoire EM2C UPR 288, CNRS, CentraleSup\'elec, Universit\'e Paris-Saclay, 3, rue Joliot-Curie 91192 Gif-sur-Yvette cedex France}
\address[fedM]{F\'ed\'eration de Math\'ematiques de CentraleSup\'elec, CNRS, 3, rue Joliot-Curie 91192 Gif-sur-Yvette cedex France}

\begin{abstract}

This paper proposes a comparison of four popular interface capturing methods : the volume of fluid (VOF), the standard level set (SLS), the accurate conservative level set (ACLS) and the coupled level set and volume of fluid (CLSVOF). 
All methods are embedded into a unified low-Mach framework based on a Cartesian-grid finite-volume discretization. 
This framework includes a sharp transport of the interface, a well-balanced surface tension discretization and a consistent mass and momentum transport which allows capillary-driven simulations with high density ratio. 
The comparison relies on shared metrics for geometrical accuracy, mass and momentum conservation which exposes the weakness and strengths of each method. Finally, the versatility and capabilities of the proposed solver are demonstrated on the simulation of a 3D head-on collision of two water droplets. Overall, all methods manage to retrieve reasonable results for all test cases presented. VOF, CLSVOF and ACLS tend to artificially create little structures while SLS suffers from conservation issues in the mesh resolution limit. This study leads us to the conclusion that CLSVOF is the most promising method for two-phase flow simulations in our specific framework because of its inherent conservation properties and topology accuracy.

\end{abstract}



\end{frontmatter}

\section{Introduction}
The numerical description of the interface between two non-miscible fluids has been widely investigated in the last decades and several approaches were developed in order to capture it accurately. For the simulation of incompressible two-phase flows with a sharp interface representation, several techniques can be found in the literature:
\begin{itemize}
\item Front tracking (FT) \cite{unverdi1992front}: FT methods explicitly transport Lagrangian markers all belonging to the interface. This allows to be highly accurate in the interface topology description such as interface normal and curvature but it requires special treatments in cases of coalescence or atomization which are non-trivial \cite{popinet1999front}\cite{Tryggvason2001}.
\item Volume-of-Fluid (VOF) \cite{hirt1981volume}: VOF methods are based on the resolution of the volume fraction of one phase in each numerical cell. This method has the advantage of being intrinsically mass-conserving. However, the interface is not explicitly known and must be retrieved through a reconstruction strategy, such as the PLIC method \cite{Rider1998}. Moreover, topology information requires a dedicated attention, as differentiating the sharp volume fraction field can lead to huge errors.
\item Level set (LS) \cite{osher1988fronts}: instead of solving for the volume fraction, such methods solve an implicit field that characterizes the distance to the interface. This field is chosen to be smoother than the volume fraction, thus being easier to solve numerically, and giving access to a better topology information. However, attention must be drawn on conservation properties.
\end{itemize}
FT methods will not be investigated in this work, as it is a totally different way of handling the interface transport both in the formulation and in the implementation. \\
In addition to the interface capturing, two-phase flow simulations are also very challenging because of the interface discontinuities which have to be treated carefully. In the literature, a large variety of methods is available to deal with the associated numerical issues \cite{popinet2009accurate,Owkes2017,Chenadec2013,Zuzio2020,Raessi2012,Ghods2013,Chiodi2017,Janodet2019}. However they are usually adapted to one particular interface capturing method and comparison to existing strategies are not systematically provided. \\
Detailed comparisons have been performed in the last decades, but they are still limited to a particular interface capturing method \cite{fuster2009numerical,Nourgaliev2007,Solomenko2017} or to a specific test case such as spurious currents \cite{Abadie2015} or rising bubbles \cite{Bazilevs2011}.
Moreover, as the field of interface capturing methods is highly active, updating comparisons studies is required to take advantage of the recent advances.
Finally, conclusions are sometimes only provided to 2D simulations \cite{Gerlach2006,Abadie2015,Bazilevs2011}. This is a serious limitation when dealing with geometrical properties as the extension to 3D is not always straightforward .
Our work is thus motivated by this observation and aims to provide a complete study of up-to-date methods in both 2D and 3D configurations. \\
Here, attention is drawn to four popular Eulerian representations of the interface : VOF with a PLIC reconstruction \cite{Rider1998}, standard LS (SLS) \cite{sussman1994level}, accurate conservative LS (ACLS) \cite{olsson2005conservative} and coupled LS-VOF (CLSVOF) \cite{Sussman2000}.
The test cases are chosen to demonstrate interface transport and surface tension modelling accuracy along with mass and momentum conservation of these methods both in 2D and 3D.
To avoid potential misleading conclusions due to solver differences, they are adapted to a same unified framework with the following specifications: a sharp transport of the interface, a well-balanced description of the surface tension and a consistent mass and momentum transport. 
The reader must keep in mind that the study is performed in a finite volume Cartesian-grid discretization and the underlying conclusions are limited to the present framework. For instance, the difficulties raised by the use of unstructured meshes would possibly change our conclusions, especially because geometrical-splitted VOF would not be applicable. \\
The paper is organized as follows: Section 2 gives an overview of the four interface capturing method considered in this work. Section 3 presents the general two-phase solver and a detailed discussion on curvature computation is presented in Section 4. Section 5 presents the results. First, a validation of the interface capturing methods on various test cases including interface transport on imposed velocity field. Then momentum conservation is investigated with high density ratio followed by test cases dedicated to the surface tension modelling. Finally solver performances are demonstrated on a 3D head-on droplet collision and a 2D shear liquid flow. The paper ends with conclusions and perspectives.

\section{Interface capturing}
\label{sec:interface}
In this section, a review of the interface capturing methods is presented followed by the description of the up-to-date versions implemented in our solver.

The Eulerian representation of the interface relies on an indicator function $\chi$
\begin{equation}
\chi(\mathbf{x}) = \left\{
    \begin{array}{ll}
        0 & \mbox{if } \mathbf{x} \in \Omega_g \\
        1 & \mbox{if } \mathbf{x} \in \Omega_l
    \end{array}
\right.
\end{equation}
where $\Omega_g$ is the gas domain, $\Omega_l$ is the liquid domain and $\mathbf{x}$ is the spatial coordinate.

Its evolution is given by
\begin{equation}
\frac{D\chi}{Dt} = \frac{\partial\chi}{\partial t} + \mathbf{u}_\Gamma \cdot \nabla  \chi = 0
\label{eq:chi}
\end{equation}
where $\mathbf{u}_\Gamma$ the interface velocity vector and $t$ the time.

To solve this last equation numerically, $\chi$ needs a discrete representation. 
The VOF method intends to solve Eq.~(\ref{eq:chi}) on the computational mesh by introducing the volume fraction of liquid $f$ \cite{tryggvason2011direct} which is the volume average of $\chi$ 
\begin{equation}
f = \frac{\int_{\mathcal{C}} \chi (\mathbf{x}) d\mathbf{x}}{\mathcal{V}_\mathcal{C}}
\end{equation}
here $\mathcal{C}$ is a computational cell of volume $\mathcal{V}_\mathcal{C}$.

The function $f$ is still very sharp and challenging to solve numerically as its gradient is only defined in a 3-cell region around the interface. As a circumvent, LS class methods solve a regular implicit function $\phi$ instead of $f$. In the SLS method \cite{sussman1994level}, $\phi$ is the minimal signed distance to the interface 
\begin{equation}
\phi(\mathbf{x}) = \left\{
    \begin{array}{ll}
        -\min\limits_{\forall \mathbf{x}_\Gamma  \in \Gamma} \lvert \mathbf{x}_\Gamma - \mathbf{x} \rvert & \mbox{if } \mathbf{x} \in \Omega_g \\
        \min\limits_{\forall \mathbf{x}_\Gamma  \in \Gamma} \lvert \mathbf{x}_\Gamma - \mathbf{x} \rvert & \mbox{if } \mathbf{x} \in \Omega_l \\
        0 & \mbox{if } \mathbf{x} \in \Gamma \\
    \end{array}
\right.
\end{equation}
where $\mathbf{x}_\Gamma$ is the interface location. 

Even if $\phi$ provides the interface location with a smooth function, it cannot strictly ensure mass conservation. To improve mass conservation, the ACLS~\cite{olsson2005conservative} method suggests to use a smoothed Heaviside $\psi$ computed from $\phi$
\begin{equation}
\psi = \frac{1}{2} \left( 1+\tanh \left(  \frac{\phi}{2 \epsilon} \right)   \right)
\label{eq:heps_sls}
\end{equation}
with $2\epsilon$ the interface thickness.

Regardless of the interface capturing method choice, they all end to a similar advection equation for the color function $c=f,\phi$ or $\psi$
\begin{equation}
\frac{\partial c}{\partial t} +  \mathbf{u} \cdot  \nabla c = 0
\label{eq:colorfadv}
\end{equation}

In the context of a divergence-free flow without phase-change, interface velocity is equivalent to the flow velocity. Eq.~(\ref{eq:colorfadv}) can thus be rewritten as a conservation equation of $c$
\begin{equation}
\frac{\partial c}{\partial t} + \nabla  \cdot  \left(  \mathbf{u} c \right) = 0
\label{eq:colorfcons}
\end{equation}

Despite the fact that the different color functions follow a similar equation, they require specific numerical treatments. In the following, details are given about flux computation and eventual additional steps in the advection process.

\subsection{VOF method}
In this work, the VOF method is based on the geometric transport of $f$ with a PLIC~\cite{Pilliod2004} representation of the interface. Note that VOF methods can also rely on algebraic fluxes~\cite{Xiao2005} but they suffer from numerical diffusion or dispersion which are incompatible with a sharp representation of the interface.
The interface is reconstructed as a plane in a mixed cell ($0<f<1$)
\begin{equation}
\mathbf{x} \cdot \mathbf{n} = d
\label{eq:plic}
\end{equation}
The interface normal  $\mathbf{n}$ is evaluated using the ELVIRA procedure which allows to produce second order normal computation \cite{Pilliod2004} while the plane parameter $d$ is deduced from the normal and the volume fraction using the analytic relations of a chopped cube~\cite{Scardovelli2000}. The fluxes are computed from the geometric reconstruction using the Weymouth and Yue  dimensional-splitting (WY) scheme \cite{Weymouth2010}. \\
Our structured Cartesian mesh allows the use of a dimensional-splitting scheme, which simplifies the flux computation through geometric reconstruction. It results in solving successive 1D advection problems
\begin{equation}
\frac{\partial f}{\partial t} + \frac{\partial \left(u_s f\right)}{\partial x_s}   = f \frac{\partial u_s}{\partial x_s} 
\label{eq:1Dadv}
\end{equation}
here is $s$ the sweep direction.
$\nabla \cdot \mathbf{u}=0$ does not imply $\frac{\partial u_s}{\partial x_s}=0$ for a given splitting direction and has to appear explicitly in the RHS of Eq.~(\ref{eq:1Dadv}). Hence, it is non trivial to have an exactly conservative scheme solving those equations successively ~\cite{scardovelli2003interface}.
In the WY scheme, $f \frac{\partial u_s}{\partial x_s} $ is replaced by $f_c \frac{\partial u_s}{\partial x_s}$.  $f_c$ is built to avoid any cell under or overfill such that 
\begin{equation}
f_c=\chi(\mathbf{x}_\mathcal{C})= \left\{
    \begin{array}{ll}
        0 & \mbox{if } f<0.5 \\
        1 & \mbox{otherwise }\\
    \end{array}
\right.
\end{equation}
After each sweep, a PLIC reconstruction is performed for flux computation of the next step. This gives the following algorithm for a timestep
\begin{enumerate}
\item Compute the compression/dilatation factor $f_c$ 
\item Perform a PLIC reconstruction by computing $\mathbf{n}$ and $d$ in all mixed cells
\item Solve (\ref{eq:1Dadv}) in a given direction using Euler Implicit scheme
\item Repeat 2 and 3 for all directions to obtain $f^{n+1}$
\end{enumerate}

This algorithm leads to exact mass conservation given a divergence-free velocity field $\mathbf{u}$. 
WY is the only method able to keep exact mass conservation in a directional-splitting fashion. Unsplit methods also achieve such conservation \cite{Owkes2014}, but the geometrical flux construction is way more demanding in terms of implementation and computation effort. Furthermore, both approaches are still limited to second order accuracy because of the geometrical nature of the fluxes based on linear reconstruction.

\subsection{SLS method}
The standard LS advection method relies on the transport of $\phi$. The discretization has a huge impact on the mass conservation as numerical diffusion leads to artificial mass loss. This subject has been widely explored, and it has been shown that LS performs well with high-order schemes. A complete comparison of ENO, WENO and HOUC schemes of different orders has been done in \cite{Nourgaliev2007}. Here, a finite volume WENO5 scheme is used for solving Eq.~(\ref{eq:colorfcons}). \\
One important requirement of the SLS method is to keep the property $\lvert \nabla \phi \rvert=1$. If this is no longer true, then the computation of topological properties from $\phi$ (such as normal and curvature) would suffer from huge errors and spurious behaviour.
The diffusion and dispersions of numerical schemes used to solve advection will lead to the loss of this property, hence an additional step is required called the redistanciation step.
It can be written as an evolution equation of $\phi$ in pseudo time $\tau$ \cite{sussman1994level}
\begin{equation}
\frac{\partial\phi}{\partial\tau} + \text{sign}(\phi_0) \left(  \lvert \nabla \phi \rvert - 1 \right) = 0
\label{eq:slsredist}
\end{equation}
This results in solving a Hamilton-Jacobi equation with the corresponding Hamiltonian $\mathcal{H}\left( \phi, \nabla \phi \right) = \text{sign}(\phi_0) \left(  1-\lvert \nabla \phi \rvert  \right) $ with $\text{sign}(\phi_0)$ a smoothed signed function based on the initial distance just after advection step $\phi_0$
\begin{equation}
\text{sign}(\phi_0) = \frac{\phi_0}{\sqrt{\phi_0^2+\Delta x^2}}
\label{eq:slssign}
\end{equation}

Eq.~(\ref{eq:slsredist}) can be again solved with high accuracy with a HJ-WENO5 scheme \cite{Jiang2000}. This leads to the following algorithm for a time step
\begin{enumerate}
\item Advance the interface by solving Eq.~(\ref{eq:colorfadv}) to obtain $\phi^*$ 
\item Compute the regularised distance sign $\text{sign}(\phi_0)$ from Eq.~(\ref{eq:slssign})
\item Perform 2 iterations of Eq.~(\ref{eq:slsredist}) to obtain $\phi^{n+1}$ with the pseudo time step $\Delta \tau = 0.5\Delta x$
\end{enumerate}

The overall method does not conserve mass as the reinitialization step is not conservative and conservation of $\phi$ does not imply conservation of mass. However, the method can be very accurate if high order schemes are applied, the implementation is straightforward even in multidimensional cases and the computational cost is expected to be fairly low. 

\subsection{ACLS method}
Accurate conservative LS method relies on a sharper function $\psi$ which is a smooth version of the indicator function $\chi$. $\psi$ can be seen as a volume fraction with a controlled interface width $2\epsilon$, usually chosen as $\epsilon=\Delta x/2$. \\
For the same reasons that $\phi$ cannot maintain the property $\lvert \nabla \phi \rvert=1$ during transport, there is no guarantee that the hyperbolic tangent profile $\psi$ will remain unchanged. This takes the form of local modifications of the interface thickness which can lead to an inaccurate representation of the interface and topology computation. \\
An additional equation has to be solved in pseudo time to overcome this problem~\cite{olsson2005conservative}
\begin{equation}
\frac{\partial\psi}{\partial\tau} + \nabla \cdot \left( \psi (1-\psi) \mathbf{n} - \epsilon \left( \nabla \psi \right)\right) =  0
\end{equation}
The stable and accurate method considered here is the ACLS of \cite{Chiodi2017} with the additional modification of \cite{Janodet2019} where the reinitialization is reformulated to 
\begin{equation}
\frac{\partial\psi}{\partial\tau} =  \nabla \cdot \left( \frac{1}{4 \cosh^2 \left( \frac{\phi_{map}}{2 \epsilon}\right)}\left(\nabla \phi_{map} \cdot \mathbf{n}_{FMM} - \mathbf{n}_{FMM} \cdot \mathbf{n}_{FMM} \right) \mathbf{n}_{FMM} \right)
\label{eq:aclsreinit}
\end{equation}
with $\phi_{map} = \epsilon \log \left( \frac{\psi}{1-\psi} \right)$.

In order to ensure $\psi$ boundness, a BHOUC5 \cite{Herrmann2006} discretization is used for Eq.~(\ref{eq:colorfadv}) which is a HOUC5 scheme with a switch to first order upwind when undershoots or overshoots occur in the transport process. The terms in equation~(\ref{eq:aclsreinit}) are discretized using second order finite differences while the normal $\mathbf{n}_{FMM}$ is computed as 
\begin{equation}
\mathbf{n}_{FMM} = \nabla \phi_{FMM}
\label{eq:nfmm}
\end{equation}
with $\phi_{FMM}$ a distance function computed from a Fast Marching Method (FMM) algorithm \cite{McCaslin2014}. 
The construction of $\phi_{FMM}$ is fundamental in the method as it removes all oscillatory behaviours of $\psi$ in the computation of normals \cite{Desjardins2008}. Note that in this formulation, normal are not normalized, which improves front merging behaviour \cite{Janodet2019}.
This leads to the following algorithm for a time step
\begin{enumerate}
\item Advance the interface by solving equation~(\ref{eq:colorfadv}) to obtain $\psi^*$ 
\item Compute the signed distance $\phi_{FMM}$ from the isocontour $\psi^*=0.5$, $\phi_{map}$ from $\psi^*$ and $\mathbf{n}_{FMM}$ with Eq.~(\ref{eq:nfmm})
\item Perform one iteration of Eq.~(\ref{eq:aclsreinit}) to obtain $\psi^{n+1}$ with $\Delta \tau = 0.25\Delta x$
\end{enumerate}

The overall method leads to a better mass conservation than the SLS because both transport and reinitialization steps are conservative and the transported color function $\psi$ represents the liquid volume in the limit $\Delta x \to 0$. Moreover, it allows to use high order schemes too. However, the FMM algorithm implies more implementation and computation effort. To improve efficiency, the FMM reconstruction and reinitialization are only performed in a narrow band of 10 cells near the interface.

\subsection{CLSVOF method}
Coupling LS and VOF makes use of both method advantages to provide a conservative and accurate representation of the interface. \\
In a dimensional splitting fashion, $f$ is advanced using WY scheme presented above, $\phi$ is also advanced in time accordingly by the 1D advection equation
\begin{equation}
\frac{\partial \phi}{\partial t} + \frac{\partial \left(u_s \phi\right)}{\partial x_s}   = \phi \frac{\partial u_s}{\partial x_s}
\label{eq:1Dsls}
\end{equation}

Then, the coupling between VOF and LS is done through reconstruction and redistanciation. The PLIC reconstruction Eq.~(\ref{eq:plic}) is performed using normal from a plane fit using $\phi$ values \cite{Sussman2000} instead of ELVIRA. The redistanciation step is modified as in \cite{Tanguy2007} by shifting the value of $\phi$ in mixed cells ($0<f<1$) to be exactly the normal distance to the PLIC reconstruction $\phi=d$ and perform redistanciation of Eq.~(\ref{eq:slsredist}) only in full or empty cells to prevent any zero-isocontour displacement. \\
Note that in the solver, no clipping is performed based on $\phi$ as in the original works \cite{Sussman2000,Tanguy2007} preventing any mass conservation issues.

This leads to the following algorithm for a time step
\begin{enumerate}
\item Compute the compression/dilatation factor $f_c$ 
\item Perform a PLIC reconstruction by computing $\mathbf{n}$ and $d$ in all mixed cells
\item Solve Eq.~(\ref{eq:1Dadv}) in a given direction using Euler Implicit scheme and Eq.~(\ref{eq:1Dsls}) using WENO5 fluxes.
\item Repeat 2 and 3 for all directions to obtain $f^{n+1}$ and $\phi^*$
\item Correct the values of $\phi^*$ in mixed cells with the PLIC reconstruction and perform 2 iterations of Eq.~(\ref{eq:slsredist}) on the remaining cells to obtain $\phi^{n+1}$
\end{enumerate}

With this method, the PLIC reconstruction is enhanced using accurate normal from $\phi$ while the redistanciation correct the mass conservation by using PLIC conservation properties. \\
Note that using $\phi=d$ introduces a second order error, the CLSVOF distance function (called $\phi_{PLIC}$ in the rest of the paper) is then expected to be less accurate than $\phi$ obtained from the SLS method. In the current implementation, the relaxation proposed in \cite{Chenadec2013} is used to reduce the error frequently introduced by the coupling between VOF and LS. This last point will be discussed further in Section~\ref{sec:curvature}.

\section{Two-phase low Mach solver}
All interface capturing methods presented above are then integrated in a unified framework described in the following section.

The general continuity equation can be written as
\begin{equation}
\deriv{\rho}{t} + \nabla \cdot \left(\rho \mathbf{u} \right) = 0
\label{eq:continuity-1}
\end{equation}
with $\rho$ the fluid density and $\uvec$ the fluid velocity.
In the incompressible form, the continuity Eq.~(\ref{eq:continuity-1}) results in the divergence-free condition
\begin{equation}
\nabla \cdot \mathbf{u} = 0
\label{eq:continuity-2}
\end{equation}

From the conservation of momentum, one can write: 
\begin{equation}
\deriv{\rho \uvec}{t} + \nabla \cdot \left( \rho \uvec \otimes \uvec \right) = -\nabla P + \nabla \cdot \left( \mu \mathbf{D} \right) + \mathbf{F}_v
\label{eq:mom-1}
\end{equation}
with $\mu$ the dynamic viscosity, $P$ the pressure, $\mathbf{F}_v$ the volume force and $\mathbf{D}$ the rate-of-deformation tensor 
\begin{equation}
\mathbf{D}=\nabla \uvec + \nabla \uvec ^\intercal 
\label{eq:defrate}
\end{equation}

In a sharp two-phase framework, the interface is considered to be thin and massless. The density and viscosity is then discontinuous between both phases 
\begin{align}
& \jump{\rho}=\rho_l-\rho_g \\
& \jump{\mu}=\mu_l-\mu_g 
\end{align}

When no phase change occurs, the interface velocity is directly the flow velocity and integration of Eq.~(\ref{eq:continuity-2}) through the interface implies normal velocity continuity between liquid and gas
\begin{equation}
\jump{\uvec \cdot \nint} = 0
\end{equation}
with $\nint$  the normal pointing outside the interface (from the liquid to the gas phase).
Integration of Eq.~(\ref{eq:mom-1}) leads to the following jump in pressure with a purely two-phase contribution related to surface tension forces acting at the interface and a contribution related to the viscosity jump
\begin{equation}
\jump{P} = \sigma \kappa + \jump{\mu \left(\mathbf{D}\cdot \nint \right) \cdot \nint}
\label{eq:pjump}
\end{equation}
with $\sigma$ the fluid surface tension and $\kappa$ is twice the interface mean curvature.

The present two-phase low-Mach solver relies on the projection method \cite{Chorin1969} with a one-fluid representation of the velocity. The one-fluid properties are defined as an average 
\begin{align}
& \rho = \rho_g + H^{\epsilon}_\Gamma \jump{\rho} \\
& \mu = \mu_g + H^{\epsilon}_\Gamma \jump{\mu} 
\end{align}
with $H^{\epsilon}_\Gamma$ a regularized Heaviside of controlled width $2\epsilon=\Delta x$.
In a VOF and ACLS framework, the natural choice is the color function ($f$ and $\psi$ respectively) while in a SLS framework, an Heaviside can be defined using a regularization of the distance function such as Eq.~(\ref{eq:heps_sls}).

\subsection{Grid arrangement}
\begin{figure}[h!]
	\centering
	\includegraphics[scale=0.5]{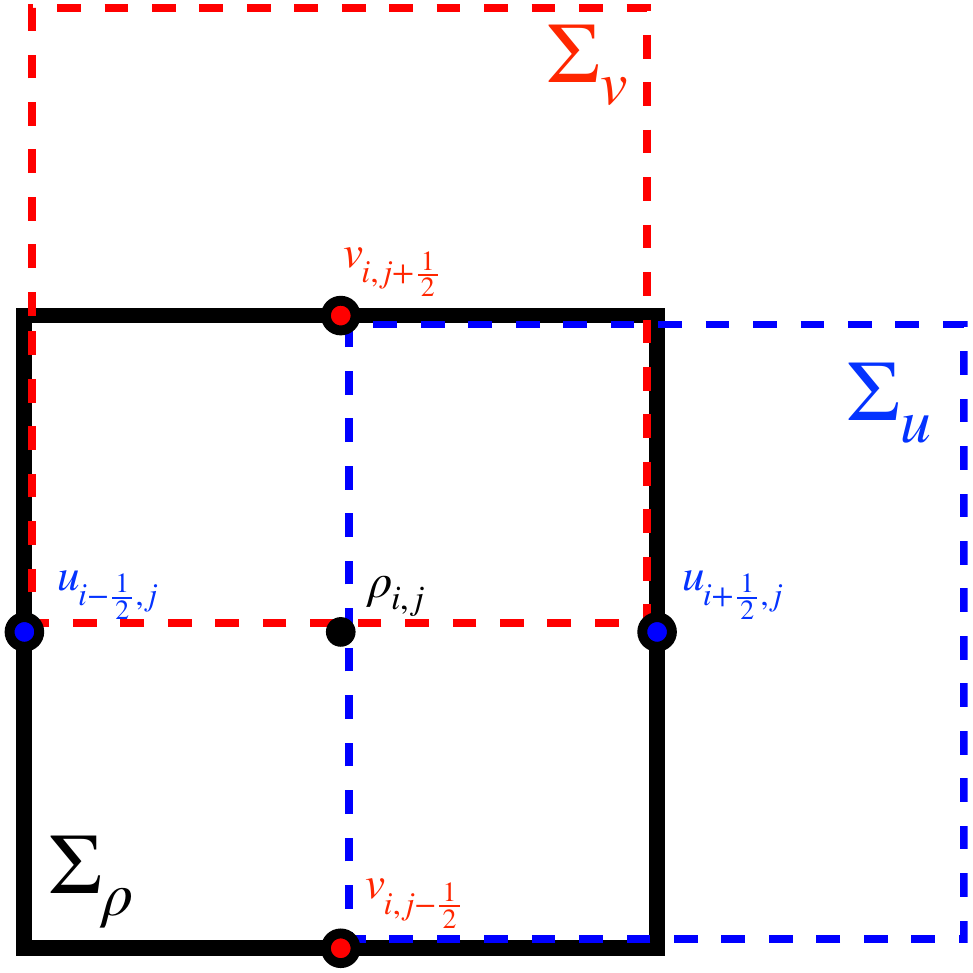}
	\caption{Mac grid arrangement on a 2D cartesian grid}
	\label{fig:mac_grid}
\end{figure}

A Marker-and-Cell grid arrangement \cite{harlow1965numerical} is used : the pressure is defined at the center of a cell while the velocity is located at the boundaries of the cell. In a two-phase flow, one can consider that all scalars are collocated (located at the center of the cell) while the velocity is staggered (located at the boundaries of the cell) as shown in Fig.\ref{fig:mac_grid} for a 2D control volume.

This grid arrangement allows to define the following operators: 
\begin{itemize}
\item A gradient operator $\nabla_f$ for collocated variables $\Phi_c$, the x-component $\nabla_u$ is defined as
\begin{equation}
\left. \nabla_u \Phi_c \right\vert_{i-\frac{1}{2},j} = \frac{\Phi_{c,i,j}-\Phi_{c,i-1,j}}{\Delta x}
\label{eq:grads}
\end{equation}
\item A gradient operator $\nabla_c$ for staggered variable is also introduced with the x-component $\nabla_c^x$ of a x-staggered variable defined as
\begin{equation}
\left. \nabla^x_c \Phi_f \right\vert_{i,j} = \frac{\Phi_{f,i+\frac{1}{2},j}-\Phi_{f,i-\frac{1}{2},j}}{\Delta x}
\end{equation}
with $\Phi_f$ a collocated scalar.
\item A divergence operator $\nabla_c\cdot\left( \Phi_c \uvec \right)$ applied on a control volume $\Sigma_\rho$
\begin{align}
\left. \nabla_c \cdot \left( \Phi_c \uvec \right) \right\vert_{i,j} &= \frac{\tilde{\Phi}_{c,i+\frac{1}{2},j}u_{i+\frac{1}{2},j}-\tilde{\Phi}_{c,i-\frac{1}{2},j}u_{i-\frac{1}{2},j}}{\Delta x} \nonumber \\ 
& + \frac{\tilde{\Phi}_{c,i,j+\frac{1}{2}} v_{i,j+\frac{1}{2}}-\tilde{\Phi}_{c,i,j-\frac{1}{2}}v_{i,j-\frac{1}{2}}}{\Delta y} 
\end{align} 
with $\tilde{\Phi}_c$ an arbitrary interpolation of $\Phi_c$ to a face of the $\Sigma_\rho$ control volume. 
\item A divergence operator $\nabla_f\cdot\left( \Phi_f \uvec \right)$ on a staggered volume control (here for $\Sigma_u$)
\begin{align}
\left. \nabla_f \cdot \left( \Phi_u \uvec \right) \right\vert_{i-\frac{1}{2},j} &= \frac{\tilde{\Phi}_{u,i,j}\bar{u}_{i,j}-\tilde{\Phi}_{u,i-1,j}\bar{u}_{i-1,j}}{\Delta x} \nonumber \\ 
& + \frac{\tilde{\Phi}_{u,i-\frac{1}{2},j+\frac{1}{2}} \bar{v}_{i-\frac{1}{2},j+\frac{1}{2}}-\tilde{\Phi}_{u,i-\frac{1}{2},j-\frac{1}{2}}\bar{v}_{i-\frac{1}{2},j-\frac{1}{2}}}{\Delta y} 
\end{align} 
with $\tilde{\Phi}_u$ an arbitrary interpolation of $\Phi_u$ and $\bar{u}$ a linear interpolation of velocity to a face of the $\Sigma_u$ control volume.
\end{itemize}
Note that the divergence operators are built such that $\nabla_c\cdot\uvec=0$ and $\nabla_f\cdot\uvec=0$ discretely thanks to the Mac grid arrangement.
In the following, general staggered operators and variables are mentioned with the subscript $f$ while staggered variable referring to a specific control volume are referred to $u$, $v$ or $w$.

\subsection{Prediction step}
\label{sec:momcons}
The prediction step is written as follows
\begin{equation}
\frac{\uvec^*-\uvec^n}{\Delta t}  = \mathcal{L}_{\text{conv}} + \mathcal{L}_{\text{visc}} + \mathcal{L}_{\text{cap}}
\label{eq:PM_pred}
\end{equation}
with $\uvec^*$ the predicted velocity field which is not divergence-free and the 3 operators defined as
\begin{align}
& \mathcal{L}_{\text{conv}} = - \uvec^n \cdot \nabla_f \uvec^n  \label{eq:lconv} \\ 
& \mathcal{L}_{\text{visc}} = \frac{1}{\rho_f^{n+1}}\nabla_f \cdot \left( \mu^{n+1} \mathbf{D}^n  \right)\\ 
&  \mathcal{L}_{\text{cap}} = \frac{1}{\rho_f^{n+1}} \mathbf{F}^{n+1}_\sigma \label{eq:lcap}
\end{align}
where $\rho_f^{n+1}$, $\mu^{n+1}$ and $\mathbf{F}^{n+1}_\sigma$ are computed using the advanced color function $c^{n+1}$. Details on the operators discretization are given hereafter.

\subsubsection{Convection term $\mathcal{L}_{\text{conv}}$ }

The prediction step given by Eq.~(\ref{eq:PM_pred}) is in a velocity form, and the convection term given by Eq.~(\ref{eq:lconv}) does not ensure any conservation on momentum. In this work, this term is rewritten in order to ensure consistent transport of mass and momentum as in \cite{Desjardins2010}. This method relies on additional continuity equations defined in the momentum control volumes $\Sigma_u$, $\Sigma_v$ (and $\Sigma_w$ in 3D) of Fig.\ref{fig:mac_grid}.
For $u$ component in $\Sigma_u$ considering only the convective contribution, the scheme is written as 
\begin{align}
& \frac{\rho_u^*-\rho_u^n}{\Delta t}  = -\nabla_f \cdot \left( \rho_u^n \uvec^n \right) \\
& \frac{\rho_u^* u^*-\rho_u^n u^n}{\Delta t}  = -\nabla_f \cdot \left( \rho_u^n u^n \uvec^n \right)
\label{eq:velupdate}
\end{align}
note that $\rho_u^*$ is only an intermediate evolution variable discarded just after prediction step. In the general case, $\rho_u^*$ will differ from the reconstruction $\rho_u^{n+1}$.

Rewriting Eq.~(\ref{eq:velupdate}) in a velocity form gives (here for $u$)
\begin{equation}
\mathcal{L}_{\text{conv}} = \frac{1}{\rho_u^*}\nabla_f \cdot \left( \rho_u^n u^n \uvec^n \right)  - \frac{u^n}{\Delta t} \frac{\rho_u^*-\rho_u^n}{\rho_u^*} 
\end{equation}
the additional term in the RHS is as a correction of momentum due to the mass change in the cell during the prediction step. 
The fluxes $\left( \rho_u^n \uvec^n \right)$ and $\left( \rho_u^n u^n \uvec^n \right)$ have to be evaluated with the same interpolation for consistency. Moreover, the scheme have to be bounded to avoid any overshoot or undershoot of density. In the present solver, a WENO5 interpolation is performed, which switches to an upwind evaluation when the stencil crosses the interface.

\begin{figure}[h!]
	\centering
	\includegraphics[scale=0.35]{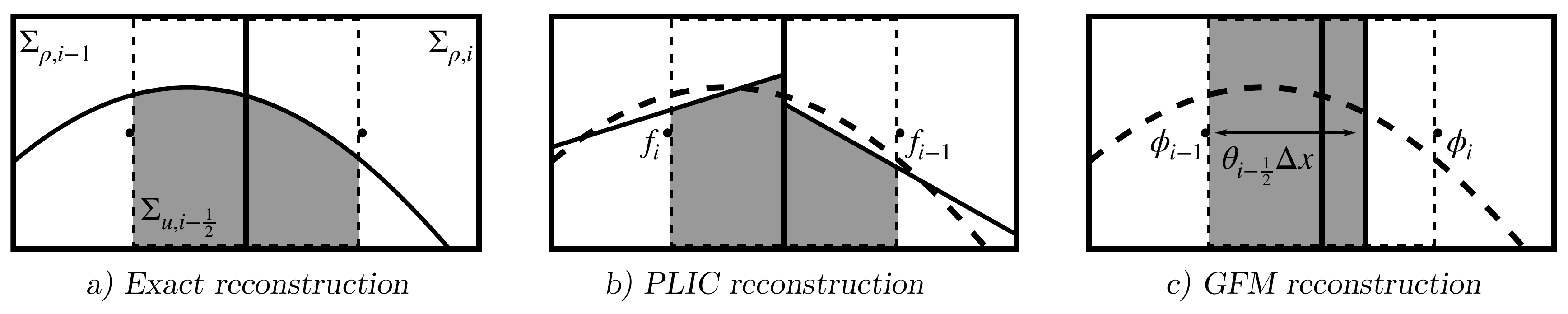}
	\caption{Reconstruction of $\rho_{u,i-\frac{1}{2}}$ on a 2D cartesian grid. The grey area represent the area of liquid evaluated in the control volume $\Sigma_{u,i-\frac{1}{2}}$.}
	\label{fig:rho_mom}
\end{figure}

In the VOF or CLSVOF framework, a second order accurate reconstruction of the density $\rho_f^n$ is performed by using the PLIC reconstruction from $f^n$ \cite{Palmore2019} as in Fig.~\ref{fig:rho_mom}. For SLS and ACLS, \cite{Desjardins2010} proposed to use the GFM expression of face interpolated density \cite{Liu2000}
\begin{equation}
\rho^{n}_{u,i-\frac{1}{2}} = \rho_g + \theta_{i-\frac{1}{2}} \jump{\rho}
\end{equation}
with $\theta_{i-\frac{1}{2}}$ defined from $\phi^n$ as
\begin{equation}
\theta_{i-\frac{1}{2}} = \left\{
    \begin{array}{lll}
        0 & \mbox{if $\phi^n_{i-1}<0$ and $\phi^n_i<0$ }  \\
        1 & \mbox{if $\phi^n_{i-1}>0$ and $\phi^n_i>0$ }  \\
        \frac{\max(0,\phi^n_{i-1})+\max(0,\phi^n_{i})}{\lvert \phi^n_{i-1} \rvert + \lvert \phi^n_{i} \rvert } & \mbox{otherwise }
    \end{array}
  	\right.
\label{eq:slsrho}
\end{equation}
this leads to a first order reconstruction of the interface as illustrated in Fig.~\ref{fig:rho_mom}.

In this work, it was found that using the regularized Heaviside $H_\Gamma^\epsilon$ as in \cite{Herrmann2008} was a better choice for momentum conservation and solver robustness in the case of ACLS. Indeed, Eq.~(\ref{eq:slsrho}) relies on geometric informations provided by a distance function. Because of the errors in $\phi_{FMM}$, this expression can lead to huge errors in density. Hence, a more robust approximation of $\rho_u^n$ is performed by using the simple $\psi$ average
\begin{equation}
\rho^{n}_{u,i-\frac{1}{2}} = \rho_g + \psi_{i-\frac{1}{2}} \jump{\rho}
\label{eq:clsrho}
\end{equation}
with $\psi_{i-\frac{1}{2}}=\frac{1}{2}\left(\psi_{i-1} + \psi_{i}\right)$.

\subsubsection{Capillary term $\mathcal{L}_{\text{cap}}$ }
Historically, surface tension was either treated as a volumetric force $\mathbf{F}_\sigma$ in the prediction step of the projection method using the Continuum Surface Force (CSF) approach or as a pressure jump in the discretization of pressure gradients using the Ghost Fluid Method (GFM). Popinet~\cite{Review2018} showed that GFM can also be expressed as a volumetric force in the prediction step leading to the same formalism as CSF. A general expression of this force is
\begin{equation}
\mathbf{F}_\sigma = \sigma \kappa \dirac \nint
\end{equation}
The Dirac function $\dirac$ is related to the Heaviside function $H_\Gamma$ such that $\nabla H_\Gamma = \dirac \nint$. Numerically, it is not possible to represent the Heaviside function at the exact interface location, and a numerical approximation is required: 
\begin{equation}
\mathbf{F}_\sigma = \sigma \kappa \nabla H^{num}_\Gamma
\label{eq:surften}
\end{equation}
with $H^{num}_\Gamma$ a discrete representation of the Heaviside.

A CSF approach would use a smooth Heaviside such as the expressions given above $H^{num}_\Gamma=H^{\epsilon}_\Gamma$. 
The expression corresponding to GFM is the sharp Heaviside defined at the cell center $H^{num}_\Gamma=H^{0}_\Gamma = H_\Gamma (\mathbf{x}_\mathcal{C})$. 

In \cite{Francois2006}, the authors noticed that discretizing $\nabla H_\Gamma$ and $\nabla P$ with the same operator provides a discrete balance between pressure gradient and capillary forces. This is crucial to obtain a stable and robust surface tension modelling. With variable density, it is also important to have the same expression for $\frac{1}{\rho^{n+1}}$ in Eq.~(\ref{eq:lcap}) to keep the well-balanced property \cite{Herrmann2008}. 
In our solver, the face gradient $\nabla_f$ is used for both surface tension force and the pressure gradient. The density is defined using the same formalism than in the correction step as $\rho_f^{n+1}=\rho_u^{n+1}$ for the x-normal faces ($\rho_v^{n+1}$ and $\rho_w^{n+1}$ for y and z-normal faces). 

The GFM discretization of $\mathcal{L}_{\text{cap}}$ used in our solver can be written (here for $u$) as
\begin{equation}
\mathcal{L}_{\text{cap}} = \frac{1}{\rho_u^{n+1}} \sigma \kappa_u^{n+1} \nabla^x_f H^{0,n+1}_\Gamma 
\end{equation}
with $\kappa_u^{n+1}$ a curvature projected to the x-normal face. Sec.~\ref{sec:kappa} is dedicated to the computation of the curvature $\kappa$ and its interpolation to the face $\kappa_f$.

\subsubsection{Viscous term $\mathcal{L}_{\text{visc}}$ }

Finally, the viscous term $\mathcal{L}_{\text{visc}}=\frac{1}{\rho^{n+1}}\nabla_f \cdot \left( \mu^{n+1} \mathbf{D}^n  \right)$ is discretized as (here for $u$)
\begin{align}
\mathcal{L}_{\text{visc}} = \frac{1}{\rho^{n+1}_u} \left( 2 \nabla_f^x \left( \mu^{n+1}\nabla^x_c u^n \right) + \nabla_c^y \left( \mu^{n+1}\nabla^y_f u^n \right) + \nabla_c^y \left( \mu^{n+1}\nabla^x_f v^n \right) \right)
\end{align}
in this expression, some $\mu^{n+1}$ values are needed at $\Sigma_\rho$ corners, they are defined with simple average of the neighbour collocated values 
\begin{equation}
\mu^{n+1}_{i-\frac{1}{2},j-\frac{1}{2}} = \frac{1}{4} \left( \mu^{n+1}_{i,j} + \mu^{n+1}_{i-1,j} + \mu^{n+1}_{i,j-1} + \mu^{n+1}_{i-1,j-1}  \right)
\end{equation}
Note that sharper choices of corner interpolation can be made based on a distance function~\cite{sussman2009stable} but no significant improvements have been observed in the test cases presented i this work.

\subsection{Correction step}
The correction step is written as 
\begin{equation}
\frac{\uvec^{n+1}-\uvec^*}{\Delta t} = -\frac{1}{\rho_f^{n+1}}\nabla_f P^{n+1}
\label{eq:PM_correc}
\end{equation}
where the pressure $P^{n+1}$ is obtained from the resolution of variable coefficient Poisson equation (here using the linear solver PETSc library \cite{petsc-efficient} with the algebric multigrid preconditioner BoomerAMG and a krylov-based GMRES method)
\begin{equation}
\nabla_c \cdot \left( \frac{1}{\rho_f^{n+1}} \nabla_f P^{n+1} \right) = \frac{1}{\Delta t}\nabla_c \cdot \uvec^*
\label{eq:PM_poisson}
\end{equation}
with $\rho_f^{n+1}=\rho_u^{n+1}$ for the x-normal faces ($\rho_v^{n+1}$ and $\rho_w^{n+1}$ for y and z-normal faces).

\section{Curvature computation}
\label{sec:kappa}
The most stringent aspect of a surface tension modelling is the computation of the curvature. This is done by using the interface information provided by the color function $c$. This section aims to give an overview of the literature works on the subject along with the choices made in our solver.

\subsection{Curvature from implicit surfaces}
A natural choice for curvature computation is the differentiation of the field $c$. In the context of implicit surfaces, Goldman \cite{Goldman2005} provided an exhaustive list of formulas for curvature computation.
From $c$, the normal pointing outward the liquid phase can be computed as
 \begin{equation}
 \nint = -\frac{\nabla c}{\lvert \nabla c \rvert}
 \end{equation}
The curvature is then deduced from one of the following expressions
\begin{equation}
\kappa = - \nabla \cdot \left(  \frac{\nabla c}{\lvert \nabla c \rvert} \right) = \frac{\text{tr}\left(\nabla\nabla c\right)-\nint \cdot \nabla\nabla c \cdot \nint^\intercal}{\lvert \nabla c \rvert} 
\end{equation}

As curvature depends on second order derivatives of the function $c$, one need to provide $c$ at least at third order to retrieve a consistent curvature using standard finite difference operators. More generally, a curvature of order $m$ is obtained using an implicit surface function of order $m+2$ \cite{Coquerelle2016}.
This limitation is important to keep in mind, as numerous strategies to reconstruct a distance function from an isocontour are of second order, which is not sufficient to ensure convergence of the curvature. This has been observed in a VOF context \cite{Cummins2005} with curvature computed from a second order reconstruced distance function (RDF). To deal with this issue, a least-square computation of operators has been presented in the context of curvature computation from an unstructured mesh \cite{Marchandise2007} or from a FMM reconstruction of a signed distance \cite{Chiodi2017}. The idea is to reduce the second order errors of the distance function by adding more points in the stencil. Even if the method leads to second order convergence in a coarse regime, the convergence is not retrieved for high resolutions as demonstrated in Sec.~\ref{sec:curvature}.

Another requirement is the smoothness of $c$ to avoid spurious higher order derivatives. With LS, this is not a problem, as $\phi$ is already a smooth well-defined function in all the domain. However, for VOF, the volume fraction is too sharp to provide non-oscillating curvature computation. Using directly the volume fraction $f$ leads to non-converging curvature. Various convolution methods were used to get a smoother version $\widetilde{f}$ (see \cite{williams2000numerical}) but even with 8$^{th}$ order convolution Kernel, curvature cannot reach mesh convergence. 

Considering the above discussion, differentiation seems to be a straightforward method for SLS, ACLS or CLSVOF. Simple second order finite differences are sufficient in the SLS case while a least-square approach is mandatory for ACLS and CLSVOF to have convergence. However, a more reliable method is needed for curvature from VOF, as $f$ can not be reasonably differentiated twice.

\subsection{Curvature from height function}
As an alternative, curvature can be deduced from height functions \cite{bornia2011properties}. 
If an interface is described by an height function $\hbar$ such that $y=\hbar(x)$ in 2D (or $z=\hbar(x,y)$ in 3D), the curvature can be deduced from 
\begin{align}
& \kappa = \frac{\hbar_{xx}}{(1+\hbar_x^2)^{\frac{3}{2}}}  & \mbox{in 2D } \\
& \kappa = \frac{\hbar_{xx}+\hbar_{yy}+h_{xx}\hbar_y^2+\hbar_{yy}\hbar_x^2-2\hbar_{xy}\hbar_x\hbar_y}{(1+\hbar_x^2+\hbar_y^2)^{\frac{3}{2}}} & \mbox{in 3D }
\label{eq:kappa_hf}
\end{align}
In \cite{Sussman2003}, an early method was proposed to compute height functions from $f$. In a cell $\mathcal{C}$, the dominant direction is determined from $\nint$. Then, an exact average height function $h_i$ can be derived  by simply summing all volume fraction of a column in this direction. For y-dominant direction, this is written as
\begin{equation}
h_i = \sum\limits_{j=-\infty}^{j=+\infty} f_{i,j}
\end{equation}
in this method $\infty$ is obviously not possible and replaced by 3~\cite{Sussman2003,popinet2009accurate,hernandez2008new}

The operators $h_x$ and $h_{xx}$ are then computed with second order accuracy using standard finite differences $\left.h_x\right\vert_i = \frac{h_{i+1}-h_{i-1}}{2\Delta x}$ and $\left.h_{xx}\right\vert_i = \frac{h_{i+1}-2h_i+h_{i-1}}{\Delta x^2}$. Finally, the curvature is retrieved with second order accuracy using Eq.~(\ref{eq:kappa_hf}). This method requires a $3\times3\times7$ stencil. However, the fixed stencil will not work in every interface configurations and lead to inconsistent curvature computations \cite{Cummins2005}.

To address this issue, each height can be constructed independently with a variable stencil which improves drastically the accuracy and robustness of the curvature computation \cite{popinet2009accurate,hernandez2008new}. Even with this improvement, height function fails to provide accurate curvature computation for low resolution \cite{popinet2009accurate}.

It is also possible to decouple height function from the mesh \cite{Liovic2010}. In \cite{Owkes2015}, the authors generalized this idea by computing height functions in a coordinate system orthonormal to the interface normal with column of parametrized width and depth. For under-resolved configuration, this has proved to give more accurate curvature compared to a mesh-aligned version. 

Finally, some attempts to apply height function in other context have been made in SLS \cite{Lopez2010} where the method leads to a less accurate curvature compared to height functions from VOF. A method has also been proposed for ACLS \cite{Owkes2013} however a stencil of 11 cells was required to achieve convergence. Thus, height function seems to be only interesting in the VOF framework.

In our solver, the use of standard height function with variable stencil is used instead of the mesh-decoupled method proposed in \cite{Owkes2015} to benefit from the Cartesian grid arrangement. With this choice, another computation method is required for under-resolved configuration following the idea in \cite{popinet2009accurate} detailed in the next section.

\subsection{Curvature from interface positions}
A last type of methods, mainly used in Front capturing methods, relies on marker positions to reconstruct an interface surfaces using a least-square approach \cite{Gois2008}. 
For the general case of an interface with normal $\mathbf{n}$ and a set of points describing the interface $\mathbf{x}_{\Gamma,i}$, one can fit a parabola $\mathcal{P}$ by using a least-square regression. 
\begin{align}
& \mathcal{P}(\mathbf{x}) = a_0 x^2 + a_1 x + a_2 & \mbox{in 2D} \\
& \mathcal{P}(\mathbf{x}) = a_0 x^2 + a_1 y^2 + a_2 xy + a_3 x + a_4 y + a_5 & \mbox{in 3D}
\end{align}

First, the set of points $\mathbf{x}_{\Gamma,i}$ is redefined as $\mathbf{x}'_{\Gamma,i}$ in a new coordinate system $[\mathbf{n},\mathbf{t}_1,\mathbf{t}_2]$ from the Cartesian coordinate system $[\mathbf{e}_x,\mathbf{e}_y,\mathbf{e}_z]$. \\
Then, the following least square minimization is performed 
\begin{equation}
R^2 = \sum\limits_{i=1}^N W_i(\mathbf{x}'_{\Gamma,i}) \left( z'_{\Gamma,i} - \mathcal{P}(x'_{\Gamma,i},y'_{\Gamma,i}) \right)^2
\end{equation}
with $W_i$ an optional weight to restrain the region of interest. \\
This resulting linear system requires at least $N=3$ (6 in 3D) to be well-conditioned. \\
Even if this method seems expensive and complex to implement,this is a good alternative to the mesh-decoupled method of \cite{Owkes2015} for robust under-resolved computations.
It has been used in a VOF context in \cite{Owkes2018} with positions from the PLIC reconstruction leading to strong improvement in under-resolved configurations compared to methods based on height function. In our solver, this parabola fit is coupled with height function to improve under-resolved robustness while keeping a second order convergence in high resolution as proposed in \cite{popinet2009accurate}.

\subsection{Curvature interpolation to the face}
Depending on the method, the curvature is computed either on the interface or at the cell center. In a MAC grid arrangement, those curvature values need to be interpolated to the face as $\kappa_f$. The way to do it depends on the location of the computed curvature.

\begin{figure}[h!]
	\centering
	\includegraphics[scale=0.5]{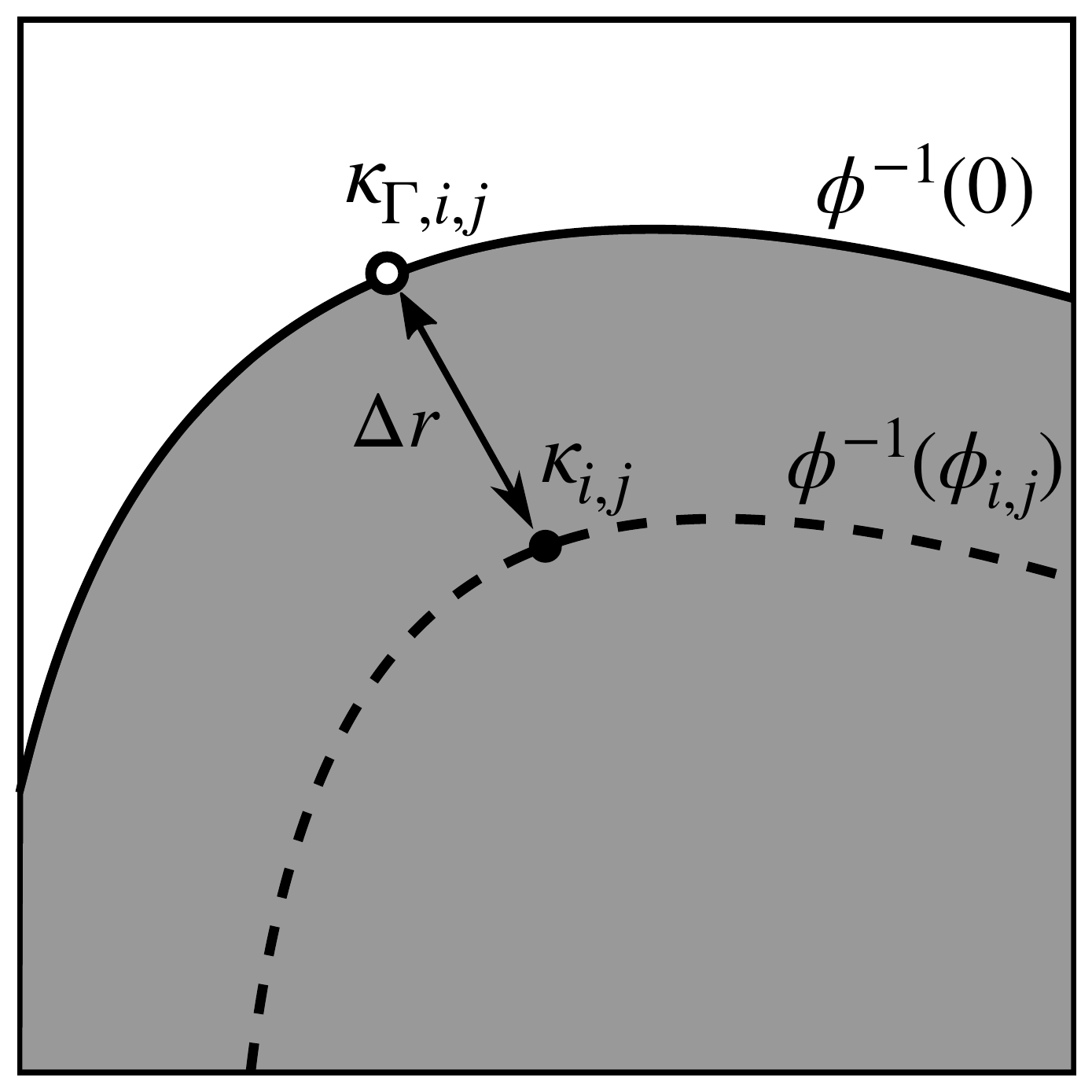}
	\caption{Position of a curvature computed from $\phi$ in a cell $\mathcal{C}_{i,j}$ with $\kappa_{i,j}$ the curvature defined at the cell center and $\kappa_{\Gamma,i,j}$ the curvature of the interface contained in the cell.}
	\label{fig:curvature_pos}
\end{figure}

When curvature is defined at the cell center ($\kappa_{i,j}$ in Fig.~\ref{fig:curvature_pos}) , the interpolation can be expressed as
\begin{align}
& \kappa_{u,i-\frac{1}{2},j} =  \frac{\kappa_{i,j}\lvert \phi_{i-1,j} \rvert + \kappa_{i-1,j} \lvert \phi_{i,j} \rvert}{\lvert \phi_{i,j} \rvert + \lvert \phi_{i-1,j} \rvert} & \mbox{for linear interpolation} \label{eq:kappalin}\\
& \kappa_{u,i-\frac{1}{2},j}=  \kappa_{i,j}  \kappa_{i-1,j}  \frac{\lvert \phi_{i,j} \rvert + \lvert \phi_{i-1,j} \rvert}{ \kappa_{i,j} \lvert\phi_{i,j} \rvert + \kappa_{i-1,j} \lvert \phi_{i-1,j} \rvert} & \mbox{for harmonic interpolation}
\label{eq:kappaharm}
\end{align}
The first one is the linear interpolation of the curvature while the second one is the linear interpolation of the radius. 
This type of interpolation can be seen as the interpolation of cell centered curvature to the interface position lying between those cells which can be different from the position of the face center. 

For a curvature directly defined at the interface ($\kappa_{\Gamma,i,j}$ in Fig.~\ref{fig:curvature_pos}), one need to average the curvatures of two adjacent cells which are not necessarily containing a curvature value. The most simple way to do it is an average of values if they both contain an interface value and keep the single available value if one of the two adjacent cells does not contain the interface
\begin{align}
& \kappa_{u,i-\frac{1}{2},j} = \frac{\kappa_{\Gamma,i-1,j}+\kappa_{\Gamma,i,j}}{2} & \mbox{for two adjacent interfacial cells} \\
& \kappa_{u,i-\frac{1}{2},j} = \kappa_{\Gamma,i,j} & \mbox{for one interfacial cell}
\end{align}
in a more general way, one can define the curvature at a cell face by weight interpolation~\cite{Renardy2002}
\begin{equation}
\kappa_{u,i-\frac{1}{2},j} =  \frac{W_{i-1,j}\kappa_{\Gamma,i-1,j}+W_{i,j}\kappa_{\Gamma,i,j}}{W_{i-1,j}+W_{i,j}} \\
\label{eq:kappaweight}
\end{equation}
with $W_{i,j} = H^\epsilon_{\Gamma,i,j}(1-H^\epsilon_{\Gamma,i,j})$. This gives importance to cell containing a large portion of interface ($H^\epsilon_\Gamma$ close to 0.5) and cancel quasi empty or full cells ($H^\epsilon_\Gamma$ close to 0 or 1) which are more prompt to curvature computation errors.

\subsection{Curvature in the solver}
From the above discussion, the following choices has been made in our solver : 
\begin{itemize}
\item For SLS, the curvature is retrieved by using second order finite differences (FD) on $\phi$ while the Least-Square minimization (LSQUAD) of \cite{Chiodi2017} is used on $\phi_{FMM}$ and $\phi_{PLIC}$ for ACLS and CLSVOF respectively. The curvature is then defined at the cell center and the harmonic interpolation Eq.~(\ref{eq:kappaharm}) is used to get face curvatures. For robustness improvement, the interpolation switches to linear interpolation Eq.~(\ref{eq:kappalin}) when curvature changes sign as in \cite{Chiodi2017}. 
\item For VOF, the curvature is computed from the general height function (HF) method \cite{popinet2009accurate} and interpolated to the face using Eq.~(\ref{eq:kappaweight}).
\end{itemize}

\section{Summary of the two-phase solver procedure}
A summary of the adaptations introduced to unify all Eulerian methods in the same low Mach solver is provided in Tab.~\ref{tab:solver_meth} and the full algorithm for a time step is given. All the implementations are integrated into the TITAN solver developed at EM2C laboratory. 
\begin{enumerate}
\item Compute $\rho_f^n$ from $c^n$;
\item Advance interface following one of the algorithm of Sec.~\ref{sec:interface} to obtain $c^{n+1}$;
\item Compute $\kappa_f^{n+1}$, $H_\Gamma^{0,n+1}$, $\rho_f^{n+1}$ from $c^{n+1}$;
\item Advance $\rho_f^*$ and $\uvec^*$ with the consistent mass and momentum scheme of Sec.~\ref{sec:momcons};
\item Solve the Pressure from Eq.~(\ref{eq:PM_poisson});
\item Correct the velocity to obtain $\uvec^{n+1}$ with Eq.~(\ref{eq:PM_correc});
\end{enumerate}
The time integration of steps 4 to 6 is performed using a RK2 SSP scheme. The timestep has a stability constraint based on the CFL, the surface tension and the viscosity :
\begin{equation}
\Delta t < \min \left( \frac{\Delta x}{2\lVert \uvec \rVert}, \sqrt{\frac{\Delta x^3 (\rho_l+\rho_g)}{(2\pi)^3 \sigma}}, \frac{\Delta x^2}{4 \max(\nu_l,\nu_g)}\right)
\end{equation}
in most of the application shown hereafter, the surface tension restriction is dominant.
\begin{table}[h!]
\center
\begin{tabular}{ |l||c|c|c|c|  }
 \hline
 method & VOF & SLS & ACLS & CLSVOF \\
 \hline
 $\kappa$   &  HF  & FD & LSQUAD & LSQUAD\\
 $\kappa_f$ & From Eq.~(\ref{eq:kappaweight}) & From Eq.~(\ref{eq:kappaharm}) & From Eq.~(\ref{eq:kappaharm}) & From Eq.~(\ref{eq:kappaharm})\\
 $H_\Gamma^\epsilon$ & $f$ & From Eq.~(\ref{eq:heps_sls}) & $\psi$ & $f$\\
 $\rho_f$ & From PLIC & From Eq.~(\ref{eq:slsrho}) &  From Eq.~(\ref{eq:clsrho}) & From PLIC \\
 \hline
 \end{tabular}
 \caption{Summary of computation choices for the four interface capturing method in the Solver}
 \label{tab:solver_meth}
 \end{table}
 
\section{Results}
The results presented in this section are organized as follows. First, the assessment of interface capturing methods is performed with imposed velocity field on 2D and 3D test cases with emphasis on mass conservation. Then, the unified solver is evaluated on momentum conservation and surface tension modelling using well-known test cases of the literature. Finally, a two applications are presented to demonstrate accuracy and robustness of the unified framework on representative two-phase flow configurations.

\subsection{Interface capturing}
\subsubsection{Error metrics}
\label{sec:case_adv}
The evaluation of the different methods is based on accuracy and mass conservation. 
To provide an equal base of comparison for accuracy, the shape error $E_{shape}$ is based on the regularized Heaviside function $H_\Gamma^\epsilon(c)$. They all are smooth versions of $\chi$ with an interface thickness of $2\epsilon=\Delta x$ such that $H_\Gamma^\epsilon(c) \underset{\Delta x \to 0}{\longrightarrow} \chi$. 
The accuracy error is then defined as
\begin{equation}
E_{shape} = \sum_{i=1}^{N_\mathcal{C}} \lvert H_\Gamma^\epsilon \left( c_{i,T} \right) - H_\Gamma^\epsilon \left( c_{i,0} \right) \rvert \mathcal{V}_i
\end{equation}
with $t= 0$ the initial time of the simulation, $t=T$ the final time, $N_\mathcal{C}$ the number of cells in the computational domain and $\mathcal{V}_i$ the volume of the cell $\mathcal{C}_i$. \\
Regarding mass conservation, VOF  and CLSVOF achieve it at machine precision and will not be displayed. The ACLS method conserves $\psi$ up to machine precision, however it does not correspond exactly to the volume enclosed in the $0.5$ isocontour. This is why a simplex decomposition is performed in each cell in order to find the intersections between the cell and the interface and compute the related volume. This method leads to a second order approximation of a volume enclosed in a given isocontour~\cite{min2008robust}. This same approach is performed for enclosed volume in the $0$ isocontour of SLS.
The mass error is then defined as 
\begin{equation}
E_{mass} = \frac{1}{V_0T}\int_{0}^T \lvert \Delta V \rvert dt
\end{equation}
with $\Delta V=V(t+dt)-V(t)$ the variation of liquid volume computed from the simplex decomposition method evaluated at time $t$ and $t+dt$. Compared to a more classical measure of mass loss based on the difference between initial and final volume $V_0$ and $V_T$, this metric gives a better overview of the mass variation of a method with respect to time. 

\subsubsection{Zalesak's disk rotation}
The Zalesak's disk \cite{zalesak1979fully} test case consists in a notched circle of radius $0.15$ initially centered at $\left( 0.5,0.75\right)$ in a $[1\times1]$ domain. The notched width is $0.05$ and notched length is $0.25$. 
The velocity field is a solid rotation defined as 
\begin{equation}
\mathbf{u} =  \left(
    \begin{array}{ll}
    2\pi(0.5-y) \\
    2\pi(x-0.5)
    \end{array}\right)
\end{equation} 
The results are given for a full rotation of the disk corresponding to a simulation time $T=1$ for a CFL number of $0.5$. \\

\begin{figure}[h!]
\centering
	 \includegraphics[width=\textwidth]{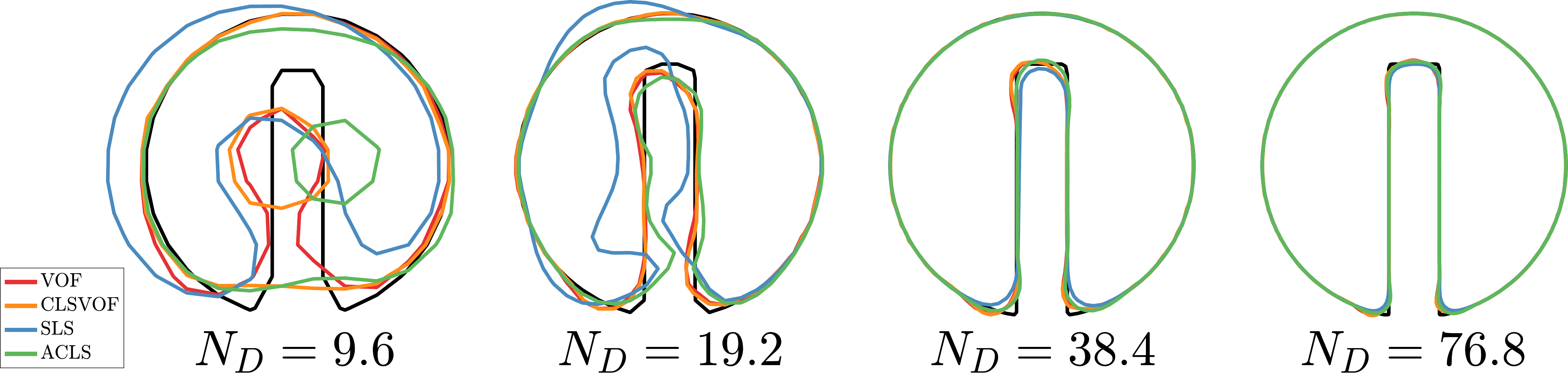}
     \caption{Initial and final shape for the Zalesak's disk rotation}
     \label{fig:zalesak_shape}
\end{figure}

The final shape is compared with the initial shape for all methods in Fig.~\ref{fig:zalesak_shape}. At the lowest resolution $32^2$, VOF keeps the notch while SLS is shifted. ACLS and CLSVOF merge the two sides of the notch. This shows the difference of normal computation between VOF and CLSVOF : while ELVIRA is able to capture poorly-resolved structures, normals from $\phi$ tend to merge fronts. From $64^2$ resolution all methods maintain the notch during the whole computation. 

\begin{figure}[h!]
\centering
     \begin{subfigure}[b]{0.48\textwidth}
     	\centering
         \includegraphics[width=\textwidth]{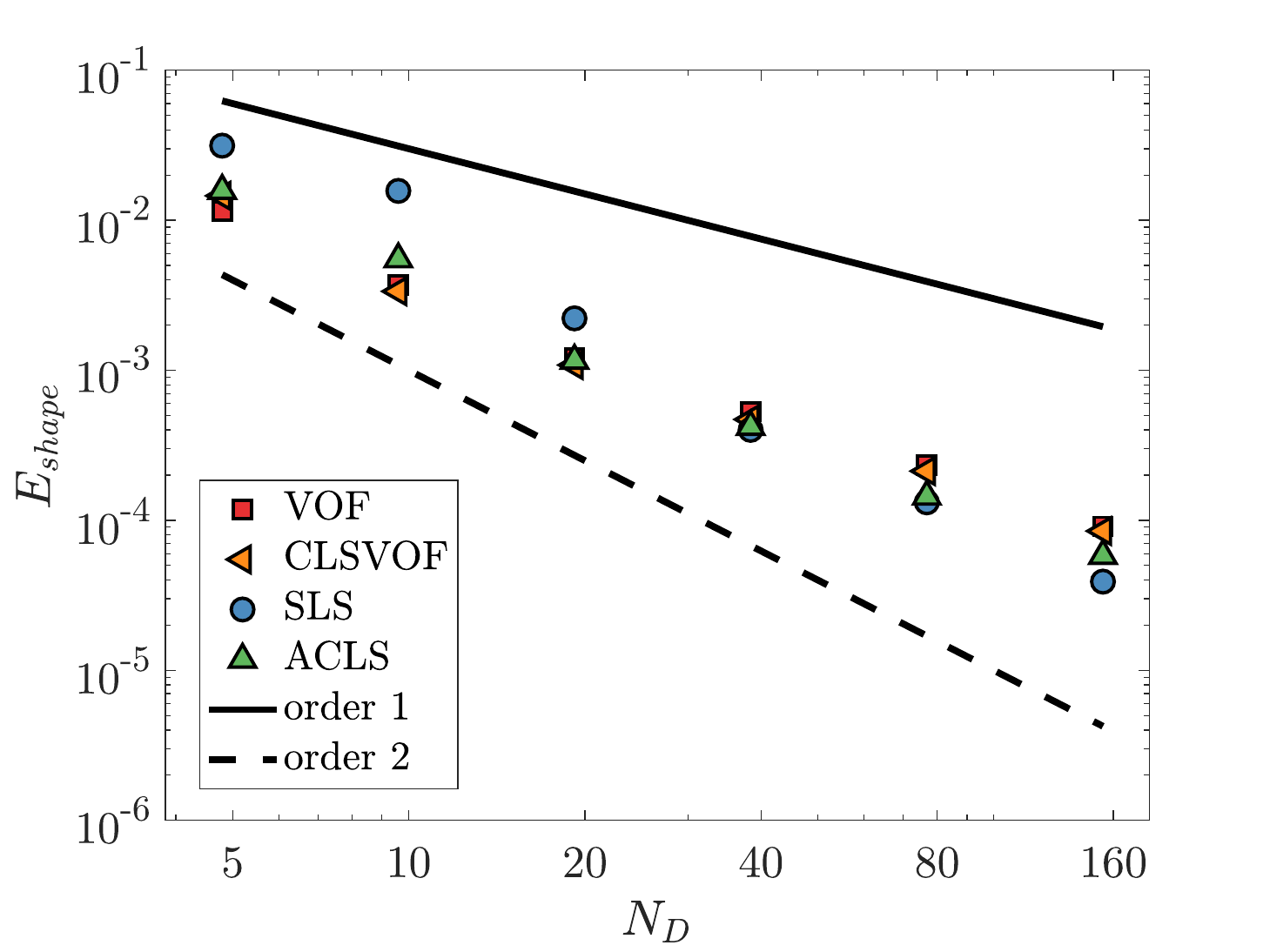}
    		\caption{$E_{shape}$ error}
    		\label{fig:zalesak_eshape}
     \end{subfigure}
     \hfill
     \begin{subfigure}[b]{0.48\textwidth}
     	\centering
         \includegraphics[width=\textwidth]{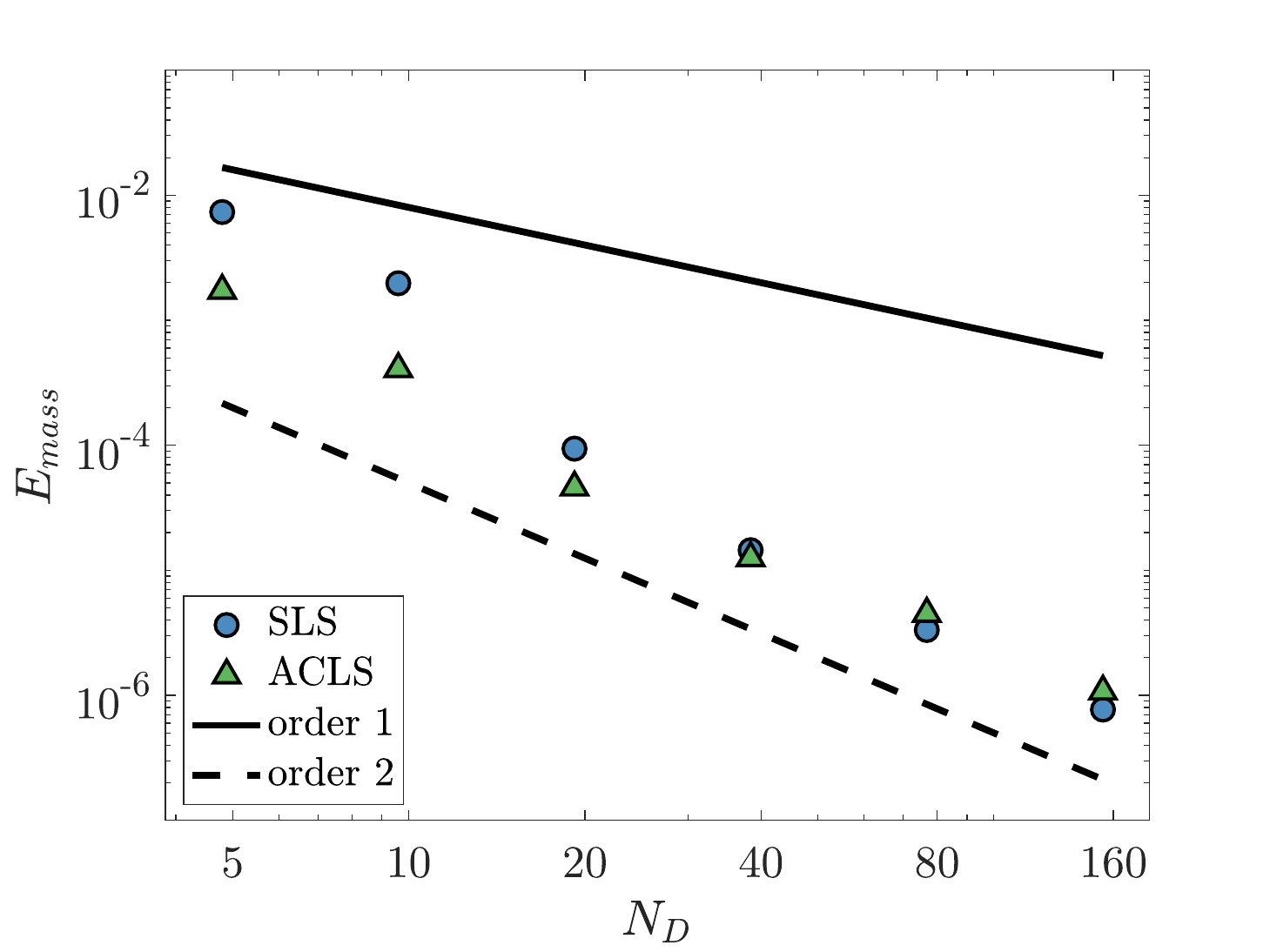}
    		\caption{$E_{mass}$ error}
    		\label{fig:zalesak_emass}
     \end{subfigure}
     \caption{Mesh error convergence for the zalesak's disk rotation}
\end{figure}

In Fig.~\ref{fig:zalesak_eshape}, the error convergence is displayed for all methods. One can notice that VOF, ACLS and CLSVOF perform well even at very low resolution while SLS and ACLS perform better for high resolution with an asymptotic second order behaviour. Regarding mass conservation, ACLS is better than SLS for the low resolution meshes while they both have the same conservation properties for the highest resolution. As pictured in Fig.~\ref{fig:zalesak_emass}.

\subsubsection{Vortex in a box}
Another classical test case is the vortex-in-a-box first used by Leveque to evaluate high order advection schemes in incompressibles flows  \cite{leveque1996high}. A circle of radius $0.15$ is initially centered at $(0.5,0.75)$ in a $[1\times1]$ domain. The velocity field is deduced from the stream function $\Psi = \frac{1}{\pi}\sin^2( \pi x ) \sin^2(\pi y) \cos \left( \pi \frac{t}{T} \right)$ such that it is reversed at $t=T/2$.
The results are given for the final time $T=8$ for an initial CFL number of $0.32$ ($\Delta t$ is kept constant for the whole simulation). \\

\begin{figure}[h!]
\centering
	\includegraphics[width=\textwidth]{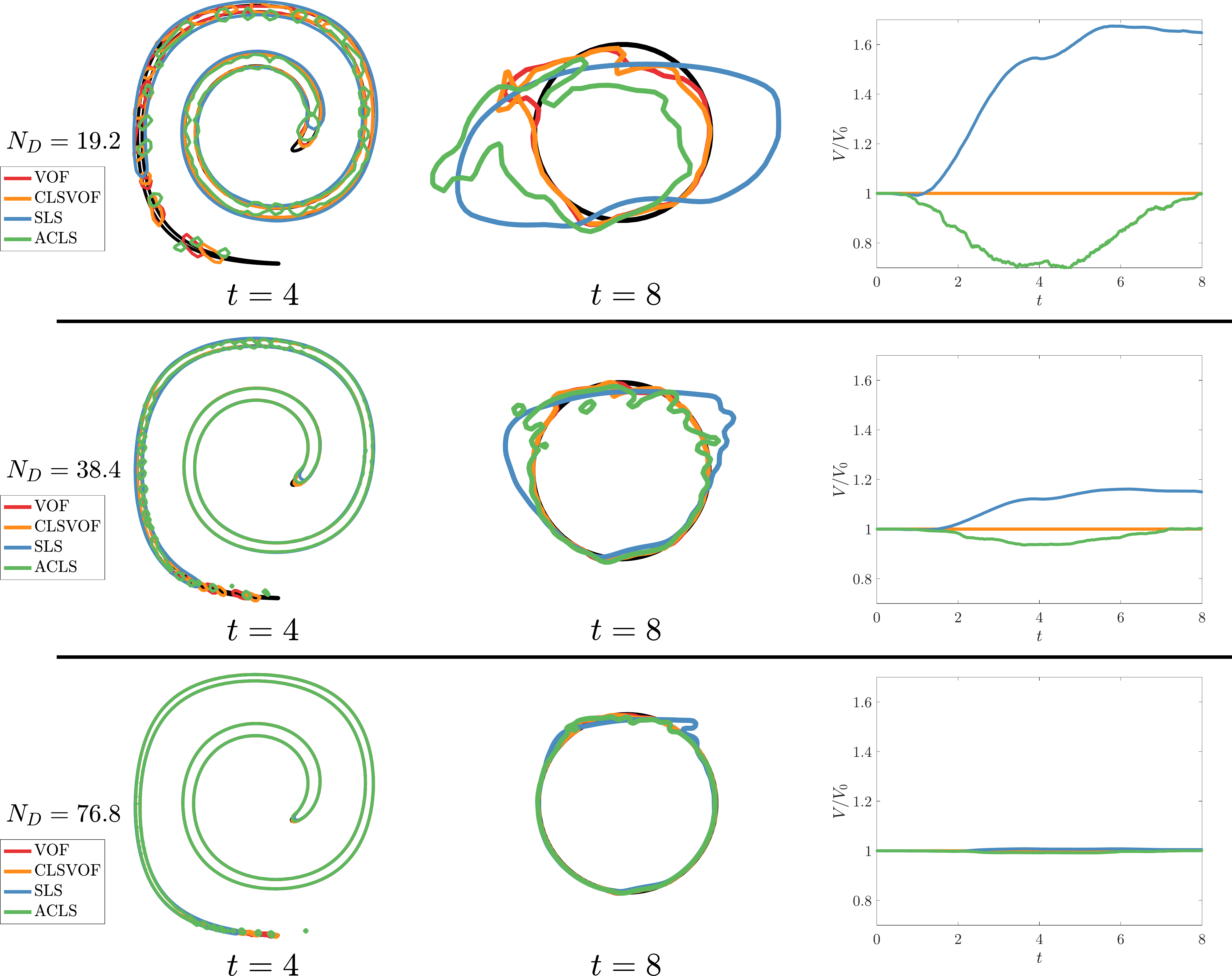}
     \caption{Vortex in a box shape at $t=4$ and $t=8$ with the temporal mass evolution}
     \label{fig:vortex_shape}
\end{figure}
In Fig.~\ref{fig:vortex_shape}, VOF, CLSVOF and ACLS tend to produce numerical atomization in the thinner structures of the serpentine while the SLS shows a more robust behaviour at the cost of mass conservation. This numerical atomization is less predominant with mesh refinement. 

\begin{figure}[h!]
\centering
     \begin{subfigure}[b]{0.48\textwidth}
     	\centering
         \includegraphics[width=\textwidth]{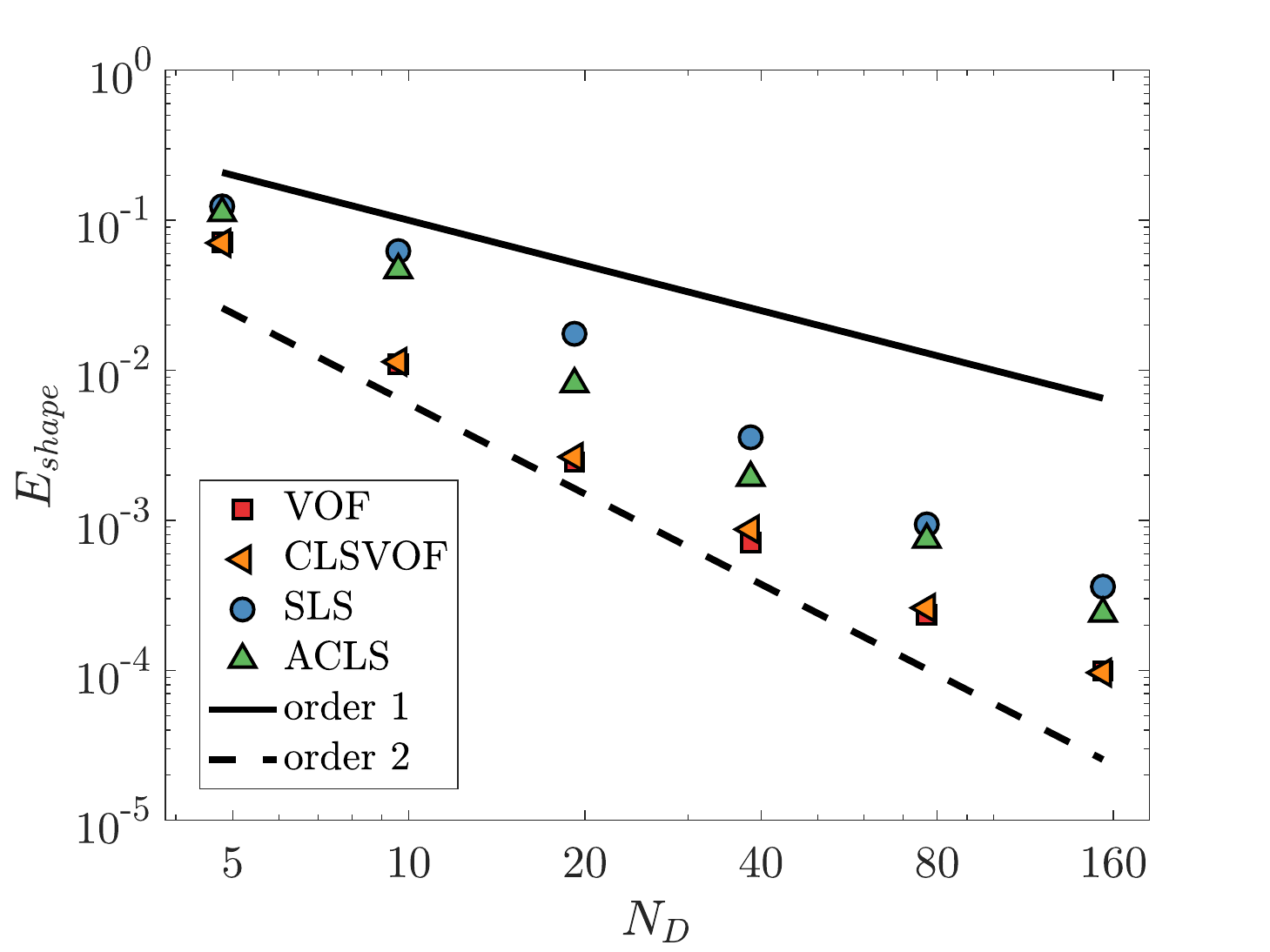}
    		\caption{$E_{shape}$ error}
    		\label{fig:vortex_eshape}
     \end{subfigure}
     \hfill
     \begin{subfigure}[b]{0.48\textwidth}
     	\centering
         \includegraphics[width=\textwidth]{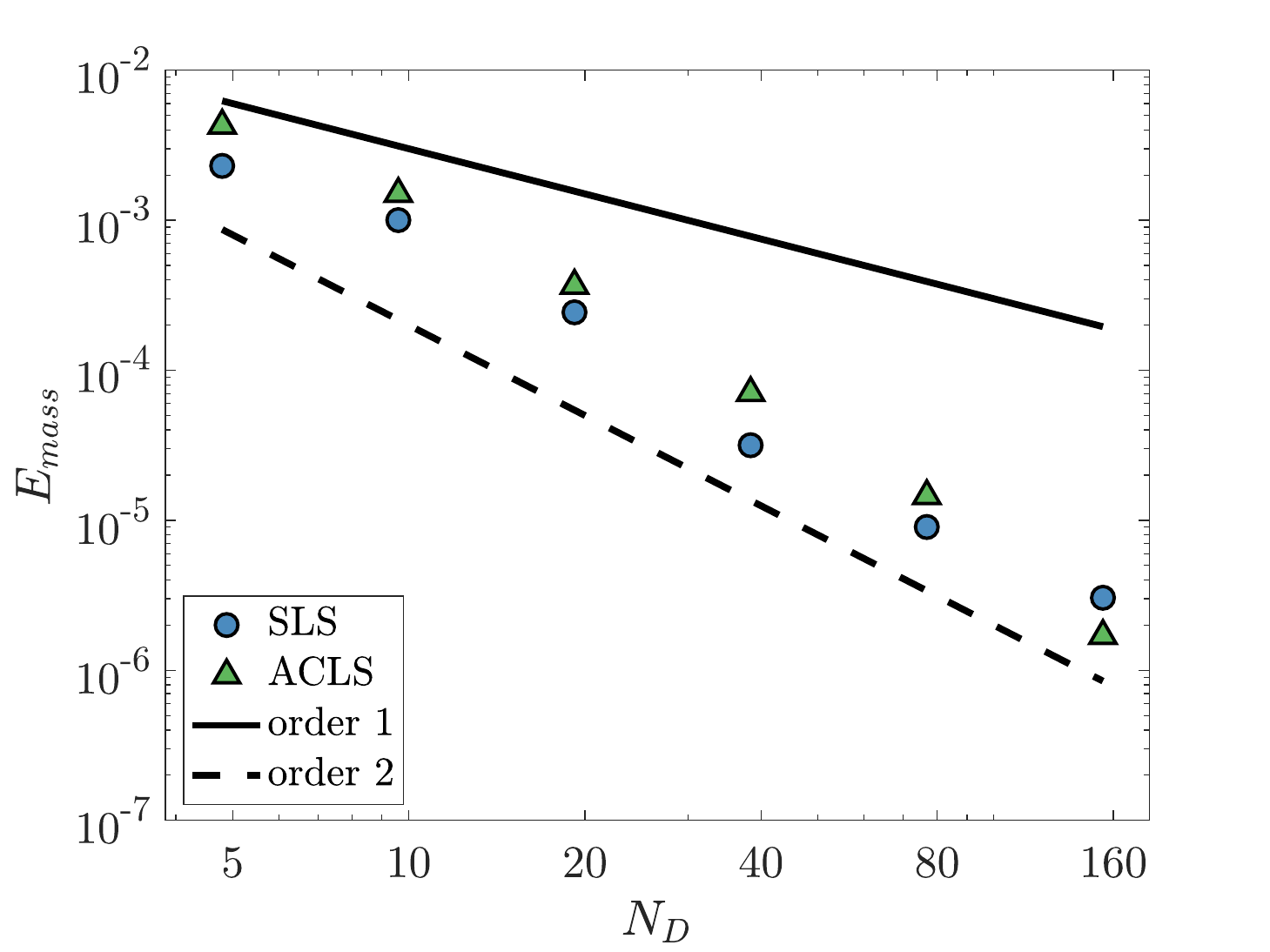}
    		\caption{$E_{mass}$ error}
    		\label{fig:vortex_emass}
     \end{subfigure}
     \caption{Mesh error convergence for the vortex in a box}
\end{figure}

In Fig.~\ref{fig:vortex_eshape}, shape and mass errors are displayed as a function of initial disk resolution. VOF and CLSVOF are performing better for all resolutions, this is expected as the $N_d=153.6$ case still implies a thin tail which is not well resolved. It is interesting to notice that the CLSVOF does not improve significantly the accuracy of the method compared to VOF. 
Surprisingly, SLS is better at conserving mass than ACLS based on our total volume variation metric. However, the ACLS method is able to retrieve a final mass close to the initial one which is not the case for SLS.

\subsubsection{Sphere deformation}
\label{sec:spheredef}
A 3D test case is the sphere deformation, also presented in \cite{leveque1996high}. A sphere of radius $0.15$ is initially centered at $(0.35,0.35,0.35)$ in a $[1\times1\times1]$ domain. It is then advected by a velocity field which induces a combination of stretching in the x-y plane and the x-z plane with an inversion at $t=T/2$.
The results are given for the final time $T=3$ for an initial CFL number of $0.32$ ($\Delta t$ is kept constant for the whole simulation). \\

\begin{figure}[h!]
	 \includegraphics[width=\textwidth]{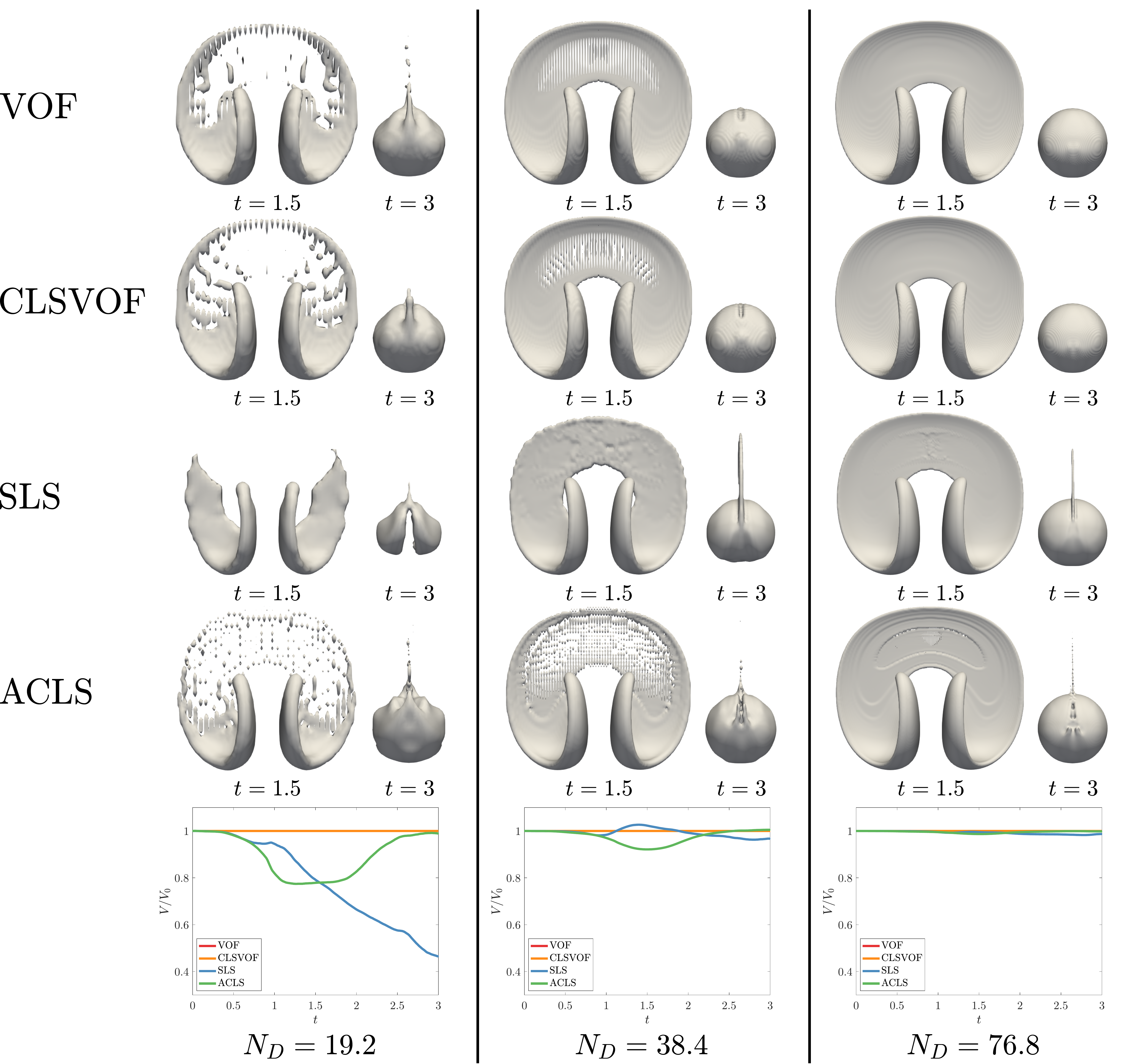}
     \caption{Shape at $t=1.5$ and $t=3$ for the sphere deformation}
     \label{fig:deformation_shape}
\end{figure}
The same conclusions as in for the vortex-in-a-box can be drawn from this 3D test case : VOF, CLSVOF and ACLS produce some numerical atomization when the interface is under-resolved as shown in Fig.~\ref{fig:deformation_shape} for $t=1.5$ at a low mesh resolution of $N_D = 9.6$. This numerical atomization disappears with mesh refinement. ACLS and SLS seems to lose a lot of mass for $N_D = 9.6$ even if ACLS is able to earn back the mass it has lost during the reversed part of the simulation. All methods exhibit a thin tail on the sphere at $t=T$ which represents less and less mass with mesh refinement even if it is still present for SLS and ACLS at $N_D=38.4$. CLSVOF seems to handle this behaviour in the best way for the smallest resolutions.

\begin{figure}[h!]
\centering
     \begin{subfigure}[b]{0.48\textwidth}
     	\centering
         \includegraphics[width=\textwidth]{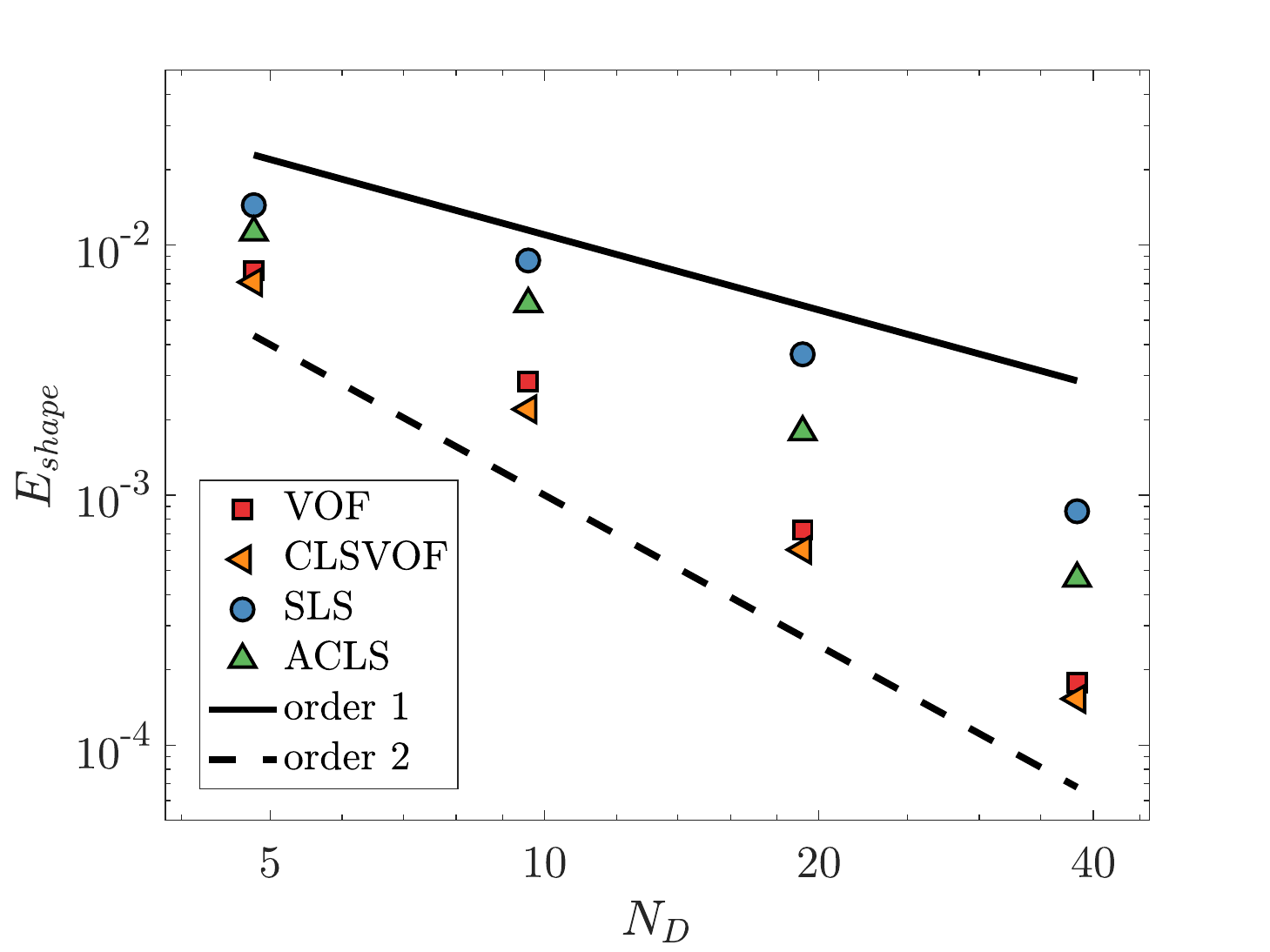}
    		\caption{$E_{shape}$ error}
    		\label{fig:deformation_eshape}
     \end{subfigure}
     \hfill
     \begin{subfigure}[b]{0.48\textwidth}
     	\centering
         \includegraphics[width=\textwidth]{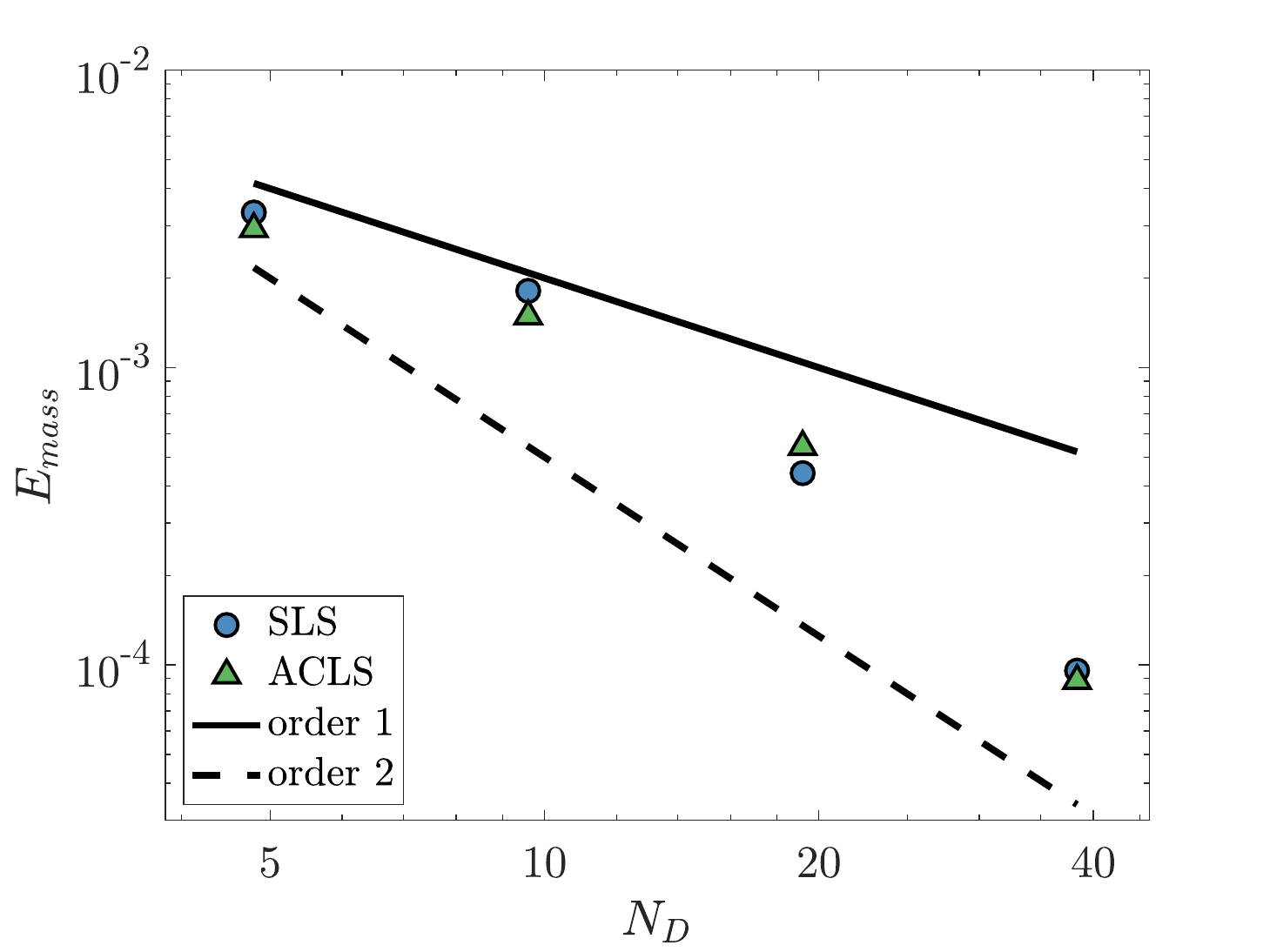}
    		\caption{$E_{mass}$ error}
    		\label{fig:deformation_emass}
     \end{subfigure}
     \caption{Mesh error convergence for the sphere deformation}
\end{figure}

From Fig.~\ref{fig:deformation_eshape}, VOF and CLSVOF are still the most accurate methods while ACLS and SLS show similar mass conservation in Fig.~\ref{fig:deformation_emass}. Apparently, the transition from 2D to 3D does not affect the overall behaviour of the methods. One slight improvement can be noticed by using CLSVOF for normal computation in 3D. This can be explained by the ELVIRA accuracy falling behind in 3D configuration if a compact stencil of $3\times3\times3$ is used. It has been shown that second order accurate normal computation is only achieved with a stencil of $5\times5\times5$ \cite{Miller2002}.

\subsubsection{Computational time}
\begin{figure}[h!]
\centering
     \begin{subfigure}[b]{0.48\textwidth}
     	\centering
         \includegraphics[width=\textwidth]{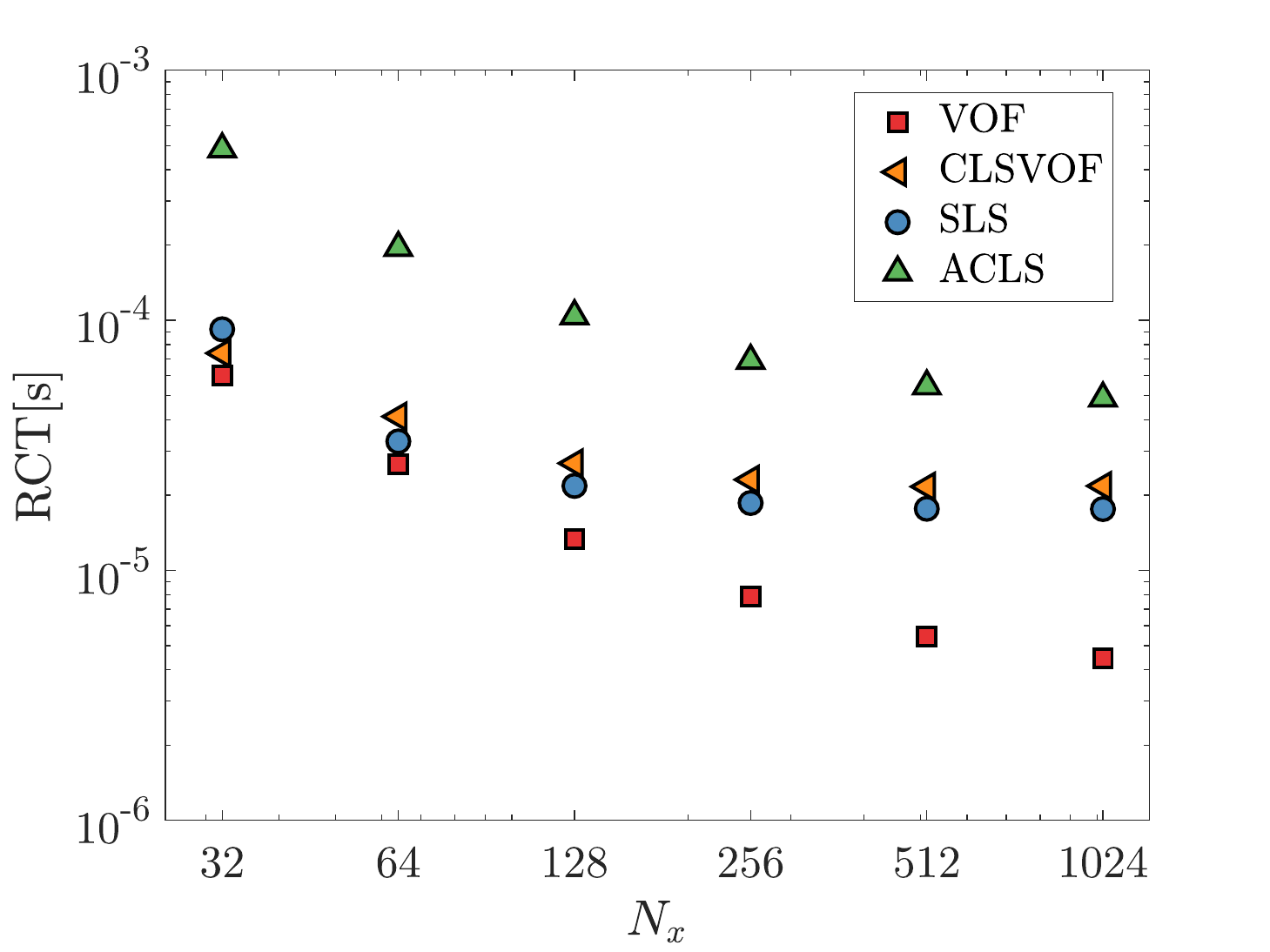}
    		\caption{Vortex in a box}
    		\label{fig:perfo_vortex}
     \end{subfigure}
     \hfill
     \begin{subfigure}[b]{0.48\textwidth}
     	\centering
         \includegraphics[width=\textwidth]{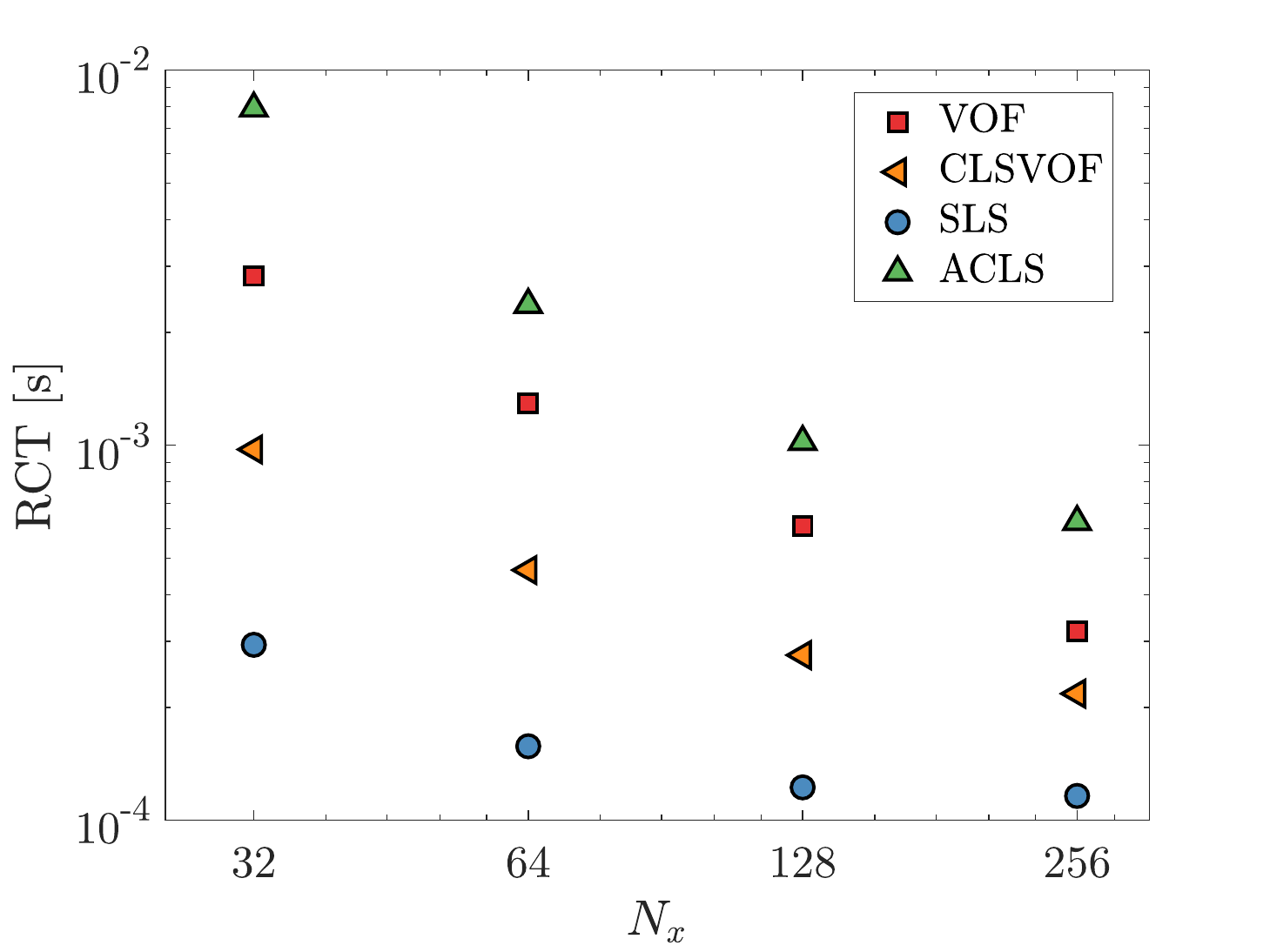}
    		\caption{Sphere deformation}
     \end{subfigure}
     \caption{RCT for 2D and 3D cases}
     \label{fig:perfo}
\end{figure}

To complete the comparison, the computational cost is compared between the methods. In Fig.~\ref{fig:perfo}, the Reduced Computation Time (RCT) is given
\begin{equation}
\mbox{RCT} = \frac{\mbox{WCT}\times N_{CPU}}{N_{\mathcal{C}}\times N_{ite}}
\end{equation}
with$N_\mathcal{C}$ the number of cell, $N_{CPU}$ the number of cores, WCT the Wall Clock Time and $N_{ite}$ the number of iterations .

The 2D vortex in a box case was run on 16 cores while the 3D deformation case was run on 64 cores. Cores used in this work are Intel Xeon Gold 6230 20C 2.1GHz.

Here, the RCT decreases as the number of elements per cores increases. This is expected as for a very coarse mesh, a more important part of the simulation time is lost in communications. The cases with more elements show an asymptotic behaviour with a constant RCT. \\
In 2D configurations, the VOF method is more efficient than CLSVOF, SLS and ACLS and seems to scale better with number of elements. This is because VOF method only requires to compute fluxes and reconstruction on the interface cells and their neighbours, hence the computational time does not scale in $N_{elem}$ but in $N_\Gamma$. Also, 2D computation of geometric flux and PLIC reconstruction is really fast in a split fashion. As expected, SLS is more efficient than ACLS because of the reinitialization which is more demanding in the case of ACLS. The CLSVOF method costs approximately VOF and SLS combined. \\
In 3D, geometry operations are more expensive, and VOF falls behind CLSVOF and SLS in efficiency.  CLSVOF  is more effective than VOF because of the normal computation which is far less expensive than ELVIRA in 3D.

\subsection{Momentum conservation}
\label{sec:mom_cons}
Now that interface capturing methods have been compared on imposed velocity fields, the coupling with the two-phase solver is explored. To assess the momentum conservation of our solver, the classic density ball test case introduced in \cite{bussmann2002modeling} is presented where a 2D droplet of radius $0.1$ and density $10^6$ is translating at a velocity $u=1$ m/s in a unity density field at rest. The density is high enough to consider the transport as a pure solid translation. The error $E_{shape}$ of Sec.~\ref{sec:case_adv} is used here to quantify the interface transport. A new error $E_{tke}$ is introduced to evaluate momentum conservation with the same formalism as $E_{mass}$
\begin{equation}
E_{tke} = \frac{1}{K_0T}\int_{0}^T \lvert \Delta K \rvert dt
\end{equation}
with $\Delta K= K(t+dt)- K(t)$ the variation of kinetic energy computed at time $t$ and $t+dt$. With $K$ computed as in \cite{Zuzio2020}
\begin{equation}
K = \frac{1}{2}\sum_{i=1}^{N_\mathcal{C}}  \rho_i \lVert \uvec_i \rVert ^2 \mathcal{V}_i
\end{equation}
with $\rho_i=\rho_g+H_{\Gamma,i}^\epsilon \jump{\rho}$ and $\uvec_i$ components defined as average of face velocities.

\begin{figure}[h!]
\centering
	 \includegraphics[width=\textwidth]{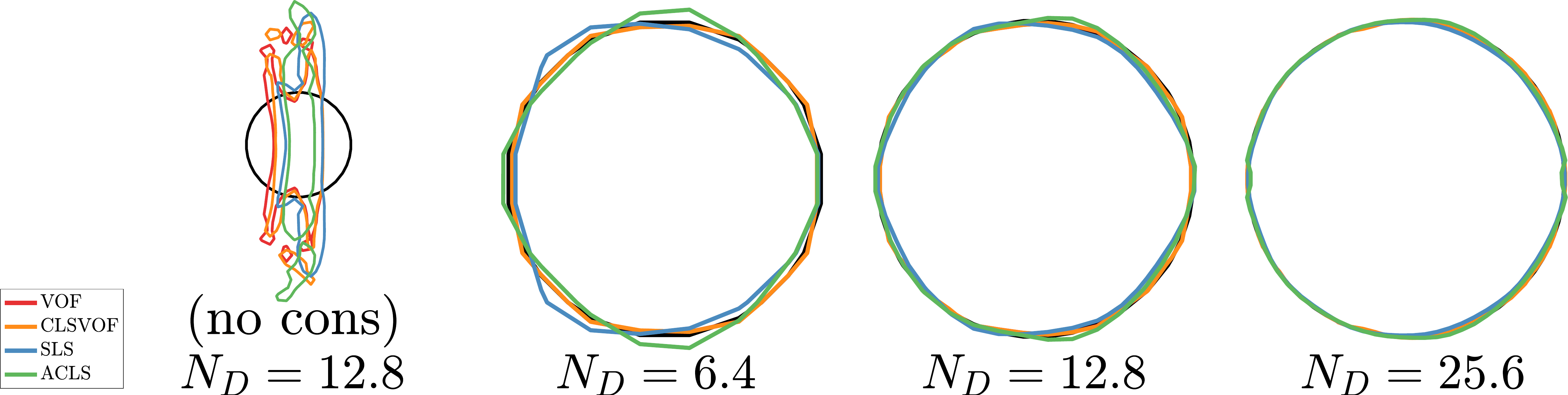}
     \caption{Initial and final shape for the density ball translation, the first image at the left shows a case without the consistent scheme while the other images are zoomed on the circle shape}
     \label{fig:density_shape}
\end{figure}

Fig.~\ref{fig:density_shape} displays the shape for different meshes. A non-conservative form of the momentum transport has been added for completeness where the shape is not conserved at all. For all other simulations, using the momentum fix, the initial circular shape is well preserved by the VOF and CLSVOF approaches even for very low resolution. This can be explained by the density approximation based on PLIC which is more accurate and less diffusive than Eq.~(\ref{eq:clsrho}). However, the circle is less distorted than the one presented in \cite{Desjardins2010} where Eq.~(\ref{eq:slsrho}) was used coupled with a ACLS approach.

\begin{figure}[h!]
\centering
     \begin{subfigure}[b]{0.48\textwidth}
     	\centering
         \includegraphics[width=\textwidth]{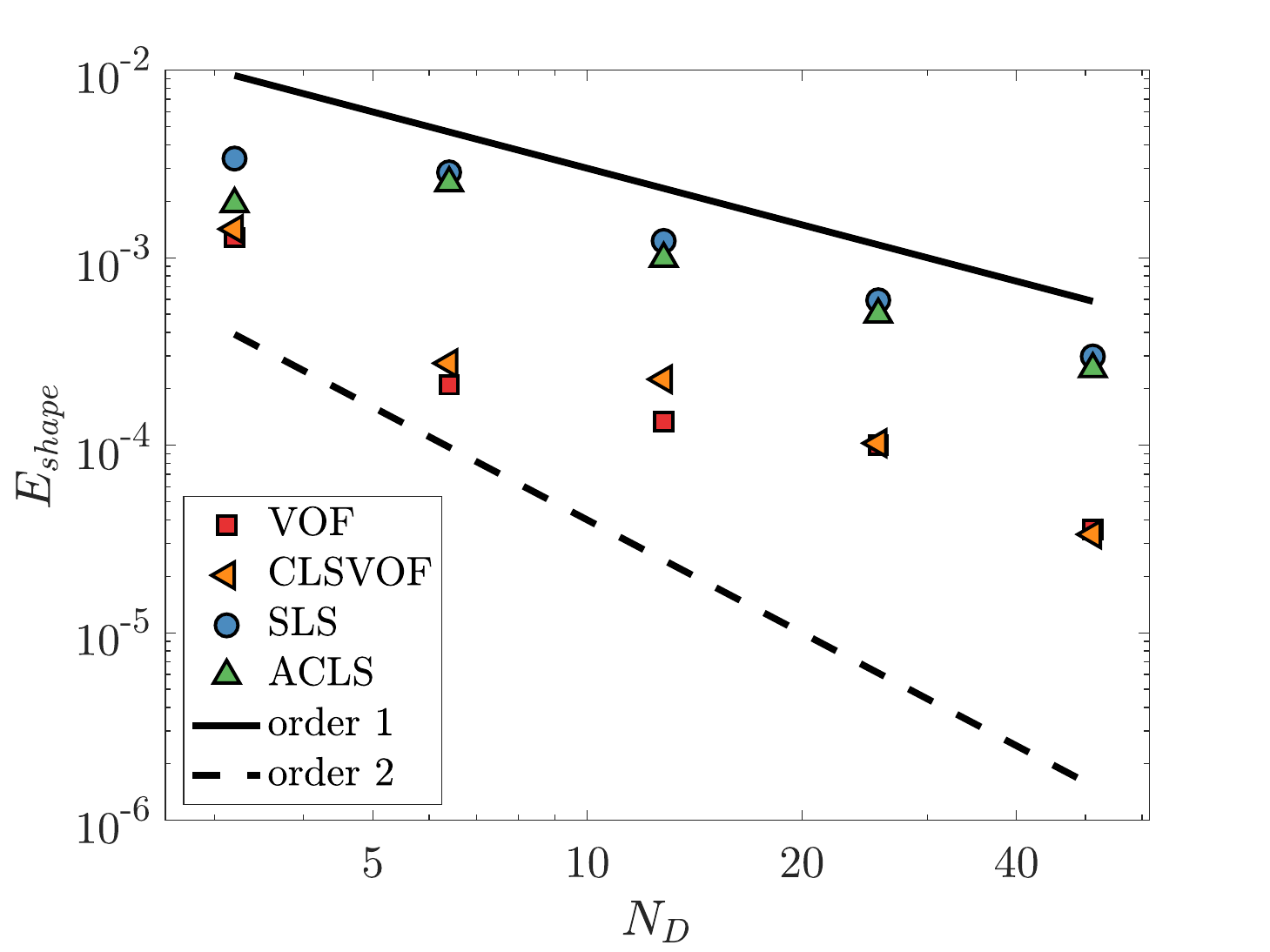}
    		\caption{$E_{shape}$ error}
    		\label{fig:density_eshape}
     \end{subfigure}
     \hfill
     \begin{subfigure}[b]{0.48\textwidth}
     	\centering
         \includegraphics[width=\textwidth]{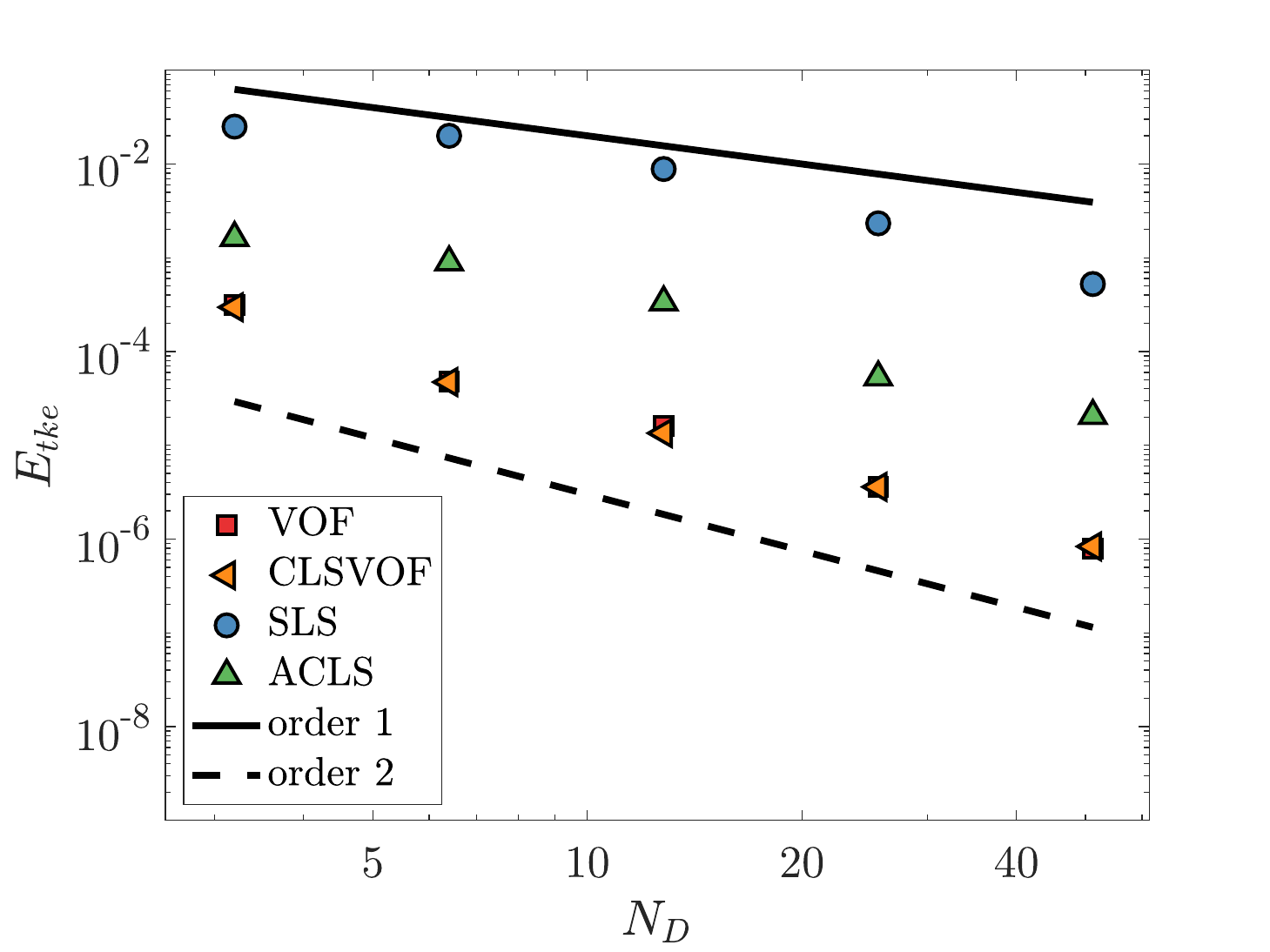}
    		\caption{$E_{tke}$ error}
    		\label{fig:density_etke}
     \end{subfigure}
     \caption{Mesh error convergence for the density ball translation}
\end{figure}

When looking at more quantitative metrics, all methods display a convergence rate between 1 and 2 for $E_{shape}$ in Fig.~\ref{fig:density_eshape} with a better accuracy for VOF and CLSVOF. This shape error is explained by the difference in momentum conservation. A huge difference of two order of magnitude is observed between VOF and SLS for $E_{tke}$ in Fig.\ref{fig:density_etke}. It is also interesting to notice that ACLS is more conservative than SLS for momentum too.  This illustrates how momentum conservation is impacted by the choice of $\rho_u^n$ computation and mass conservation.

\subsection{Surface tension modelling}
\begin{figure}[h!]
\centering
	\begin{minipage}{0.48\textwidth}
     \begin{subfigure}[b]{\textwidth}
     	\centering
         \includegraphics[width=0.87\textwidth]{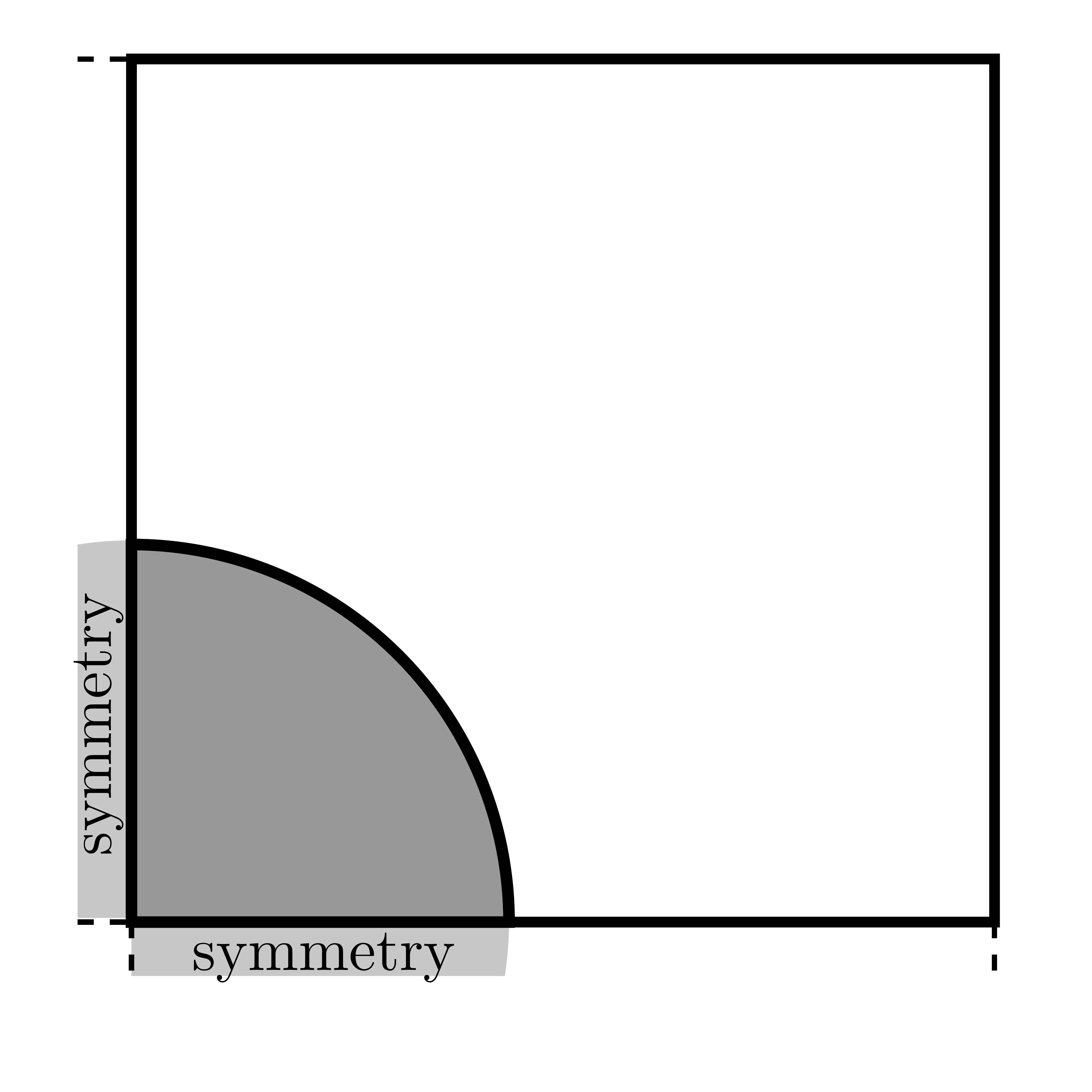}
    		\caption{Static test case}
     \end{subfigure}
     \hfill
     \begin{subfigure}[b]{\textwidth}
     	\centering
         \includegraphics[width=0.87\textwidth]{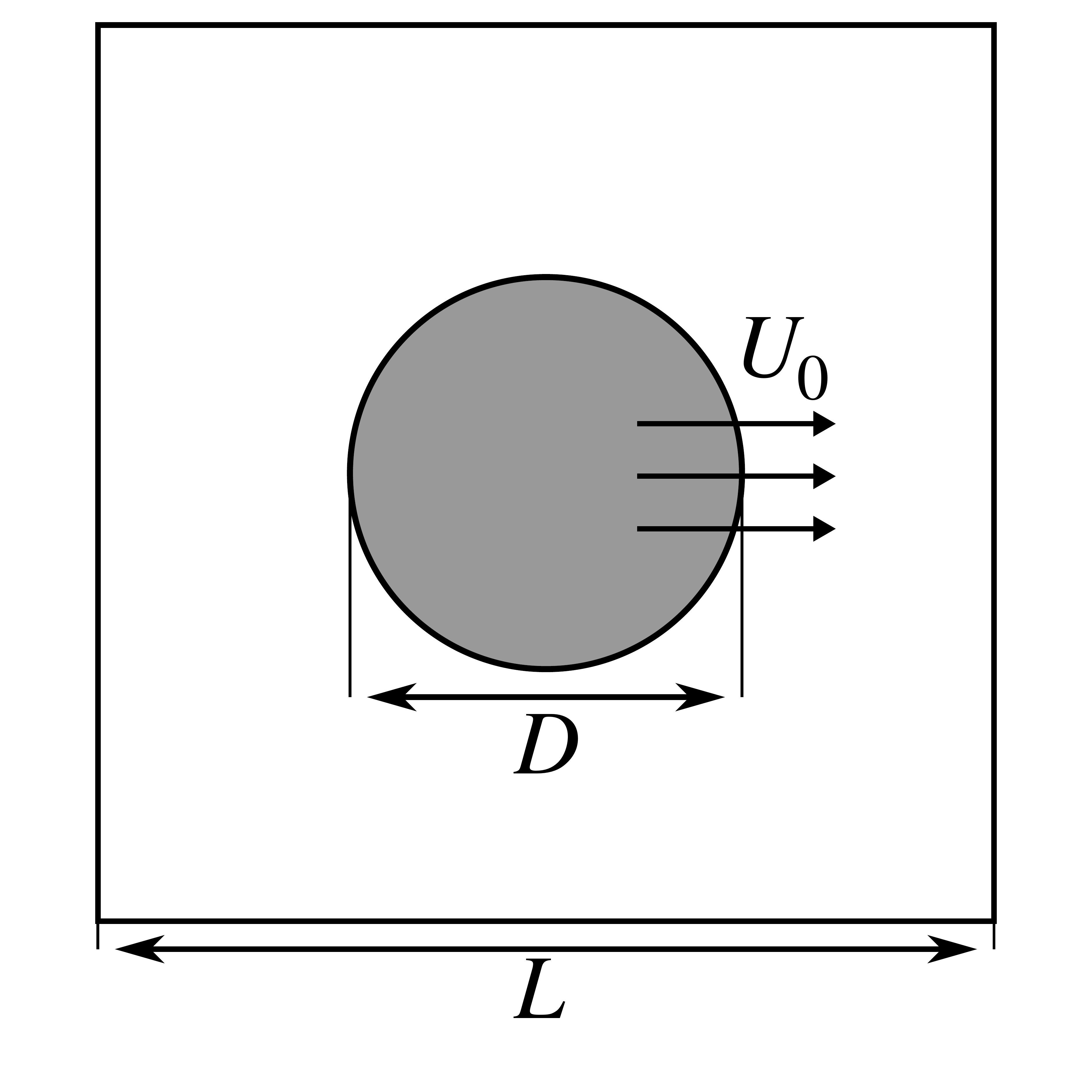}
    		\caption{Dynamic test case}
     \end{subfigure}
     \caption{Test case set up for the spurious currents quantification}
     \label{fig:tc_case}
     \end{minipage}\hfill
   	\begin{minipage}{0.48\textwidth}
		\centering
		\includegraphics[width=\textwidth]{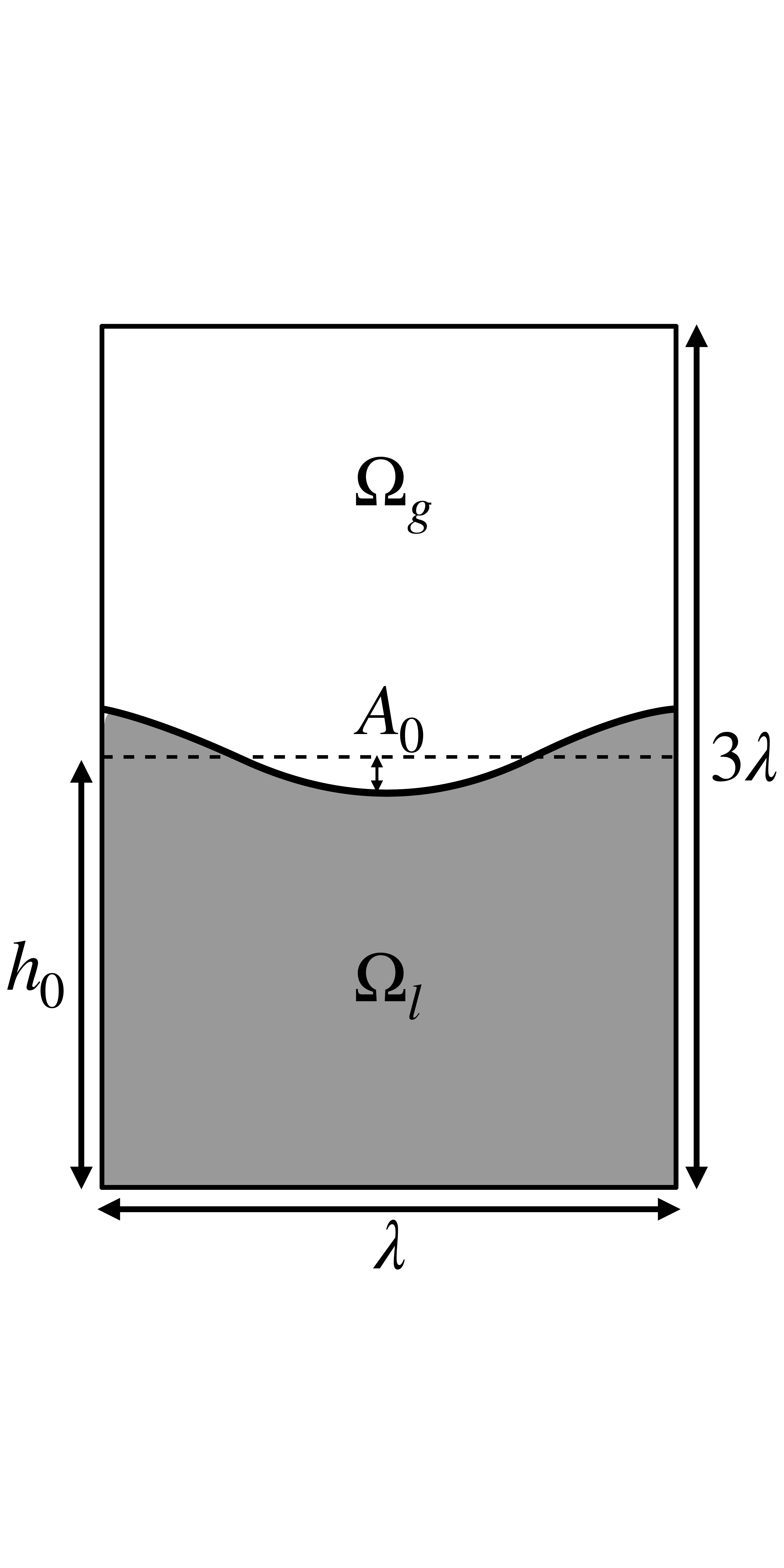}
     	\caption{Planar damping wave simulation set up}
     	\label{fig:planar_config}
    \end{minipage}
\end{figure}

The other focus of the solver focuses on the surface tension modelling implying both curvature computation and surface tension force discretization. In a first test, the curvature computation accuracy is evaluated without taking into account the transport errors induced by the velocity field. When an interface capturing method is coupled with a two-phase solver, the curvature errors act as a vorticity source in the momentum equation scaling with $\nabla \kappa$ and produce well-known parasitic currents \cite{Abadie2015}. To enlighten this behaviour two additional cases are considered and illustrated in Fig.\ref{fig:tc_case}: a static test case where only curvature error acts as a source of error and a dynamic case where the interface transport acts as a second source of error. \\ 
Finally, the canonical damping wave test case is considered to assess the solver on capillary-diven flows.

\subsubsection{Curvature computation}
\label{sec:curvature}

First, the accuracy of the curvature is demonstrated for the different representations of the interface. The choices of $\kappa$ computation summed up in Tab.\ref{tab:solver_meth} are considered. Note that for ACLS, $\phi_{FMM}$ is used while for CLSVOF $\phi_{PLIC}$ is used instead of the exact distance function $\phi$. 

The relative curvature errors are defined as in \cite{Chiodi2017} : 
\begin{align}
& L_2(\kappa) = \frac{\sqrt{\frac{1}{N_\Gamma}\sum_{i=1}^{N_\Gamma}\left( \kappa_{exact} - \kappa_{f,i}\right)^2}}{\kappa_{exact}} \\
& L_\infty(\kappa) = \frac{\max\limits_i{\vert \kappa_{exact} - \kappa_{f,i} \rvert}}{\kappa_{exact}}
\end{align}
with $\kappa_{exact}$ the exact curvature and $\kappa_f$ an interpolated curvature to the faces where $\nabla_f H_\Gamma^0$ is non-zero. In practice, $\kappa_f$ are the only curvature values used for surface tension modelling regardless of the interface capturing method. \\
For a range of $N_D=3.2$ to $N_D=409.6$, the errors are evaluated on 100 circles randomly located in the domain in order to meet as much configurations as possible. The final metric is $\left<L_2(\kappa)\right>$ the mean of $L_2(\kappa)$ and $\max\left(L_\infty(\kappa)\right)$ the maximum of $L_\infty(\kappa)$ over all those configurations.

\begin{figure}[h!]
\centering
     \begin{subfigure}[b]{0.48\textwidth}
     	\centering
         \includegraphics[width=\textwidth]{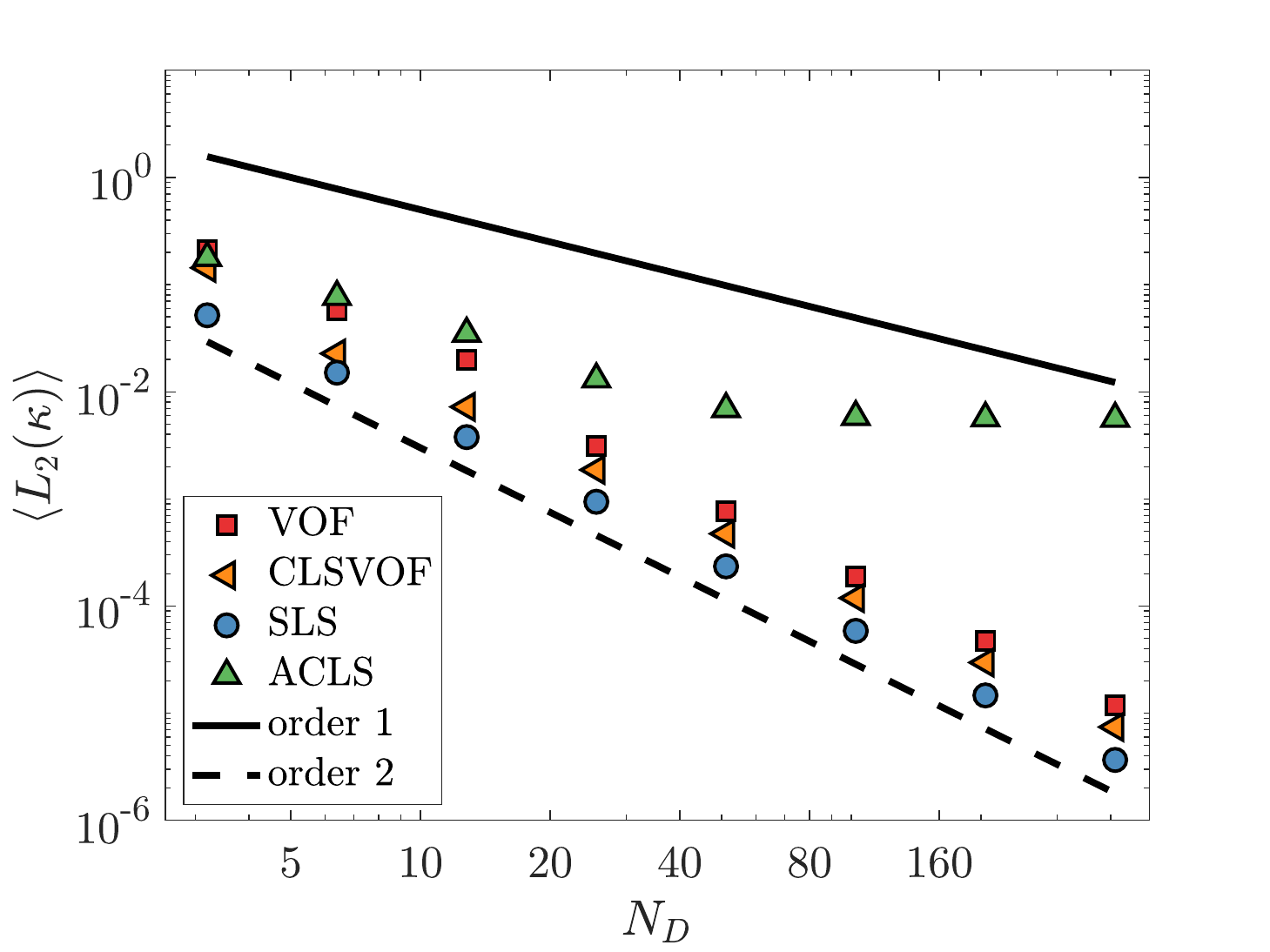}
    		\caption{Mean of $L_2(\kappa)$}
    		\label{fig:tc_curvature_l2}
     \end{subfigure}
     \hfill
     \begin{subfigure}[b]{0.48\textwidth}
     	\centering
         \includegraphics[width=\textwidth]{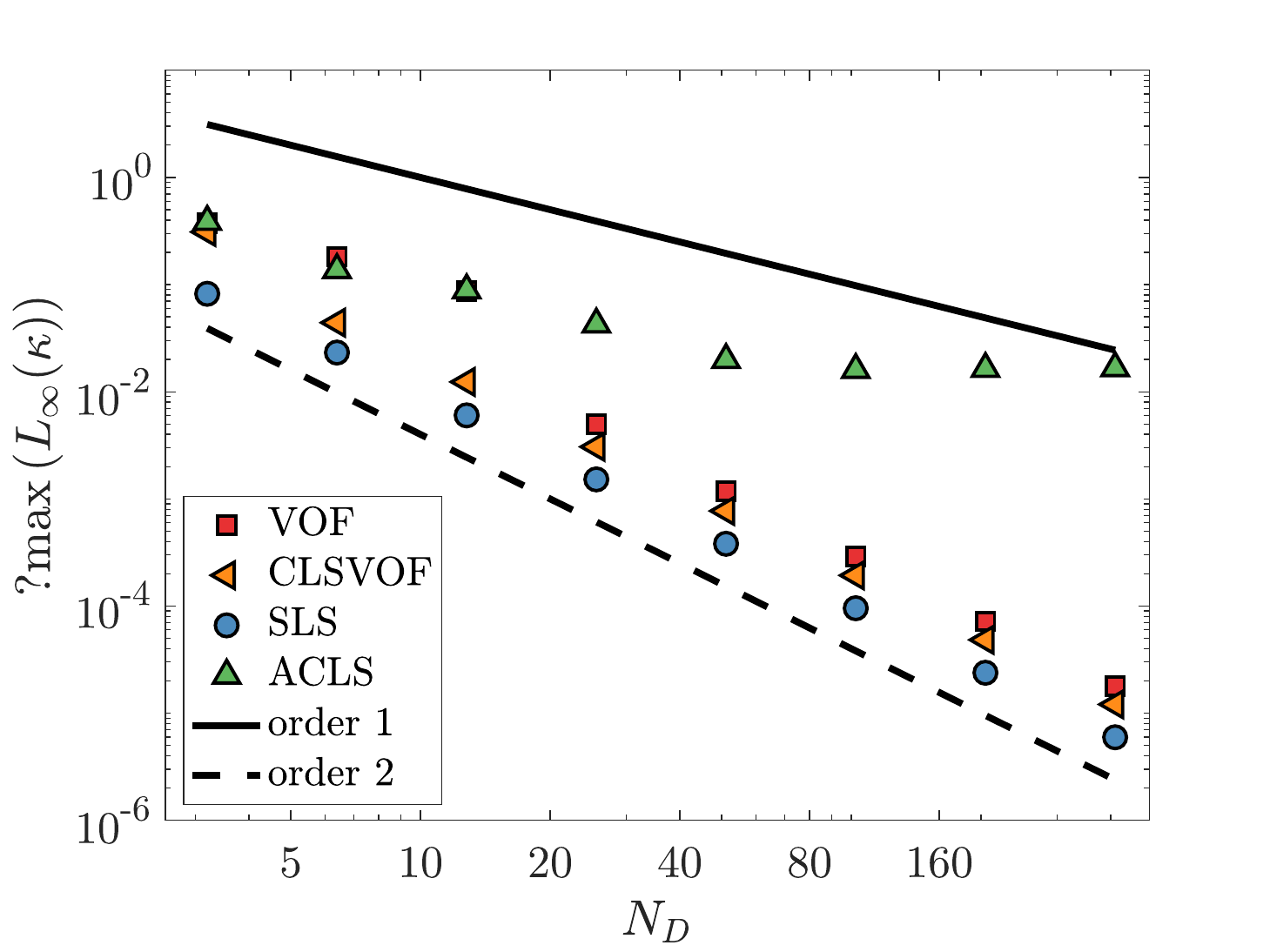}
    		\caption{Maximum of $L_\infty(\kappa)$}
    		\label{fig:tc_curvature_linf}
    	\end{subfigure}
     \caption{Mesh convergence of $\left<L_2(\kappa)\right>$ and $\max\left(L_\infty(\kappa)\right)$ for the 4 methods}
\end{figure}

\begin{figure}[h!]
\centering
     \begin{subfigure}[b]{0.48\textwidth}
     	\centering
         \includegraphics[width=\textwidth]{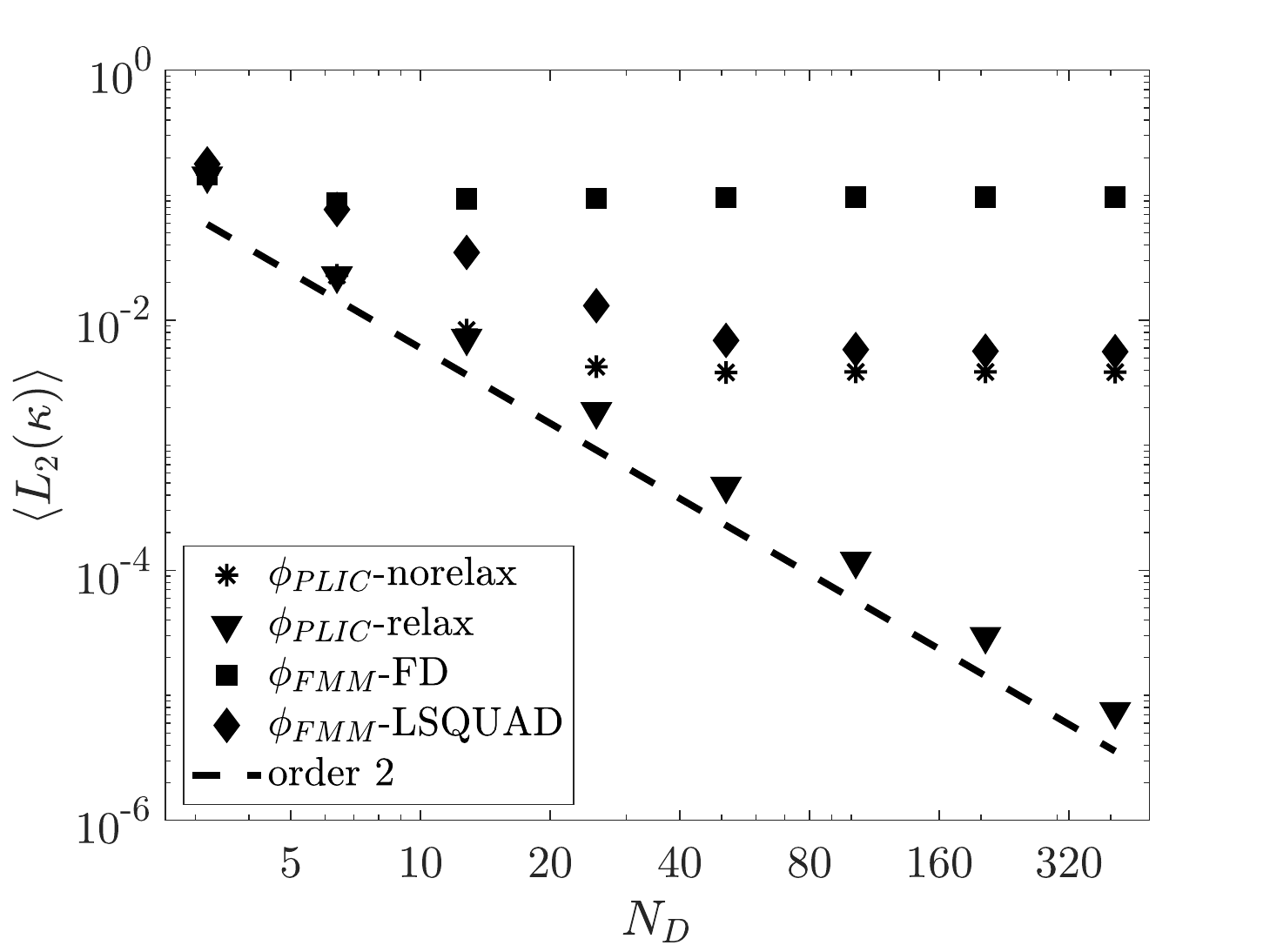}
    		\caption{Mean of $L_2(\kappa)$}
    		\label{fig:tc_kappa_l2}
     \end{subfigure}
     \hfill
     \begin{subfigure}[b]{0.48\textwidth}
     	\centering
         \includegraphics[width=\textwidth]{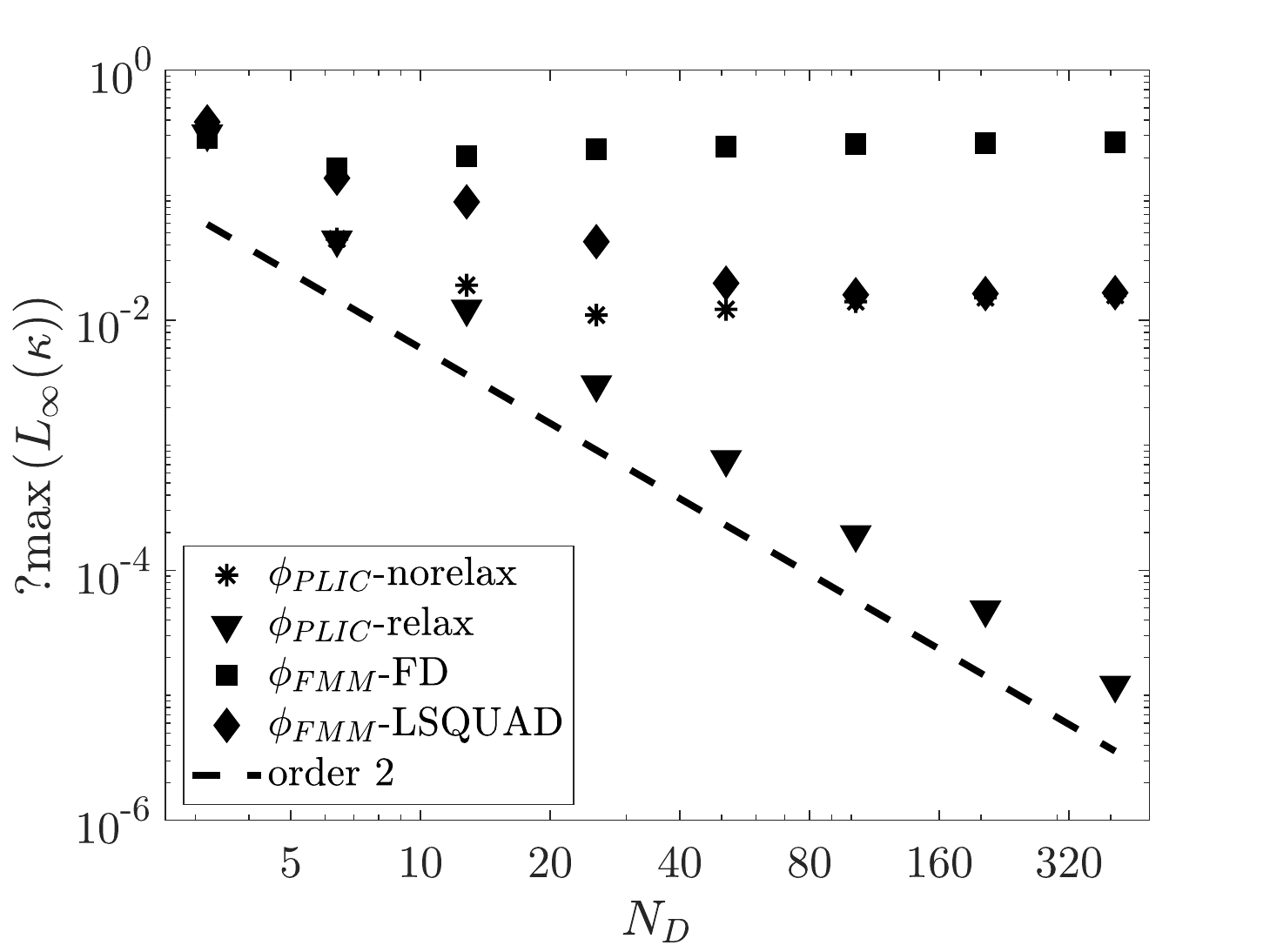}
    		\caption{Maximum of $L_\infty(\kappa)$}
    		\label{fig:tc_kappa_linf}
     \end{subfigure}
     \caption{Mesh convergence of $L_2(\kappa)$ and $L_\infty(\kappa)$ error for other choices of $\kappa$ computation}
\end{figure}

Fig.~\ref{fig:tc_curvature_l2} and \ref{fig:tc_curvature_linf} show convergence in $L_2$ and $L_\infty$ for all methods but ACLS. The ACLS method shows a saturation of the error convergence for high resolutions because of the second order nature of the $\phi_{FMM}$ distance function. Note that for completeness, Fig.~\ref{fig:tc_kappa_l2} and \ref{fig:tc_kappa_linf} illustrate the difference between LSQUAD and FD approach applied on $\phi_{FMM}$. It can be seen that the FD method does not show any convergent behaviour while LSQUAD, the method used here for CLSVOF and ACLS, manages to decrease the error until $N_D=51.2$. \\
This saturation is not observed for the CLSVOF approach because of the relaxation applied on the $\phi_{PLIC}$. This takes the form of $\phi_{PLIC}=\omega \phi + (1-\omega) d$ and $\omega$ is a function of the difference between $\phi$ and $d$. As observed in  Fig.~\ref{fig:tc_kappa_l2} and \ref{fig:tc_kappa_linf}, if no relaxation is performed, the curvature computation does not converge at high resolution for the same reasons as for ACLS. If the relaxation is activated, $\omega$ will tend to one for high resolution as the local curvature in a given cell is closer to a plane, so the discrepancy between $\phi$ and $d$ is lower. This allows CLSVOF curvature to show convergence even for high resolutions. \\
Finally, HF method has a transient convergence regime in the lowest resolutions as already pointed out in \cite{popinet2009accurate}.

\subsubsection{Static test case}
An infinite cylinder of diameter $D=0.4$ is centered in a $[1 \times 1]$ domain without gravity force. Only a quarter of the domain is considered here where right and top boundary conditions are no-slip walls and bottom and left boundary conditions are symmetry. 
The cylinder is supposed to stay at rest as the pressure force is expected to balance exactly the capillary forces. The methods are not able to compute a constant $\kappa$ and spurious currents are present in the domain. This behaviour is quantified here with the maximum Capillary number $Ca_{max}=\frac{\rho \nu  \lVert \uvec \rVert_{max}}{\sigma}$. The time scale is defined as $t_\sigma=\sqrt{\frac{\rho D^3}{\sigma}}$ and the Laplace number $La = \frac{\rho \sigma D}{\mu^2}$. The fluid properties are the same in both phases such that $La=12000$. 

\begin{figure}[h!]
\centering
     \begin{subfigure}[b]{0.48\textwidth}
     	\centering
         \includegraphics[width=\textwidth]{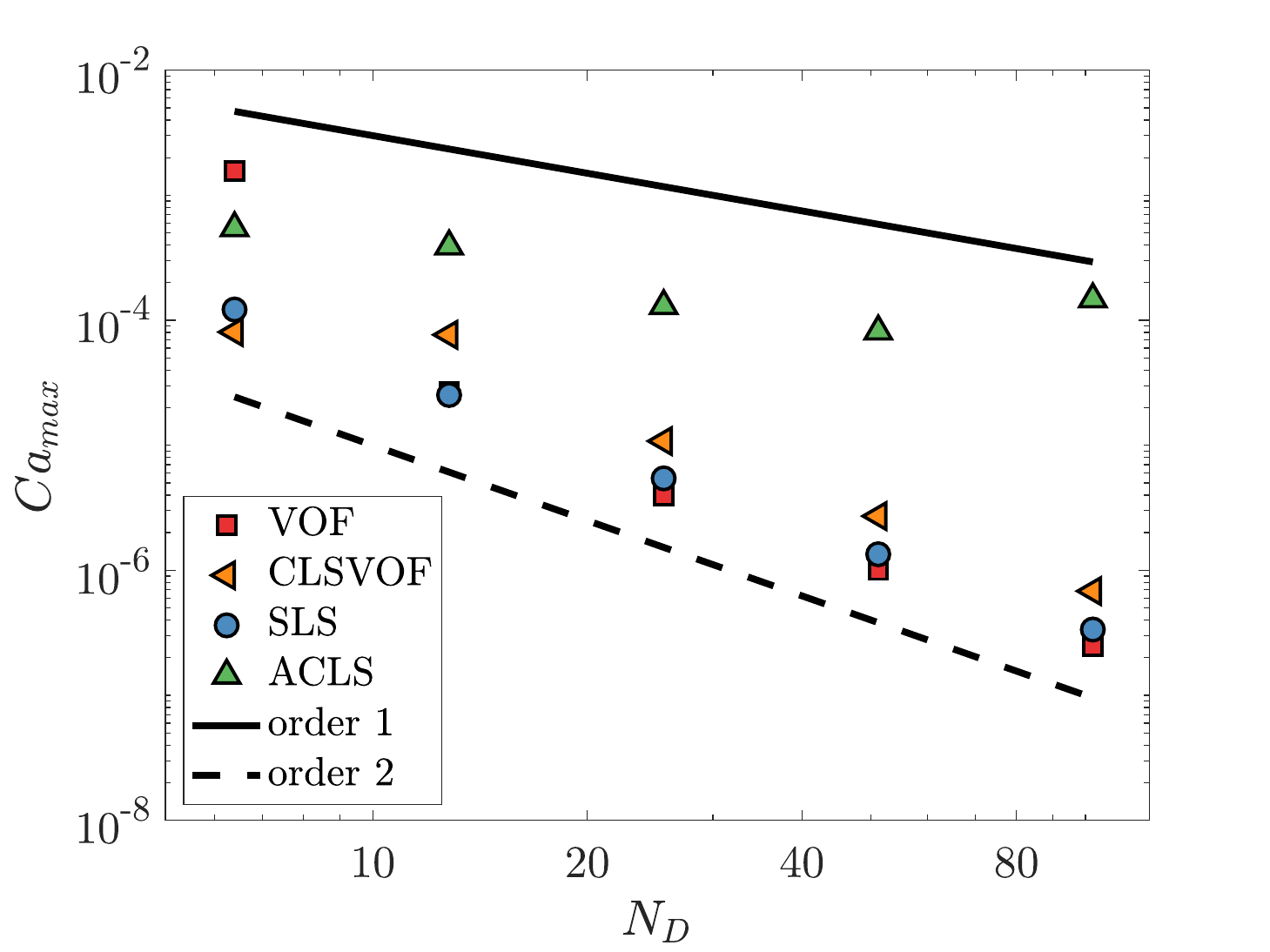}
    		\caption{$Ca_{max}$ for a full run}
    		\label{fig:tc_static_Camax}
     \end{subfigure}
     \hfill
     \begin{subfigure}[b]{0.48\textwidth}
     	\centering
         \includegraphics[width=\textwidth]{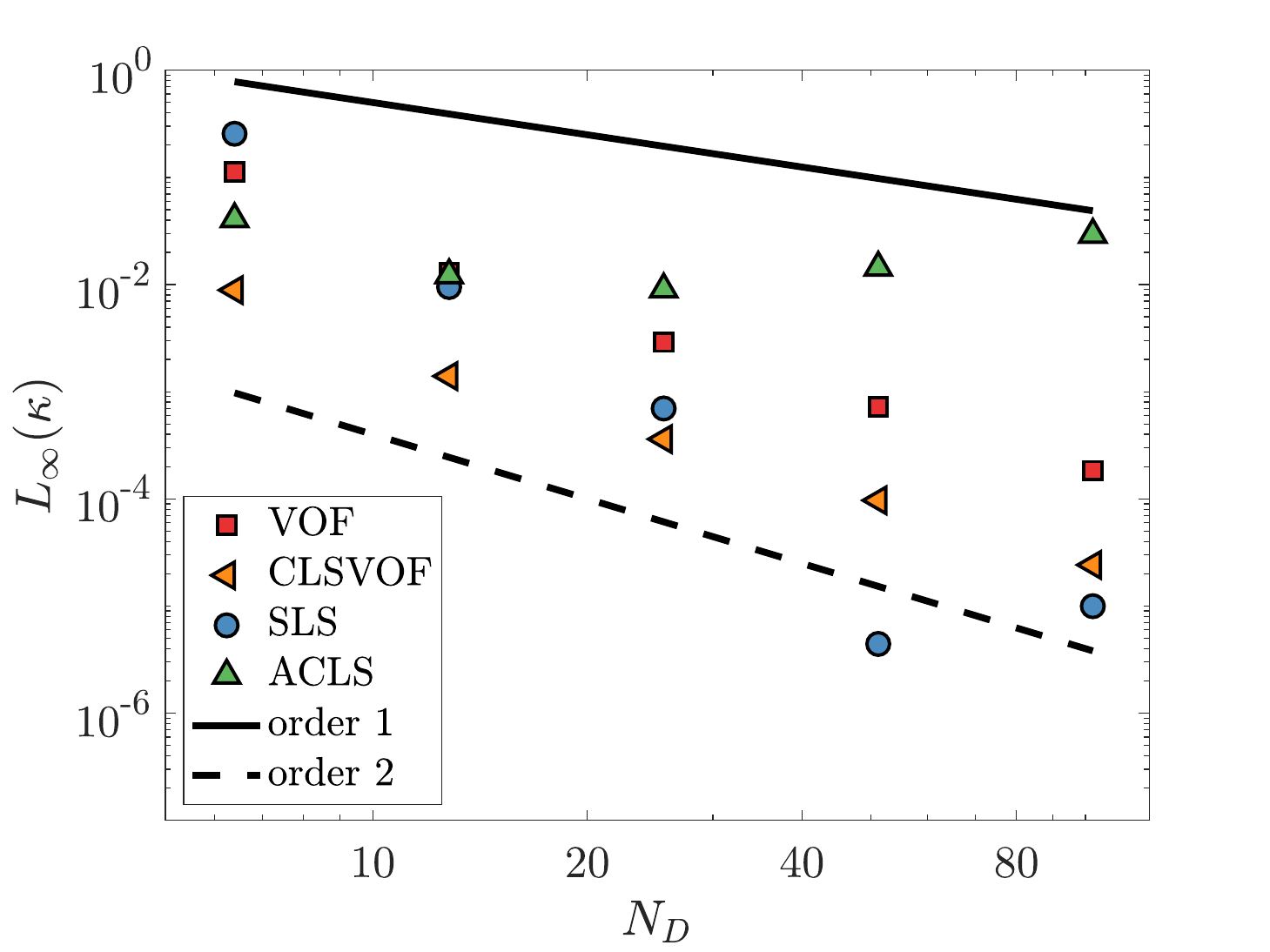}
    		\caption{$L_\infty(\kappa)$ error at $t/t_\sigma=20$}
    		\label{fig:tc_static_linf}
     \end{subfigure}
     \caption{Mesh convergence of $Ca_{max}$ and $L_\infty(\kappa)$ error for the static case}
\end{figure}

In Fig.~\ref{fig:tc_static_Camax} is displayed the mesh convergence of the maximum $Ca_{max}$ during a static simulation. One can see that all methods but ACLS converge with mesh resolution. This is because of the non-converging behaviour of $\kappa$ which has been demonstrated from previous section and from the $L_\infty(\kappa)$ error at the end of the simulation in Fig.\ref{fig:tc_static_linf}. This has been previously observed in \cite{Chiodi2017} in 3D where the $Ca_{max}$ was closely the same for the $40^3$ and $80^3$ meshes.

\begin{figure}[h!]
\centering
     \begin{subfigure}[b]{0.48\textwidth}
     	\centering
         \includegraphics[width=\textwidth]{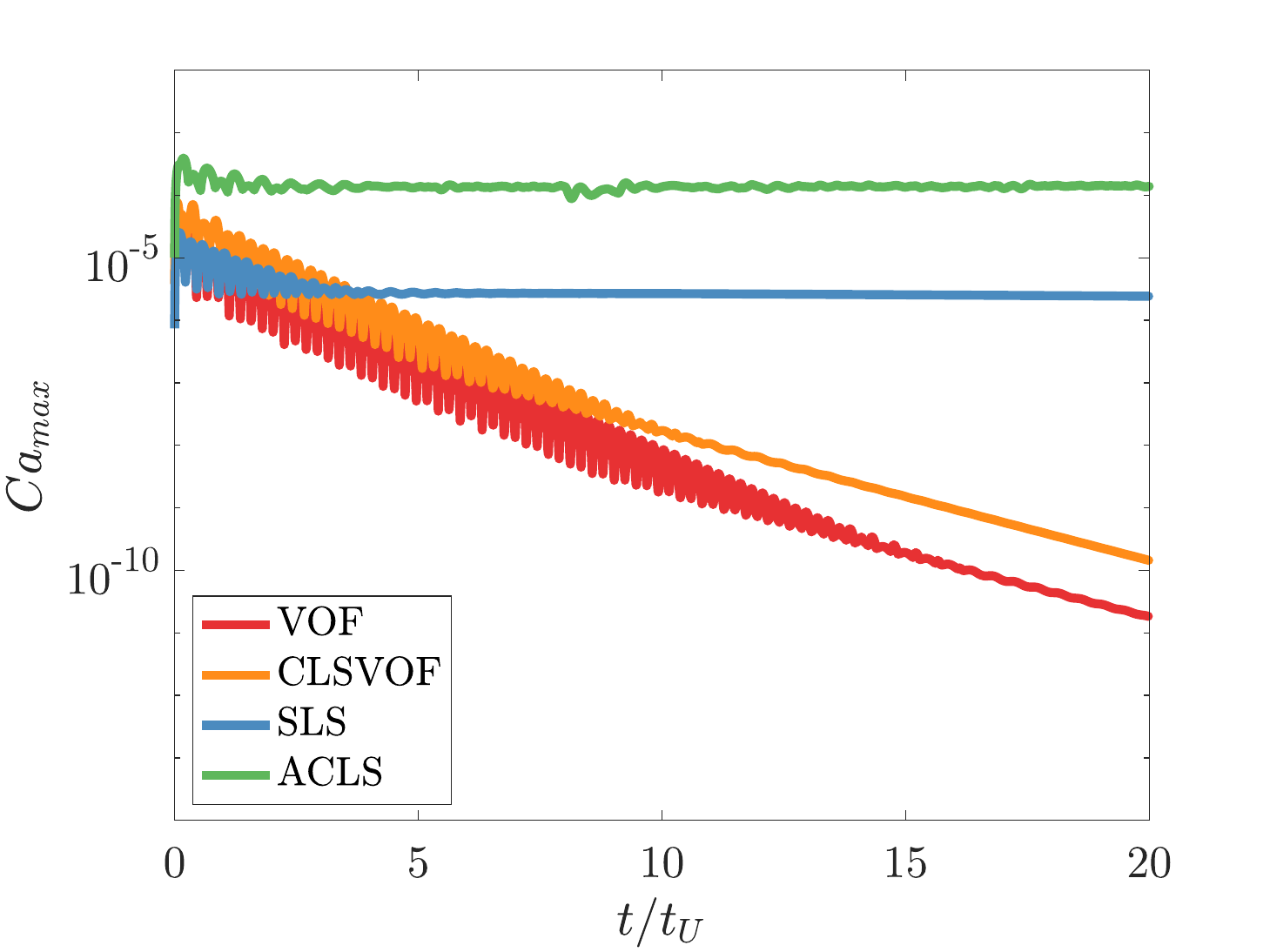}
    		\caption{$N_D=12.8$}
     \end{subfigure}
     \hfill
     \begin{subfigure}[b]{0.48\textwidth}
     	\centering
         \includegraphics[width=\textwidth]{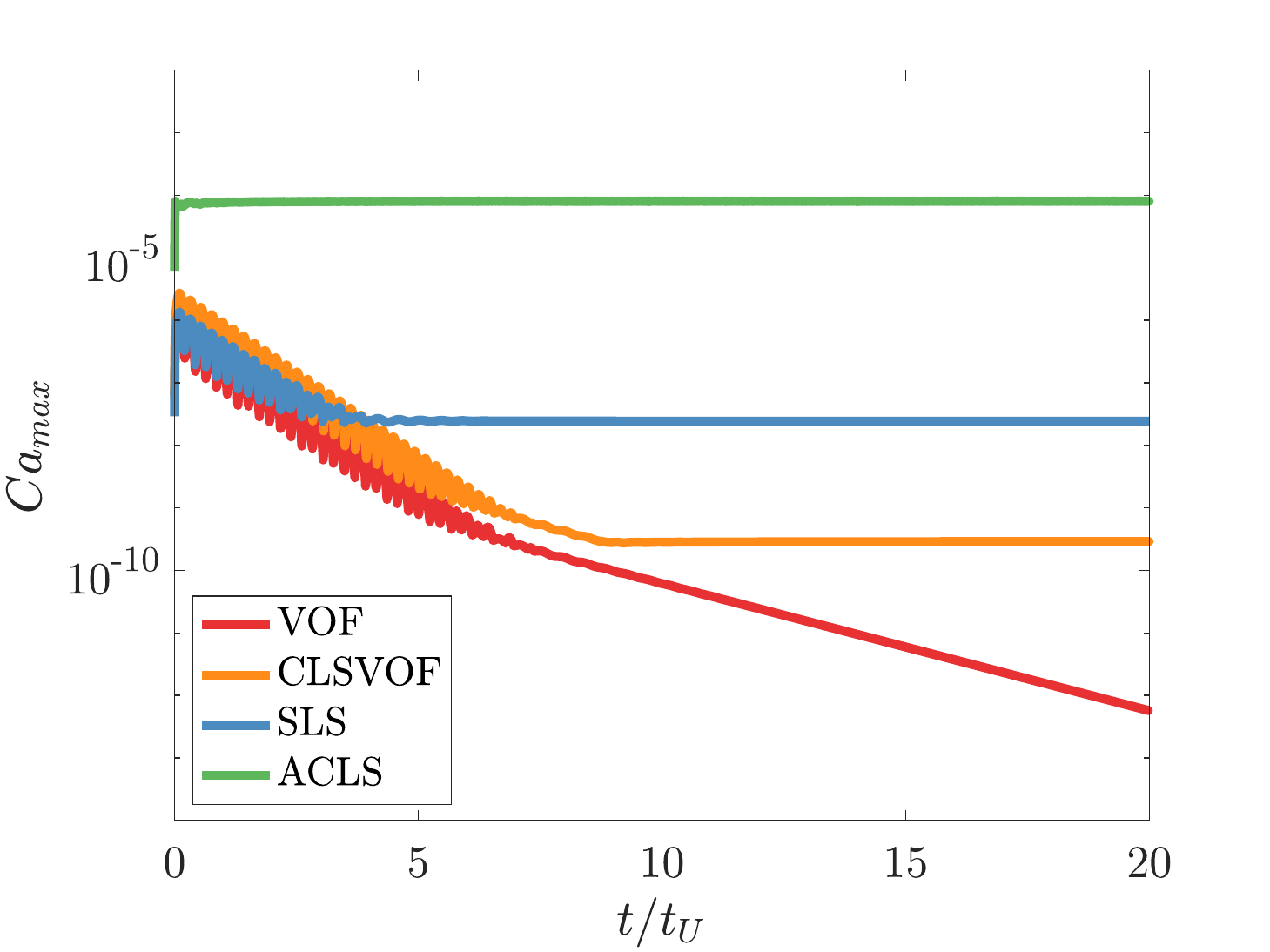}
    		\caption{$N_D=51.2$}
     \end{subfigure}
     \caption{Temporal evolution of $Ca_{max}$ for the static case}
     \label{fig:tc_static_damp}
\end{figure}

The damping of the spurious currents in Fig.~\ref{fig:tc_static_damp} can be explained by the numerical curvature computation. The initial spurious currents are a direct consequence of the initial error introduced by the curvature computation.  VOF is able to reach an equilibrium close to zero machine as shown in \cite{popinet2009accurate}. The other methods also reach a steady $Ca_{max}$ value which is not zero because of the reinitialization step introducing new errors at each iteration \cite{Abadie2015}. While this is very noticeable for ACLS, the magnitude decreases using SLS and decreases even more with CLSVOF.

\subsubsection{Dynamic test case}
In order to quantify the impact of the flow dynamic on the spurious currents, the following test case is considered. An infinite cylinder of diameter $D=0.4$ is centered in a $[1 \times 1]$ domain with a uniform horizontal velocity $U_{0}$ where boundary conditions are periodic in the velocity direction and free slip conditions are imposed on the top and bottom boundaries.
\\
From the new velocity scale $U_0$, a new time scale can be defined as $t_U = \frac{D}{U_0}$ and the Weber number $We=\frac{\rho U_0^2 D}{\sigma}$. The fluid properties are the same as in the static case with $We = 0.4$.

\begin{figure}[h!]
\centering
     \begin{subfigure}[b]{0.48\textwidth}
     	\centering
         \includegraphics[width=\textwidth]{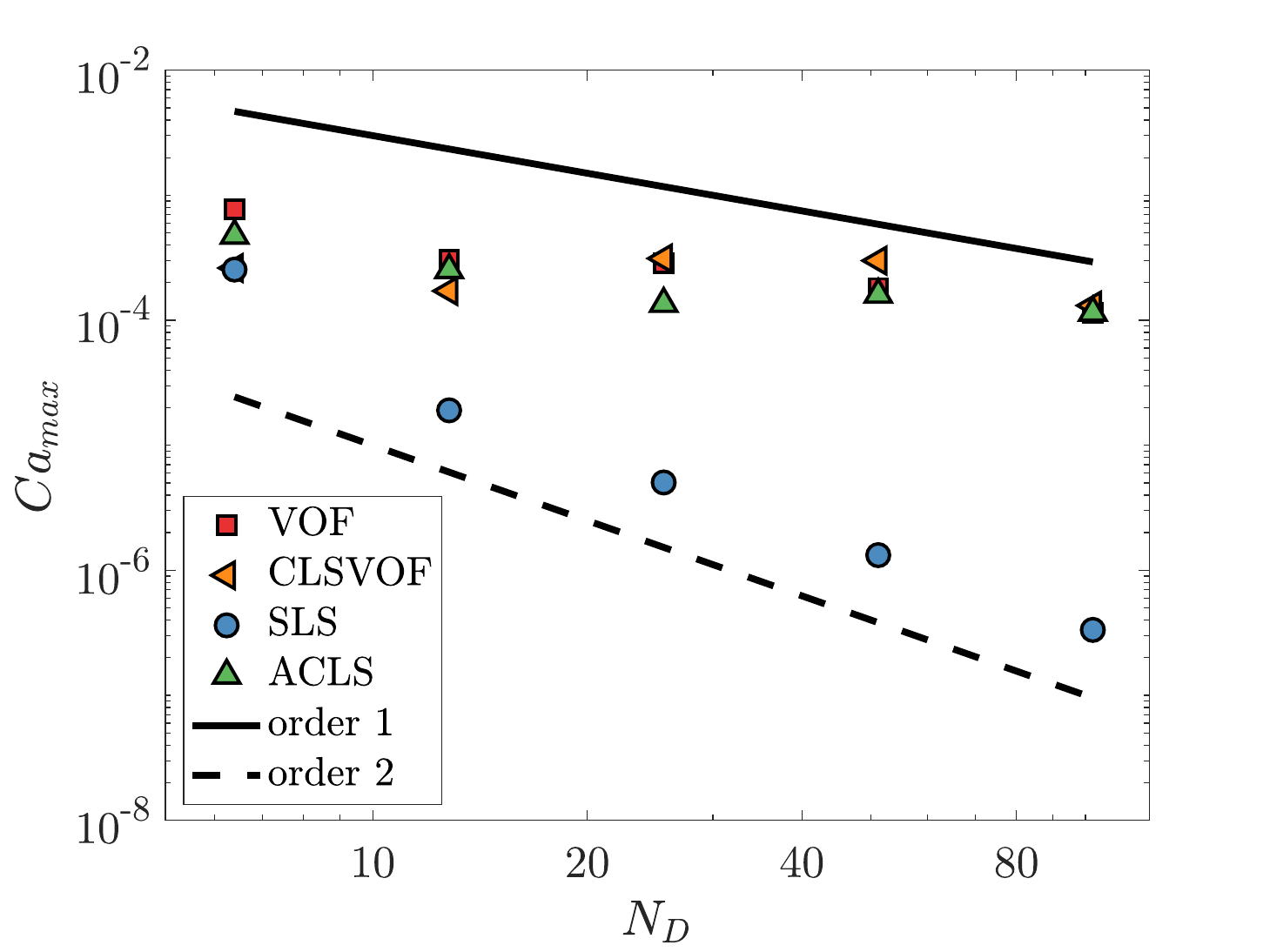}
    		\caption{$Ca_{max}$ for a full run}
    		\label{fig:tc_dynamic_Camax}
     \end{subfigure}
     \hfill
     \begin{subfigure}[b]{0.48\textwidth}
     	\centering
         \includegraphics[width=\textwidth]{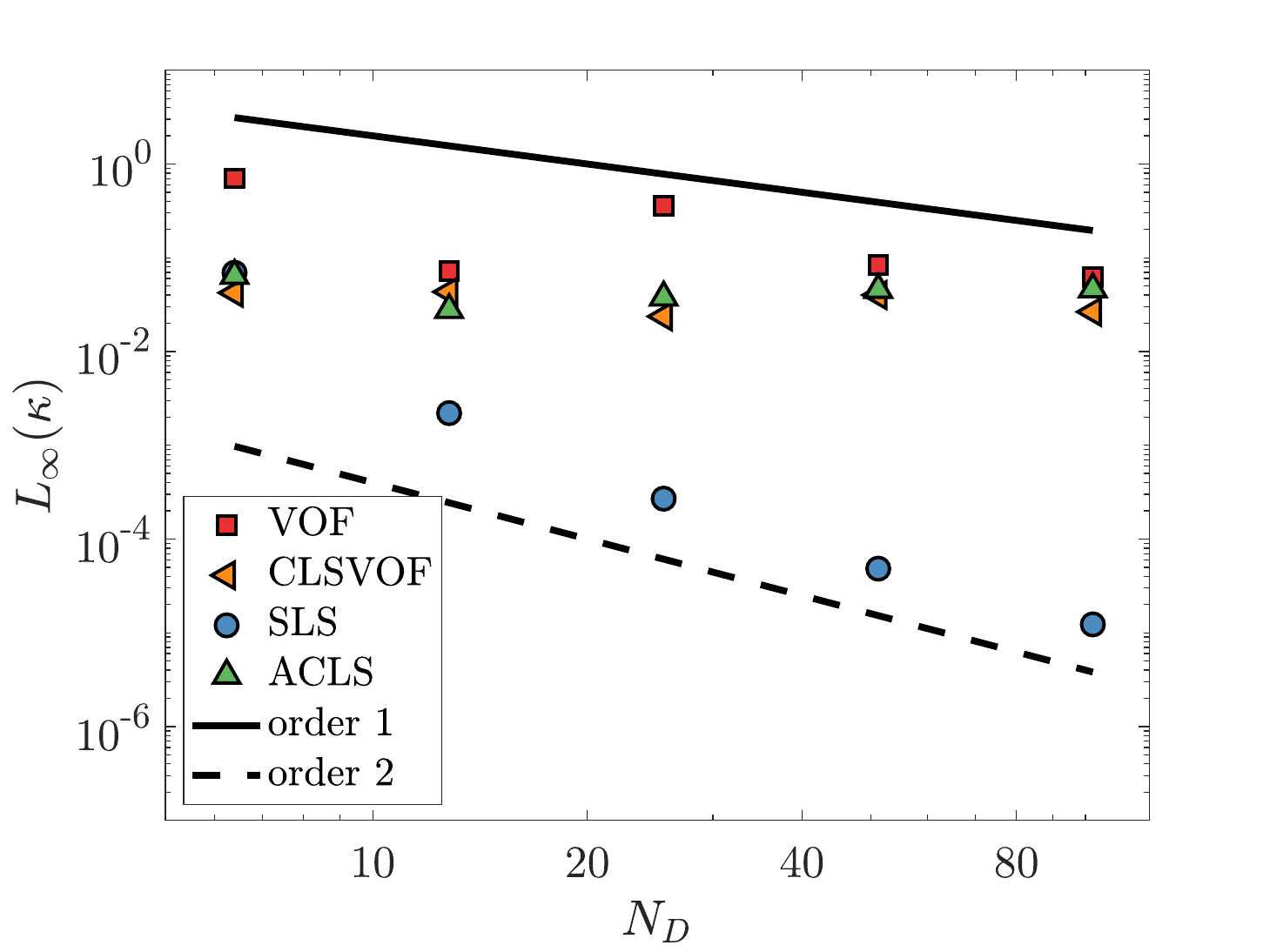}
    		\caption{$L_\infty(\kappa)$ error at $t/t_U=2.5$}
    		\label{fig:tc_dynamic_linf}
     \end{subfigure}
     \caption{Mesh convergence of $Ca_{max}$ and $L_\infty(\kappa)$ error for the dynamic case}
\end{figure}
The mesh convergences of $Ca_{max}$ and $L_\infty(\kappa)$ after one revolution are illustrated in Fig.~\ref{fig:tc_dynamic_Camax} and \ref{fig:tc_dynamic_linf}. As in \cite{popinet2009accurate} and \cite{Abadie2015}, no convergence is observed for VOF because of the second-order errors introduced by the transport step. This is also true for CLSVOF and ACLS. However, SLS exhibits a huge reduction of spurious currents explained by the higher accuracy of curvature and transport.

\begin{figure}[h!]
\centering
     \begin{subfigure}[b]{0.48\textwidth}
     	\centering
         \includegraphics[width=\textwidth]{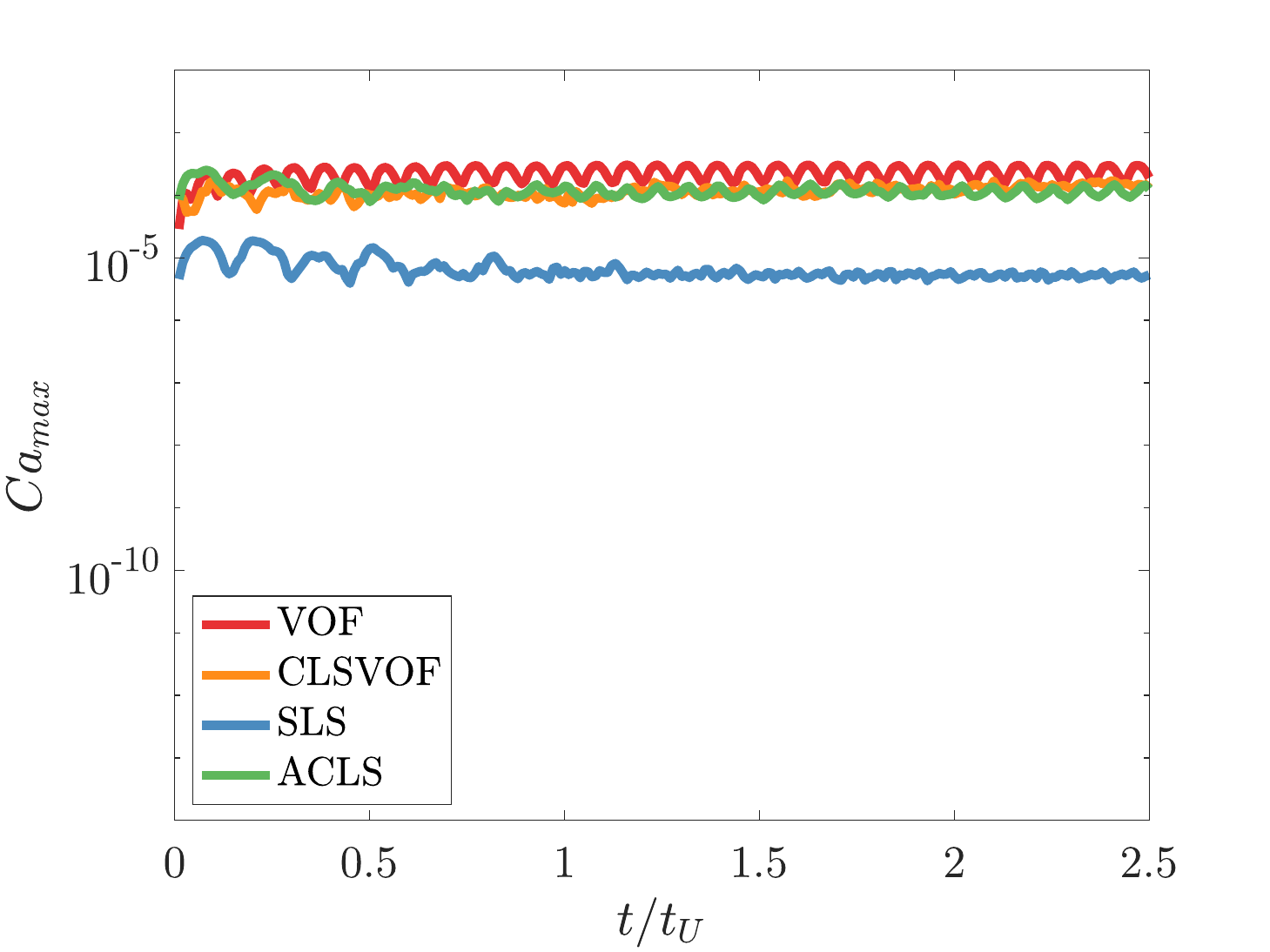}
    		\caption{$N_D=12.8$}
     \end{subfigure}
     \hfill
     \begin{subfigure}[b]{0.48\textwidth}
     	\centering
         \includegraphics[width=\textwidth]{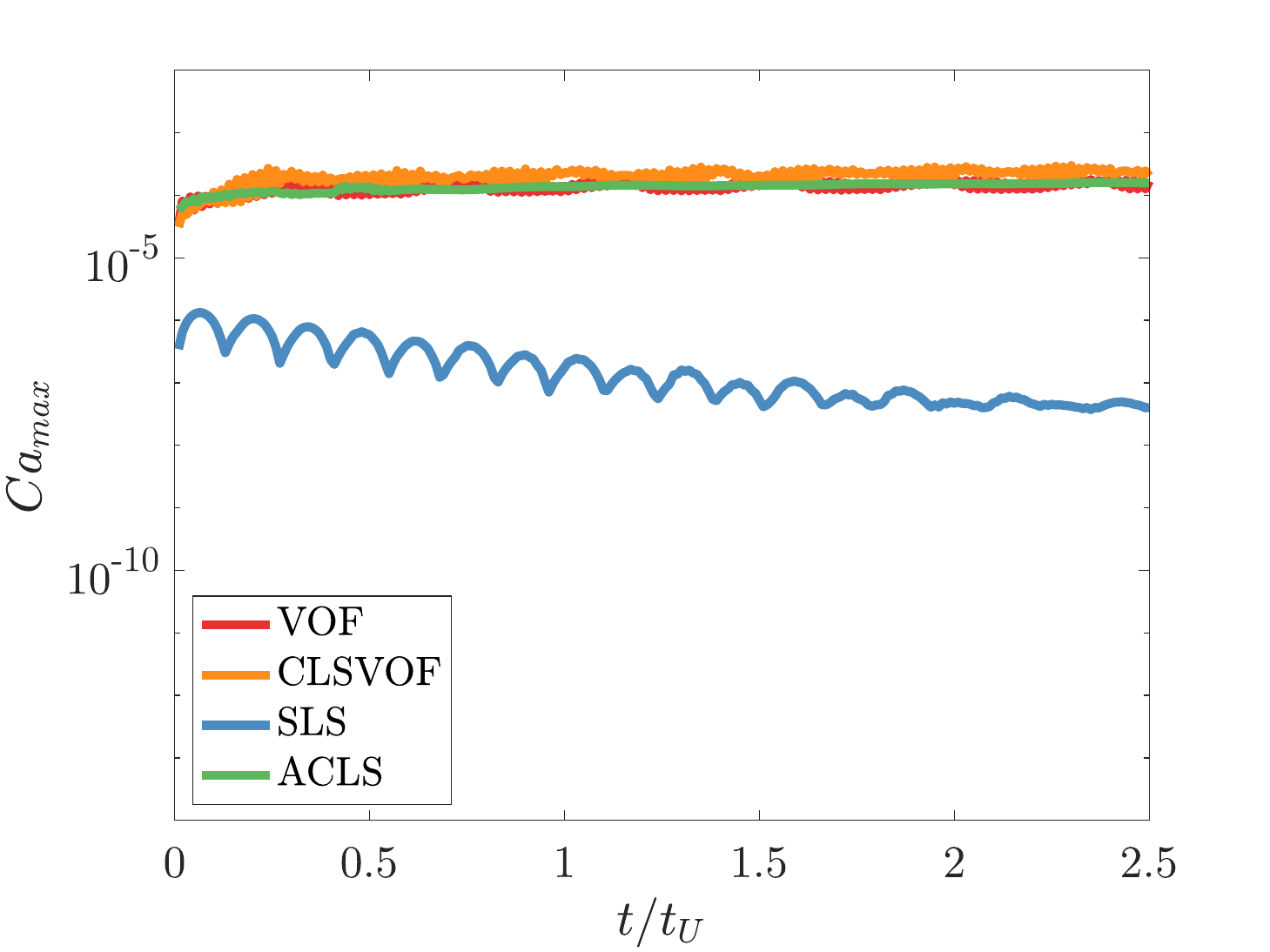}
    		\caption{$N_D=51.2$}
     \end{subfigure}
     \caption{Temporal evolution of $Ca_{max}$ for the dynamic case}
     \label{fig:tc_dynamic_damp}
\end{figure}

\begin{figure}[h!]
\centering
 	\begin{subfigure}[b]{0.48\textwidth}
     	\centering
         \includegraphics[width=\textwidth]{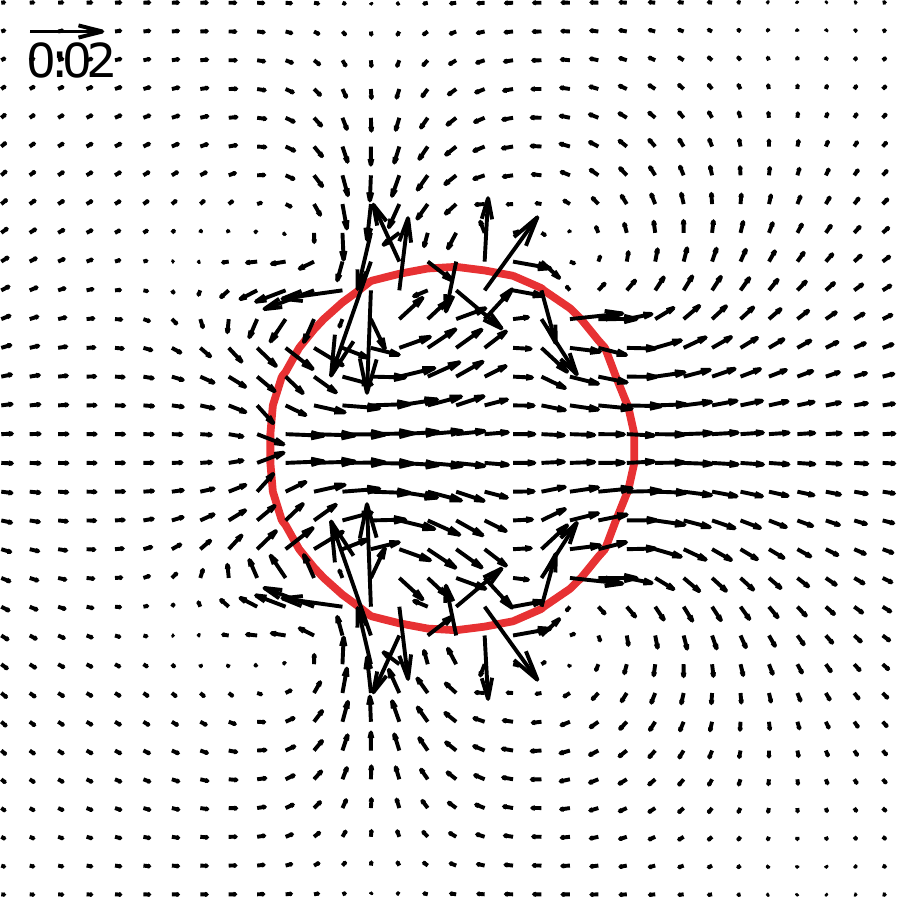}
    		\caption{VOF}
     \end{subfigure}
     \hfill
     \begin{subfigure}[b]{0.48\textwidth}
     	\centering
         \includegraphics[width=\textwidth]{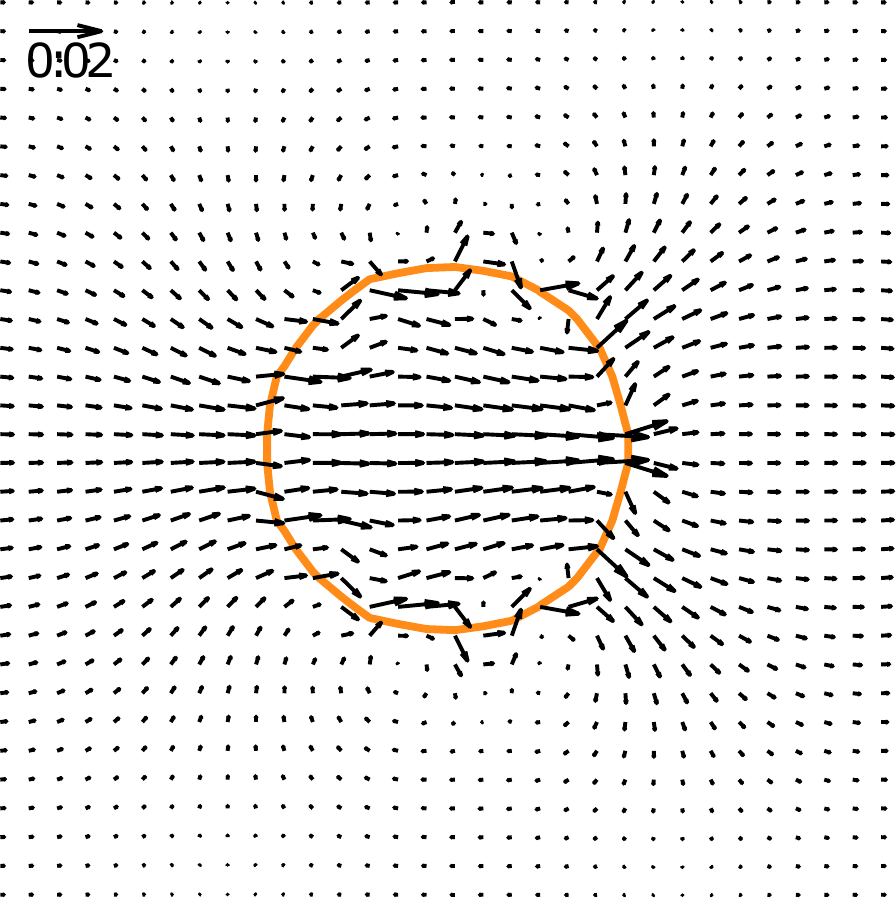}
    		\caption{CLSVOF}
     \end{subfigure}
     \hfill
     \begin{subfigure}[b]{0.48\textwidth}
     	\centering
         \includegraphics[width=\textwidth]{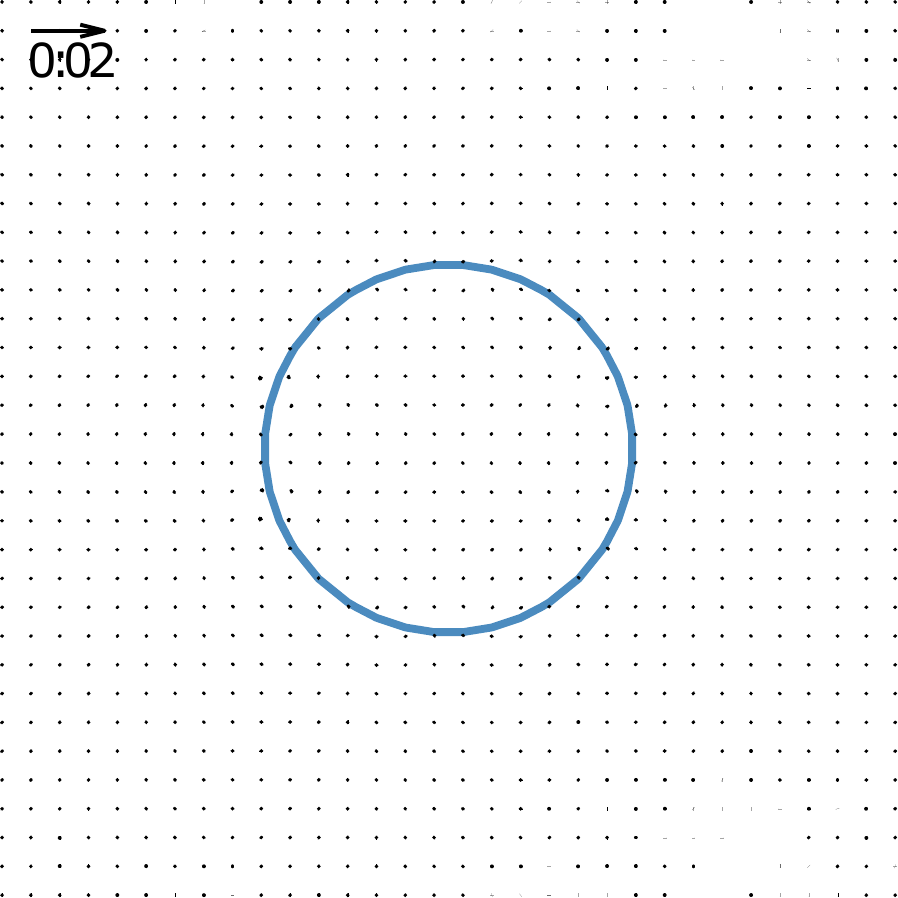}
    		\caption{SLS}
     \end{subfigure}
     \hfill
     \begin{subfigure}[b]{0.48\textwidth}
     	\centering
         \includegraphics[width=\textwidth]{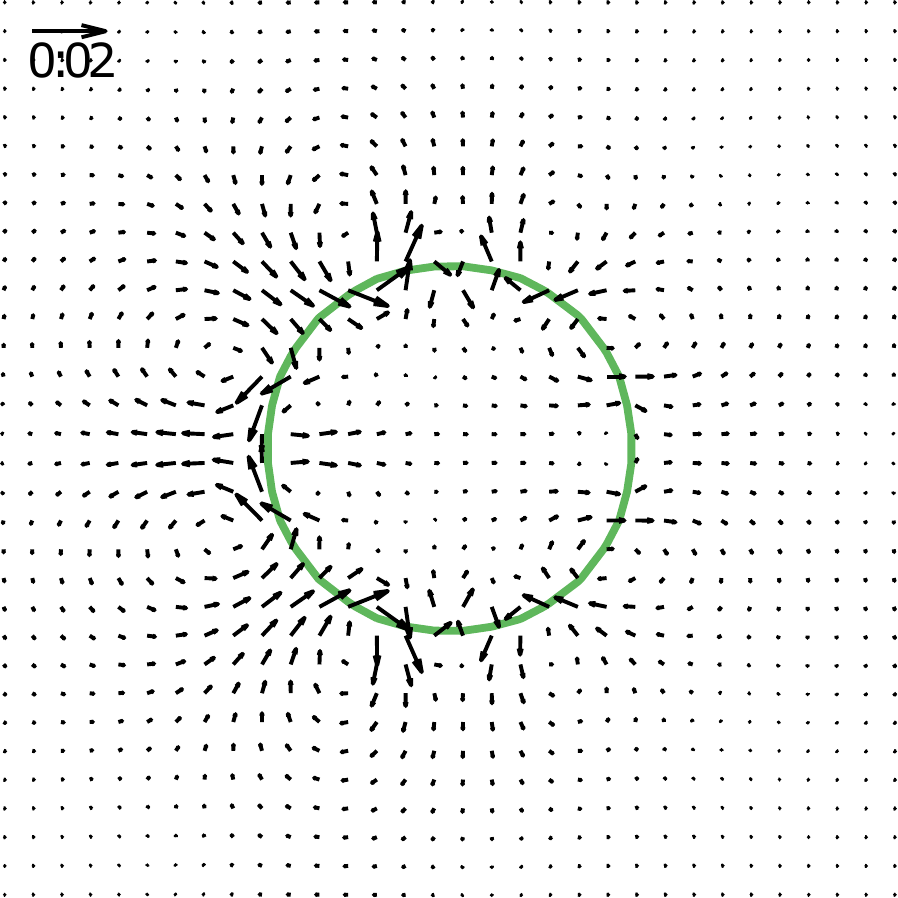}
    		\caption{ACLS}
     \end{subfigure}
     \caption{Relative velocity field ($U_0$ is subtracted for visualization) for the dynamic case at $N_D=12.8$}
     \label{fig:tc_field}
\end{figure}

Compared to the static case, no damping of spurious currents is observed in Fig.~\ref{fig:tc_dynamic_damp}. This is because of the permanent introduction of transport errors acting as an imbalance in the curvature computation. For a more complete visualization, the velocity field in the reference of the translating droplet is given for a medium resolution of $N_D=12.8$ in Fig.~\ref{fig:tc_field}. CLSVOF is slightly improving VOF curvature computation but it is not as accurate as SLS because of the $\phi_{PLIC}$ perturbations introduced by the LS-VOF coupling. ACLS performs better than VOF and CLSVOF in the medium resolution range while SLS maintains a fairly low amount of spurious currents. The spurious intensity $ \lVert \uvec \rVert_{max}$ represents about $1\%$ of $U_0$ for VOF, CLSVOF and ACLS against $0.001\%$  for SLS. 

\subsubsection{Planar damping wave}
The planar damping wave is an interesting test case as an analytical solution is available in the literature. The test case has been widely investigated as a solver validation  \cite{Gueyffier1999,popinet1999front,Gerlach2006,Desjardins2009,popinet2009accurate,Ghods2013,Janodet2019}.
Here, a planar wave is initialized with a small harmonic perturbation of amplitude $A_0$ and both fluids are at rest with the same density and viscosity properties.
The initial interface height can be described by
\begin{equation}
y_0 = h_0 + A_0 \cos \left( \frac{2\pi x}{\lambda}\right)
\end{equation}
with $\lambda$ the wavelength of the perturbation and $h_0=3\lambda/2$ the vertical interface position. $\lambda$ is taken to unity and $A_0=\lambda /100$. 
In this problem, non dimensional time and viscosity are defined as
\begin{equation}
\tau = t\omega_0, \quad \xi=\nu \lambda^2/\omega_0
\end{equation}
with the perturbation frequency $\omega_0=\sqrt{ \sigma \lambda^3 / (\rho_l+\rho_g)}$. 

The perturbation amplitude is then deduced from the analytical solution derived by Prosperetti et al. \cite{prosperetti1981motion} 
\begin{align}
A(\tau) = &A_0\frac{4(1-4\beta)\xi	^2}{8(1-4\beta)\xi^2+1}\erfc{\left(\sqrt{\xi\tau}\right)} \nonumber \\ 
& + \sum_{i=1}^4 A_0\frac{z_i}{Z_i}\frac{\omega_0^2}{z_i^2-\xi \omega_0}  \exp\left( \frac{\left( z_i^2-\xi\omega_0\right)\tau}{\omega_0} \right) \erfc{\left(z_i\sqrt{\frac{\tau}{\omega_0}}\right)} 
\end{align}
with $\beta=\rho_l\rho_g/(\rho_l+\rho_g)^2$, $Z_i=\prod\limits_{{\forall j \neq i}} (z_j-z_i)$ and $z_i$ the four complex roots of the following quartic equation in $z$
\begin{align}
z^4-4\beta\left(\xi\omega_0\right)^{1/2}z^3 &+2(1-6\beta)\xi\omega_0 z^2 \nonumber \\ 
& + 4(1-3\beta)\left(\xi\omega_0\right)^{3/2}z+(1-4\beta)\left(\xi\omega_0\right)^2+\omega_0^2=0
\end{align}

In the special case of same density, momentum and viscosity jumps at the interface cancel and the numerical errors are only due to curvature computation and interface transport. The densities are $\rho_l=\rho_g=1$ which leads to $La = 3000$, $\xi=0.0647$ and $\beta=0.25$. The solution holds for infinite domain in the y-direction while x is periodic. The box is taken as $[\lambda \times 3 \lambda ]$ to limit boundary effects with wallslip imposed at the top and bottom as shown in Fig.~\ref{fig:planar_config}.

The error is defined as the RMS of relative amplitude error over time
\begin{equation}
L_2(A) = \sqrt{\frac{1}{T\omega_0}\int_{0}^T \frac{\lvert( A_{exact}(\tau)-A(\tau) \rvert}{A_0} d\tau}
\end{equation}
with $T\omega_0=25$ which corresponds to approximately 4 oscillations. 

\begin{figure}[h!]
\centering
 	\begin{subfigure}[b]{0.48\textwidth}
     	\centering
         \includegraphics[width=\textwidth]{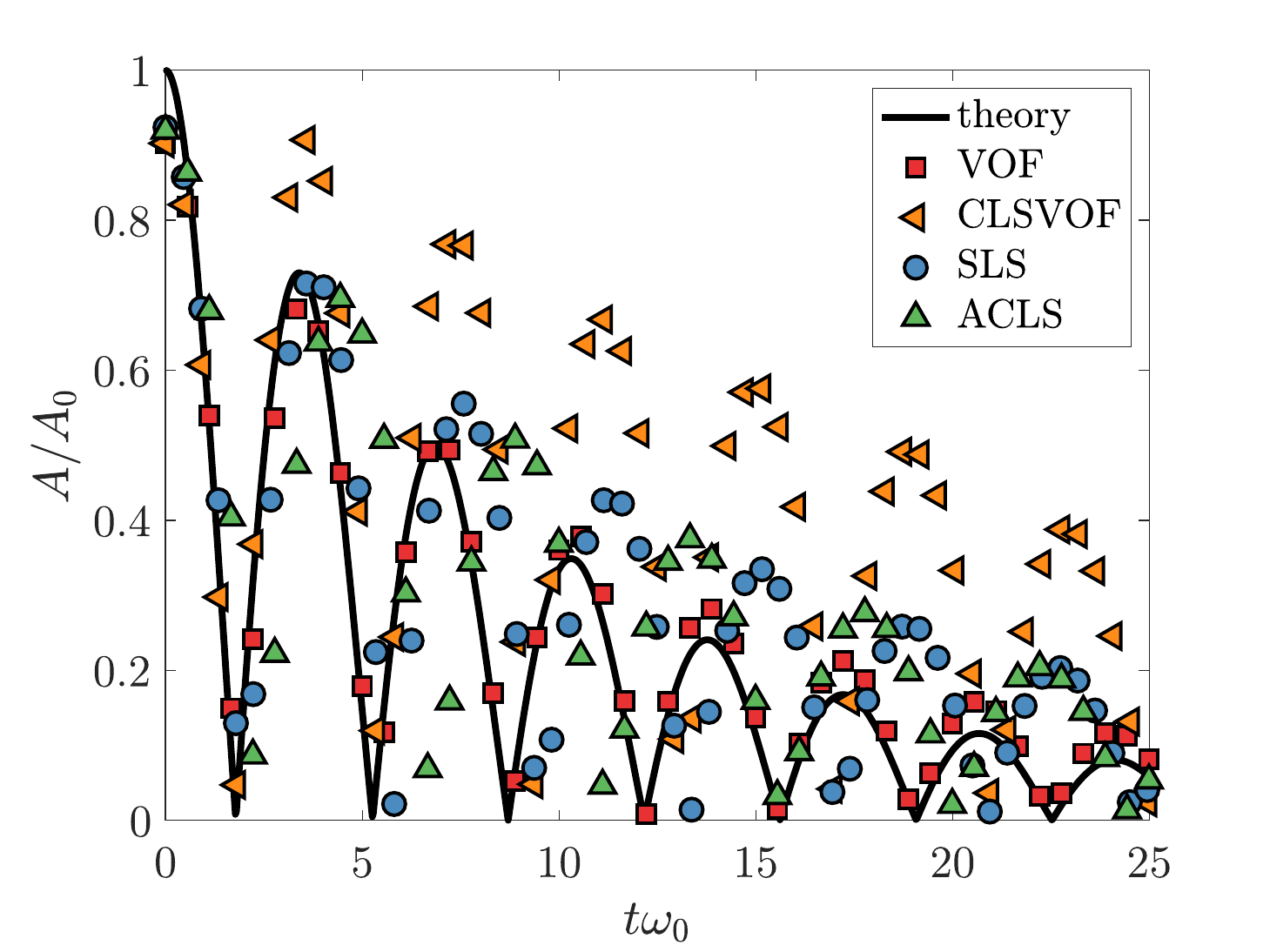}
    		\caption{$N_\lambda=8$}
     \end{subfigure}
     \hfill
     \begin{subfigure}[b]{0.48\textwidth}
     	\centering
         \includegraphics[width=\textwidth]{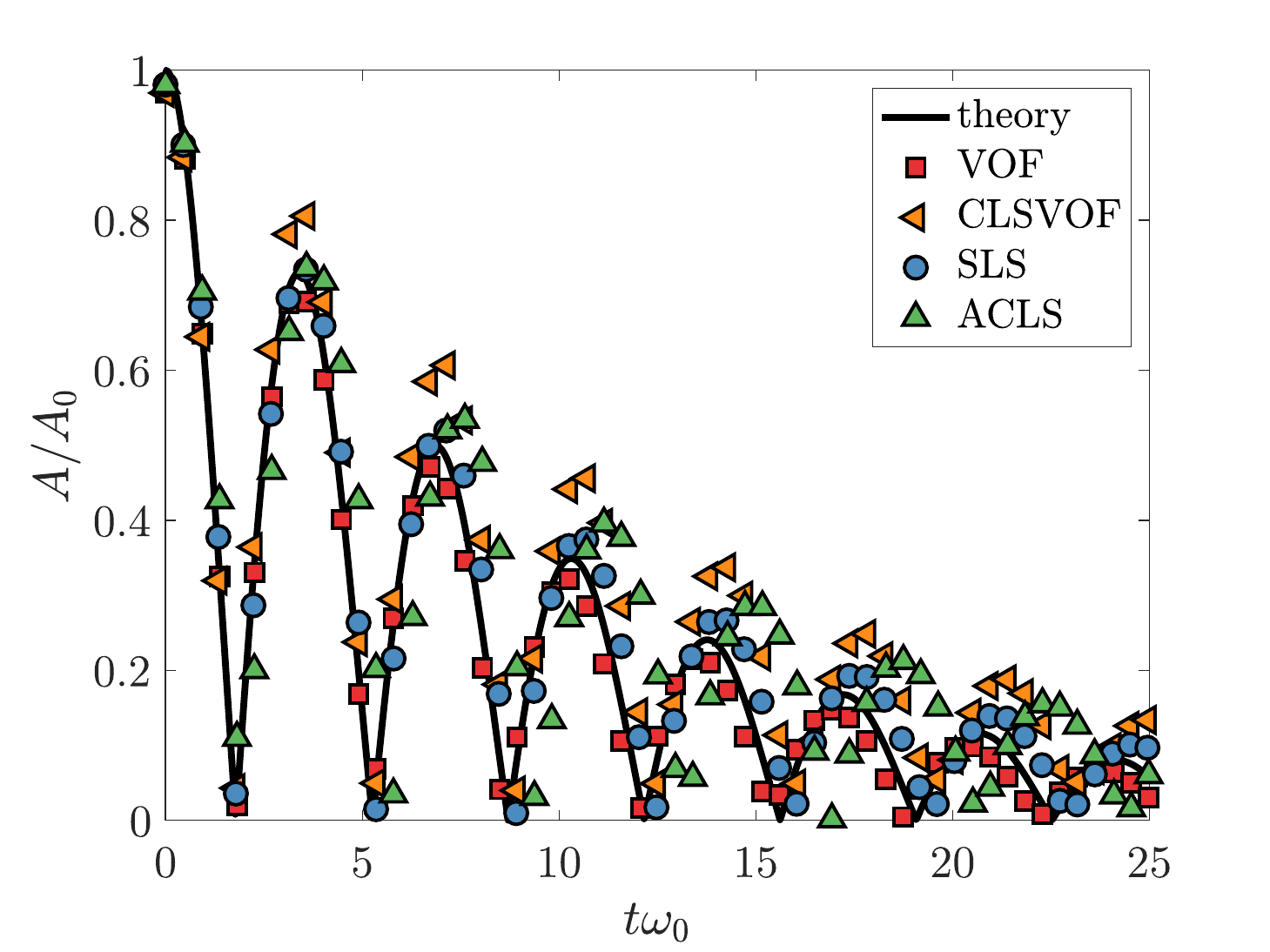}
    		\caption{$N_\lambda=16$}
     \end{subfigure}
     \hfill
     \begin{subfigure}[b]{0.48\textwidth}
     	\centering
         \includegraphics[width=\textwidth]{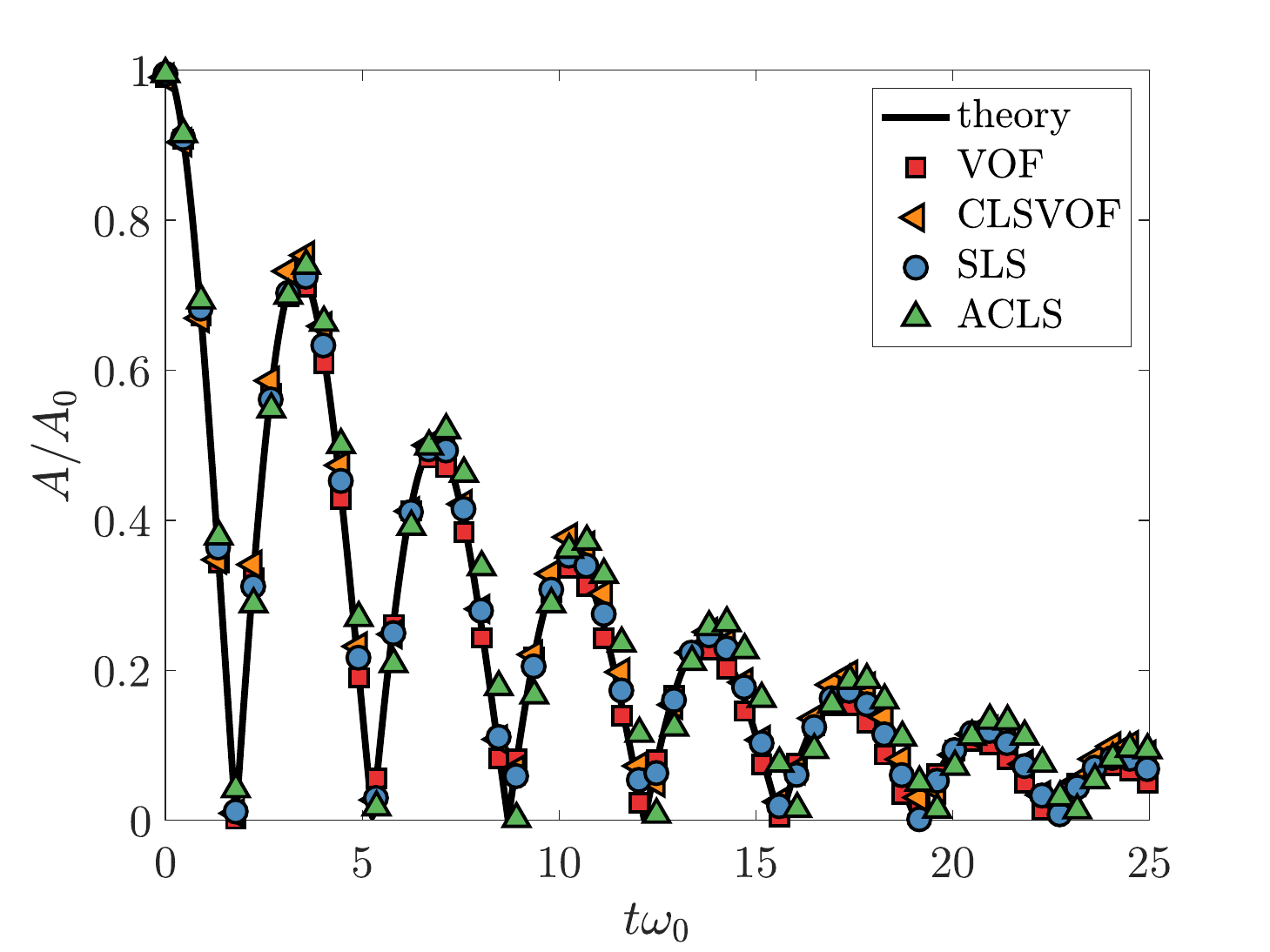}
    		\caption{$N_\lambda=32$}
    		\label{fig:damping_temporal}
     \end{subfigure}
     \hfill
     \begin{subfigure}[b]{0.48\textwidth}
     	\centering
         \includegraphics[width=\textwidth]{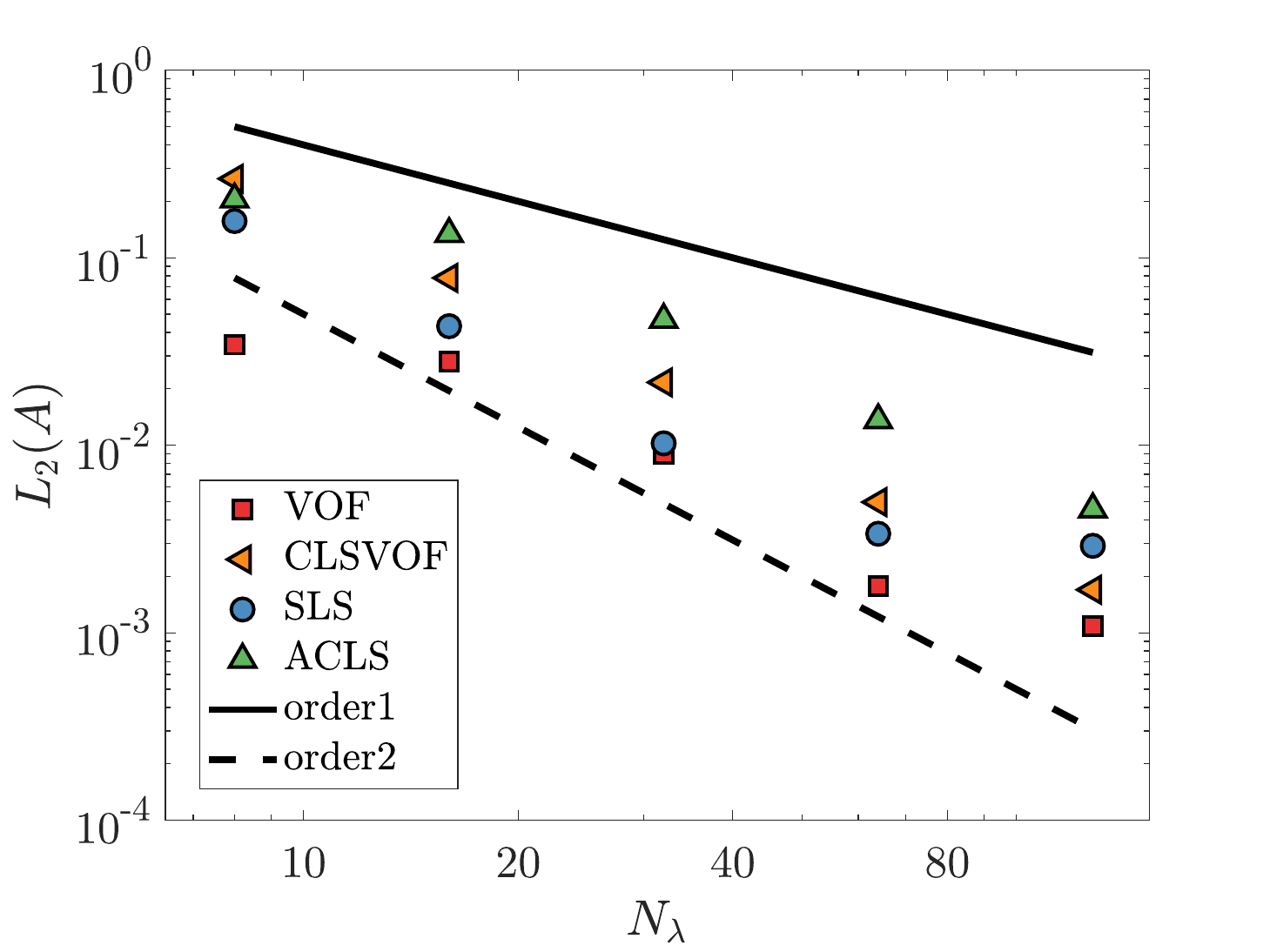}
    		\caption{$L_2(A)$ error}
    		\label{fig:damping_error}
     \end{subfigure}
     \caption{Temporal evolution of $A/A_0$ for different mesh resolutions and mesh error convergence of $L_2(A)$}
\end{figure}

The error $L_2(A)$ is displayed in Fig.~\ref{fig:damping_error} with a number of points in the wavelength $N_{\lambda}$ from $8$ to $128$. All methods converge at second order with a better accuracy for VOF. A convergence saturation is observed for SLS which is due to parasitic perturbations appearing for the highest resolution. A temporal evolution of the amplitude for $N_{\lambda}=32$ is also presented in Fig.~\ref{fig:damping_temporal}, where VOF is already very accurate while SLS, CLSVOF and ACLS are a little bit shifted. 
Note that this test case is specifically well suited for VOF as the HF method is the most accurate in mesh-aligned configurations. This explains why it performs better than the other methods on the damping wave.

\subsection{Applications}

\subsubsection{Droplet collision}
\label{sec:collision}
\begin{figure}[h!]
\centering
     \begin{subfigure}[b]{0.48\textwidth}
     	\centering
         \includegraphics[width=\textwidth]{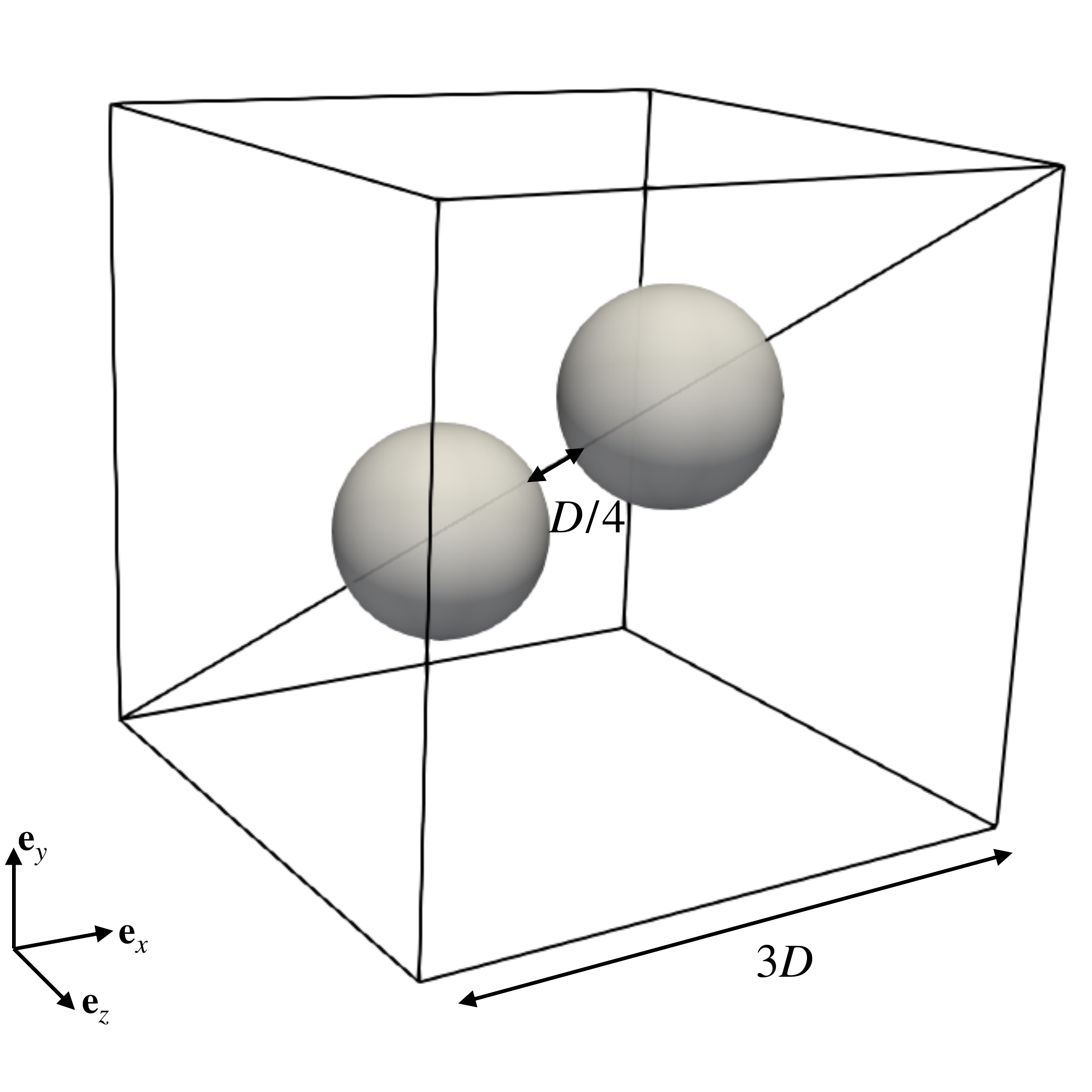}
    		\caption{Initial setup}
    		\label{fig:colli_setup}
     \end{subfigure}
     \hfill
     \begin{subfigure}[b]{0.48\textwidth}
     	\centering
         \includegraphics[width=\textwidth]{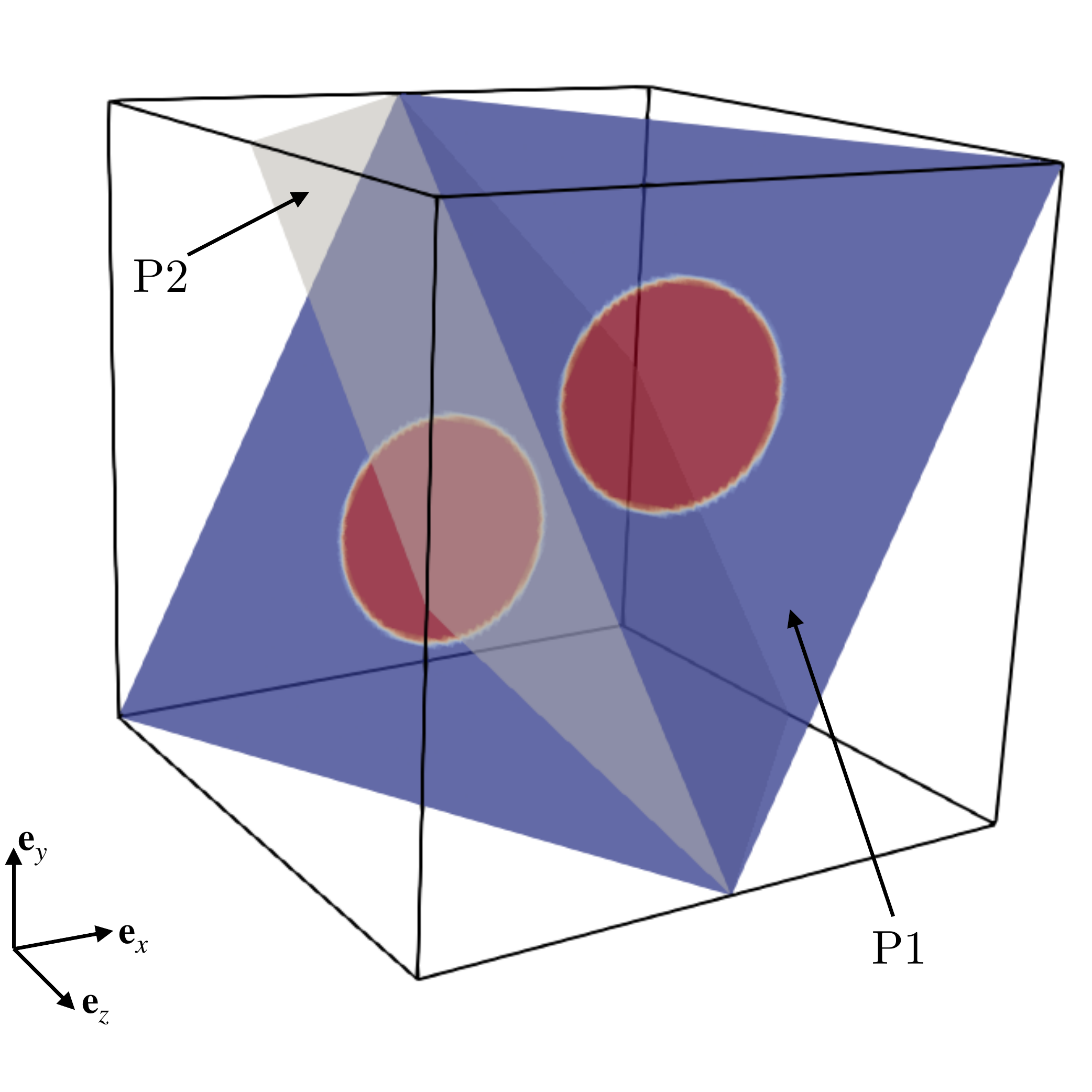}
    		\caption{Plane definition}
    		\label{fig:colli_plane}
     \end{subfigure}
     \caption{Numerical configuration for the head-on collision}
\end{figure}

A collision between two water droplets in quiescent air is presented. The solver have to handle large density ratio and strong capillary effects accurately in order to retrieve the correct collision regime. Indeed, a bad prediction of curvature or momentum will lead to a modification of the Weber number $We=\frac{\rho D u^2}{\sigma}$ and the Ohnesorge number $Oh=\frac{\mu}{\rho D \sigma}$ which drive the collision regime. Here, the head-on collision of two equal-sized droplet with $We=40$ and $Oh=0.0047$ is considered. The expected outcome is a reflexive collision with one satellite observed experimentally \cite{Ashgriz1990}. This regime includes coalescence and break up which are the most challenging behaviour to capture as they will always happen in the mesh resolution limit. Thus, this last test case is the most discriminating for interface capturing methods as they behave differently in this limit cases. Special attention is drawn to the topology transitions. \\
The set up is the same as in \cite{Finotello2017} : the two droplets are located in the body diagonal of of $[3D\times 3D\times 3D]$ cube with a distance of $D/4$ between each other as illustrated on Fig.~\ref{fig:colli_setup}.

\begin{figure}[h!]
	 \includegraphics[width=\textwidth]{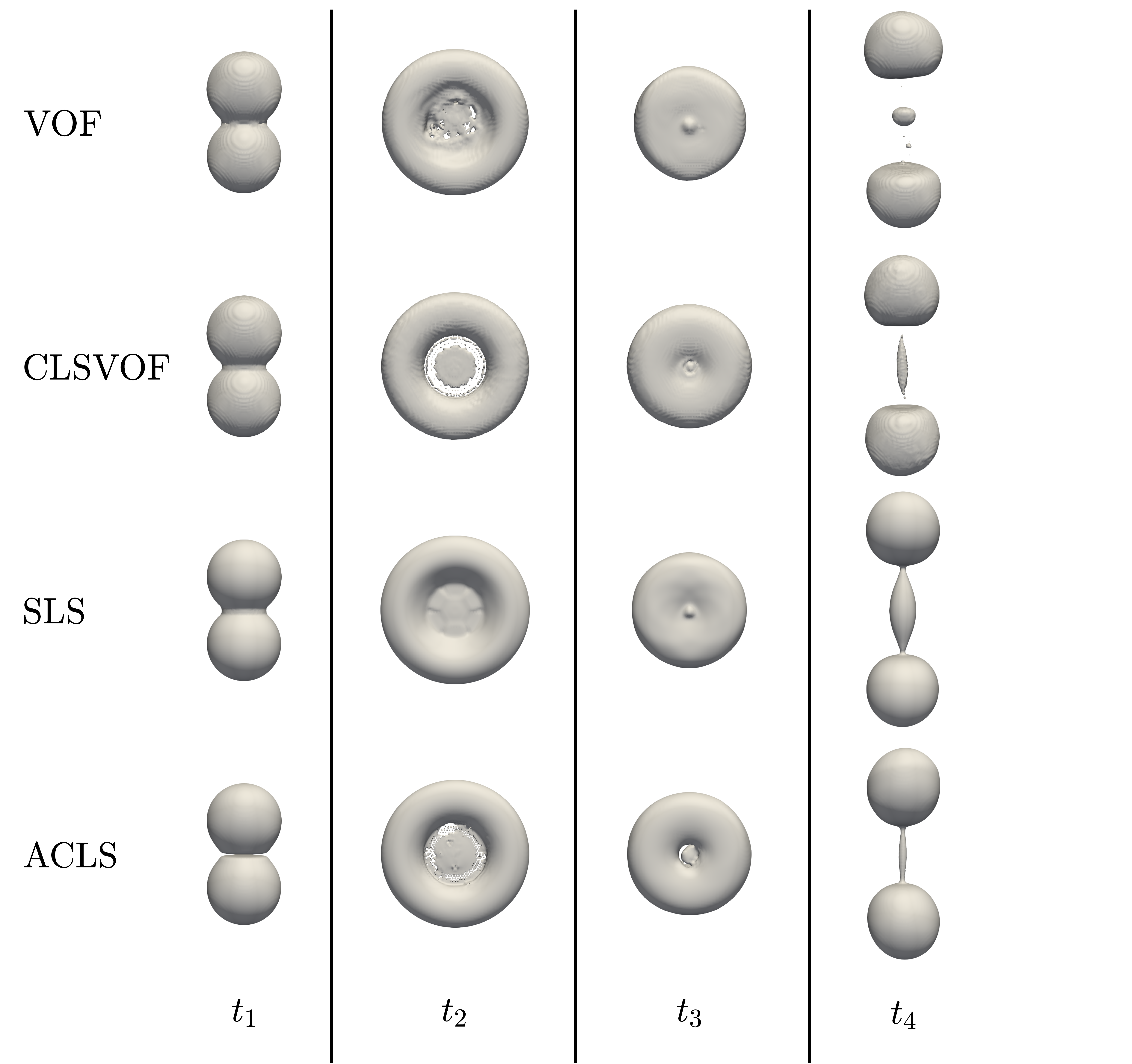}
     \caption{Shape for the four topology changes during the head on collision, $t_1$ and $t_4$ are represented on P1 while $t_2$ and $t_3$ are on P2 (see Fig.~\ref{fig:colli_plane} for plane definition)}
     \label{fig:collision_shape}
\end{figure}

A resolution of $N_D=40$ is chosen and the boundary conditions are free-slip walls in all directions. This diagonal droplet trajectory have two main interests : it avoids any favourable alignments with the mesh and it allows to take a smaller domain to reduce computational time. \\
In Fig.~\ref{fig:collision_shape} are displayed the four topology changes :
\begin{itemize}
\item Coalescence of the droplets at $t_1=150$ µs 
\item Film break up of the disk at $t_2=1250$ µs 
\item Torus coalescence at $t_3=1850$ µs 
\item Thin cylinder break up at $t_4=4700$ µs 
\end{itemize}

\begin{figure}[h!]
\centering
     \begin{subfigure}[b]{0.48\textwidth}
     	\centering
         \includegraphics[width=\textwidth]{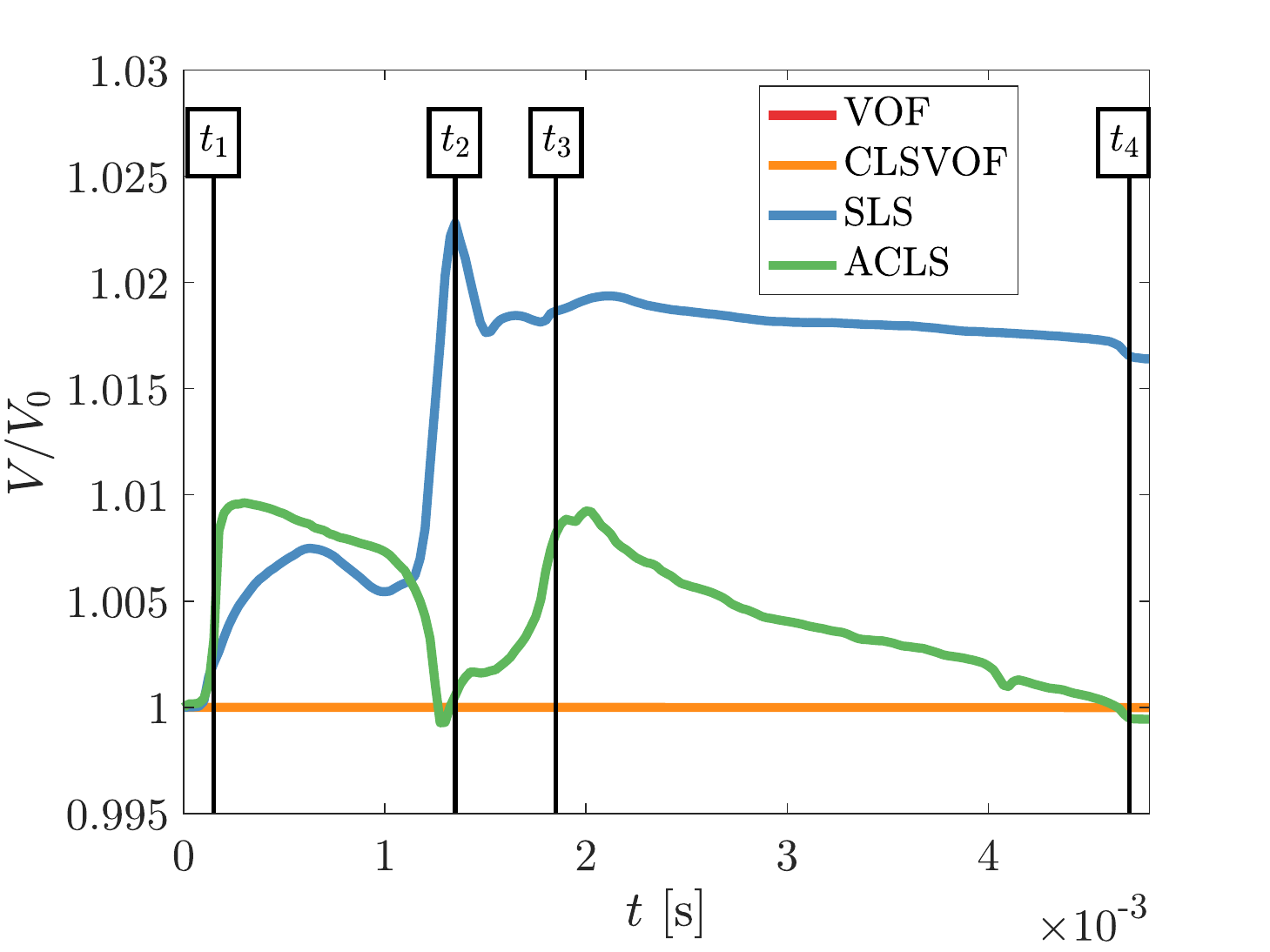}
    		\caption{$V/V_0$}
    		\label{fig:collision_mass}
     \end{subfigure}
     \hfill
     \begin{subfigure}[b]{0.48\textwidth}
     	\centering
         \includegraphics[width=\textwidth]{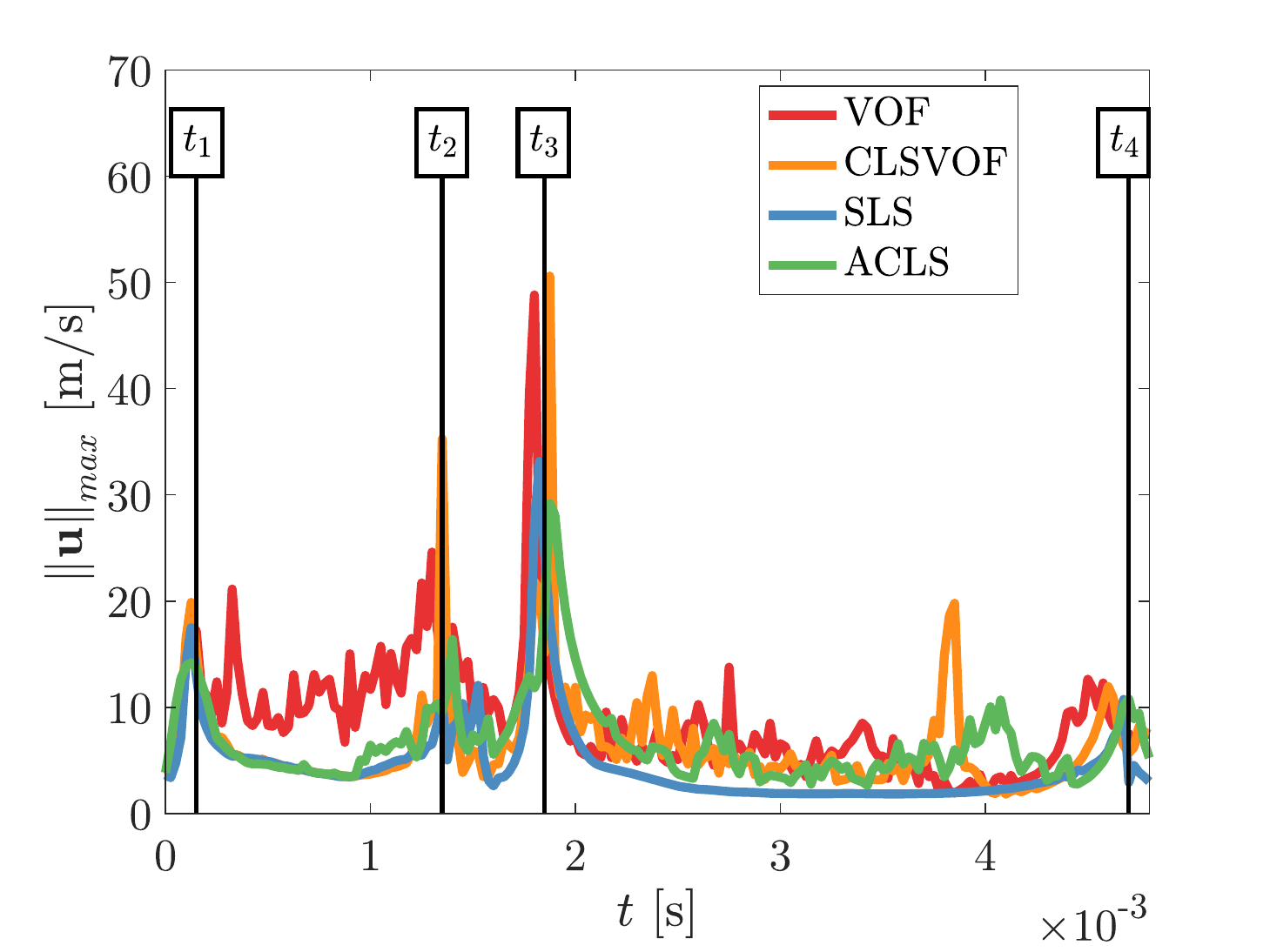}
    		\caption{$\lVert \uvec \rVert_{max}$}
    		\label{fig:collision_umax}
     \end{subfigure}
     \caption{Temporal evolution of $V/V_0$ and $\lVert \uvec \rVert_{max}$ during the head on collision}
     \label{fig:collision_metric}
\end{figure}

\begin{figure}[h!]
\centering
     \begin{subfigure}[b]{0.24\textwidth}
     	\centering
         \includegraphics[width=\textwidth]{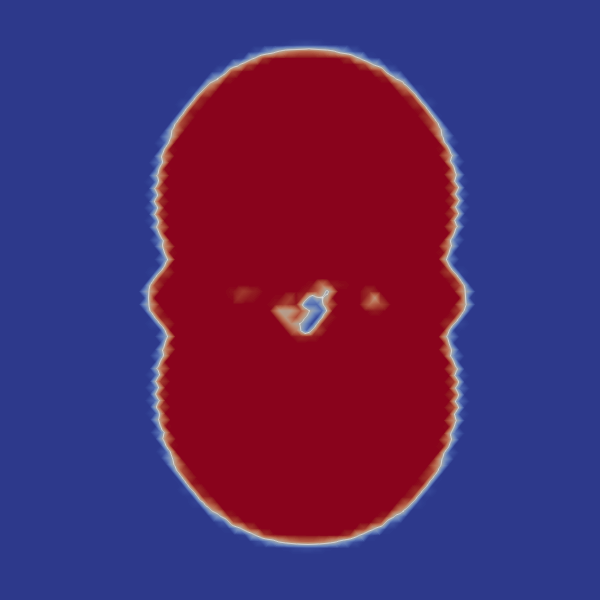}
    		\caption{VOF}
     \end{subfigure}
     \hfill
     \begin{subfigure}[b]{0.24\textwidth}
     	\centering
         \includegraphics[width=\textwidth]{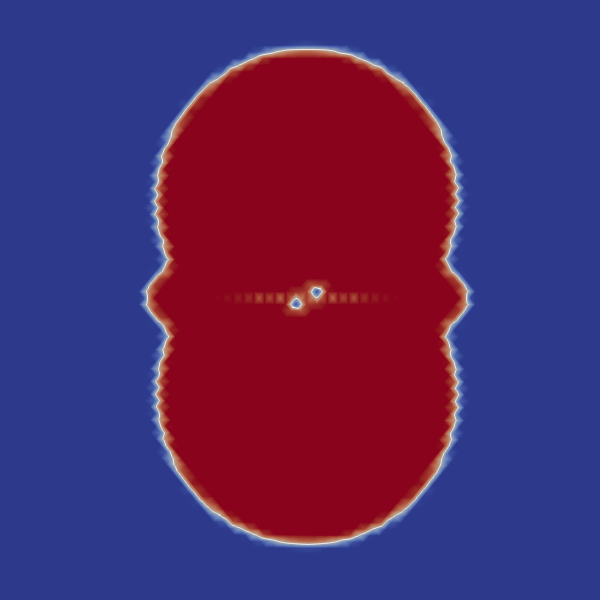}
    		\caption{CLSVOF}
     \end{subfigure}
     \begin{subfigure}[b]{0.24\textwidth}
     	\centering
         \includegraphics[width=\textwidth]{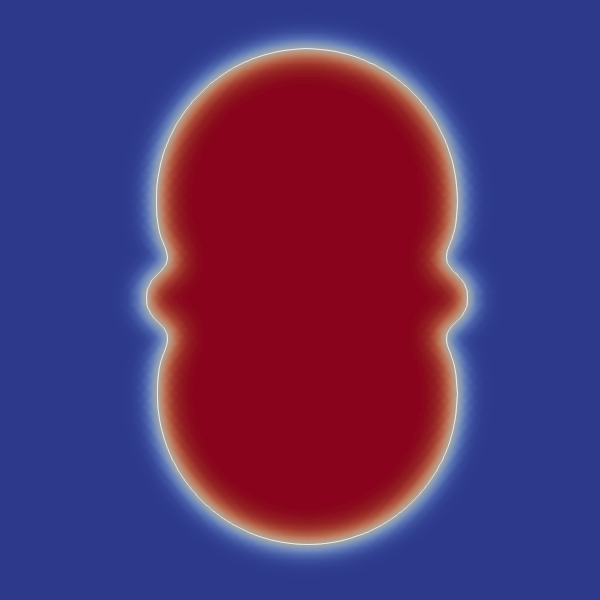}
    		\caption{SLS}
     \end{subfigure}
     \hfill
     \begin{subfigure}[b]{0.24\textwidth}
     	\centering
         \includegraphics[width=\textwidth]{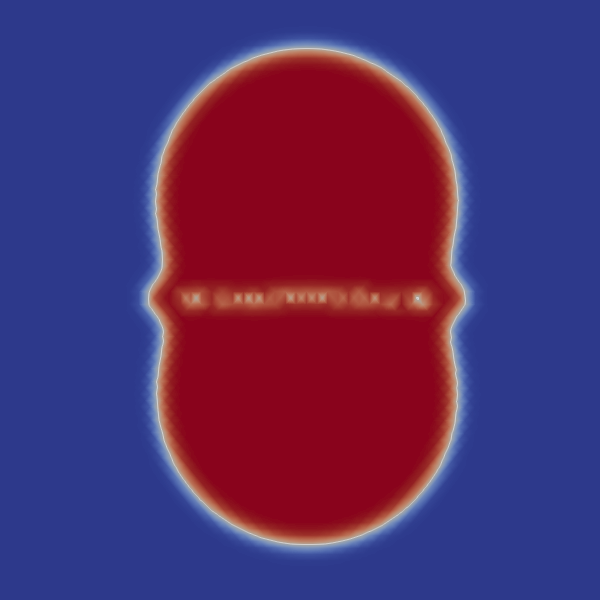}
    		\caption{ACLS}
     \end{subfigure}
     \caption{Collision outcome at $t=2500$ µs represented here with $H_\Gamma^\epsilon$ on P1 }
     \label{fig:collision_bubble}
\end{figure}

Each topology change is characterized by a mass change for SLS and ACLS and a peak of $\lVert \uvec \rVert_{max}$ for all methods (see Fig.~\ref{fig:collision_metric}). This is because of the bad curvature and normal computations when the mesh limit is reached, which is always the case in topology changes. The expected satellite is retrieved at the end of the simulation even if the size vary depending on the method used. \\
The first coalescence at $t_1$ introduces small bubbles at the stagnation point for VOF and CLSVOF because of the normal computation. As the mass is conserved with VOF and CLSVOF, those bubbles will be trapped in the liquid (see Fig.~\ref{fig:collision_bubble}). It is also interesting to notice that ACLS also creates bubbles as previously observed in \cite{Janodet2019}. However, they represent less gas mass compared to VOF and CLSVOF. The presence of those bubbles can cause numerical problem in the simulation as it is not well-resolved by the mesh and produces very bad curvature evaluations. This effect can be related to the spurious $\lVert \uvec \rVert_{max}$ behaviour for VOF in Fig.~\ref{fig:collision_umax} and the use of CLSVOF seems to reduce this phenomenon. In the case of SLS, those bubbles are rapidly turning to liquid as they go under mesh resolution, this corresponds to the first mass creation at $t_1$ in Fig.~\ref{fig:collision_mass}.

At $t_2$, the film is so thin that it goes under mesh resolution which creates break up. This break up causes a loss of mass for ACLS while the SLS manage to maintain the film for a longer time by creating even more liquid mass. VOF, CLSVOF and ACLS create small under-resolved structures at the mesh limit when the film is too thin. This behaviour is inherent to the methods and is the same as for the sphere deformation case of Sec.~\ref{sec:spheredef}.

The torus coalescence at $t_3$ implies the highest peak in $\lVert \uvec \rVert_{max}$ as a lot of small structures created during the film break up are merging and creating multiple source of momentum trough curvature contribution.

Finally all this event history of curvature computation failure and mass change has an impact on the final outcome observed at $t_4$. The satellite is of different size and for VOF, not centred any more.

\begin{figure}[h!]
\centering
	 \includegraphics[width=\textwidth]{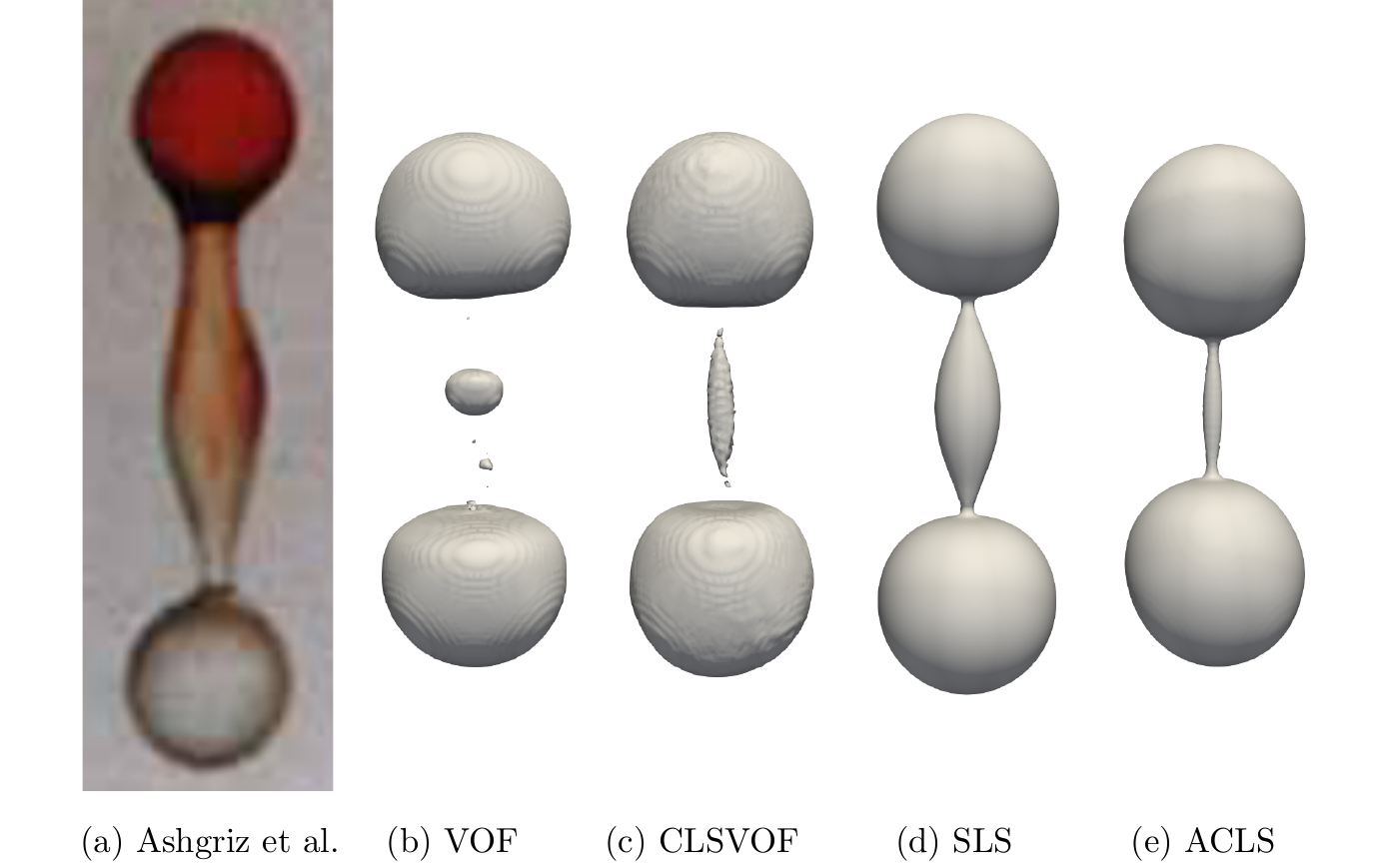}
     \caption{Satellite break up comparison with experiment \cite{Ashgriz1990} }
     \label{fig:colli_comp}
\end{figure}

In Fig.\ref{fig:colli_comp} are compared the satellite break up obtained with the different methods and an experimental acquisition of \cite{Ashgriz1990}. The satellite mass is under-predicted while the two droplet mass is over-predicted as observed in \cite{Tanguy2005} using SLS. This can be caused by capillary instabilities causing premature break-up compared to experiment. The SLS seems produces the biggest satellite but this is partly due to the mass gain of around $1.5$ \% observed in Fig.\ref{fig:collision_mass}. The CLSVOF seems to provide a good trade off between conservation and accuracy. 

\subsubsection{Shear layer}
\begin{figure}[h!]
\centering
	 \includegraphics[width=0.5\textwidth]{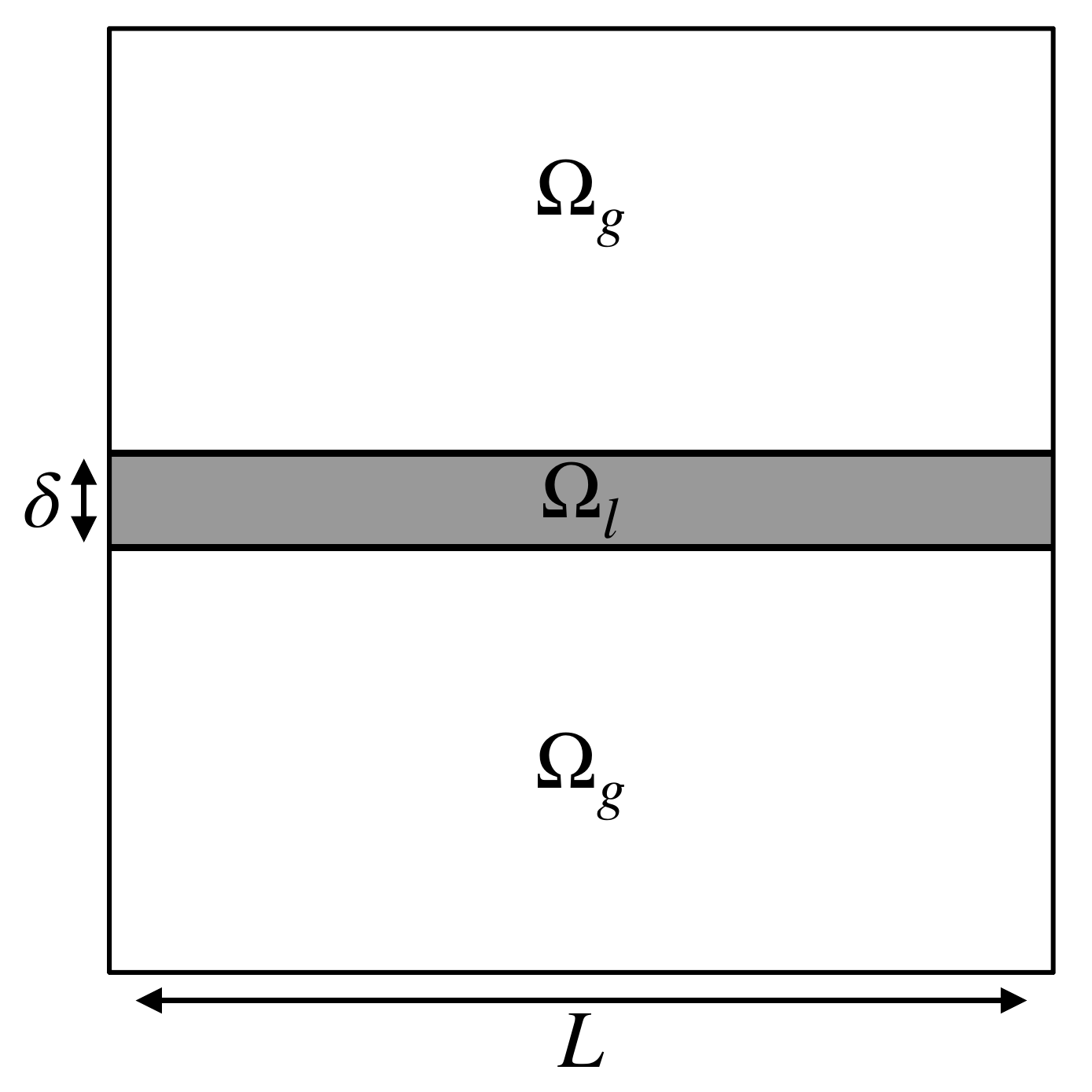}
     \caption{Shear layer simulation set up}
     \label{fig:shear_config}
\end{figure}

This last test case is a planar shear flow similar to \cite{AsuriMukundan2020} and reproduces critical aspects of complex atomization configurations. Without regularizing effect such as viscosity or surface tension ($We=\infty$ and $Oh=0$), the jet will break up in thin structures only impacted by convection. Small errors in the computation of $\rho_f$ lead to error in velocity which can cause severe stability issues. In \cite{Fuster2018}, some of the methods could not reach a long physical time before the simulation breaks down. The set up is illustrated in Fig.~\ref{fig:shear_config} with the periodic box length $L=1$ mm and the liquid shear layer thickness $\delta=L/10$.
Here, the densities are chosen such that $\rho_l / \rho_g = 1000$ and the initial divergence-free velocity is defined as
\begin{align}
& u = U_0 - 0.04 \frac{L}{2 \pi} \frac{-4y}{\delta^2} \cos \left( \frac{2\pi x }{L}\right) \exp \left( -2 \left(\frac{y-h_0}{\delta}\right)^2 \right) \mbox{m/s}  \\
& v = 0.04 \sin \left( \frac{2\pi x }{L} \right) \exp \left( -2 \left(\frac{y-h_0}{\delta}\right)^2  \right) \mbox{m/s}
\end{align} 
with $U_0$ is $2$ m/s in the liquid and $30$ m/s in the gas and $h_0=L/5$ the liquid shear center position.

The simulation is performed until one of the methods fails. In our framework, all methods seemed to be robust but the ACLS which breaks down after $t=1.65$ ms, while the others manage to reach the same physical time $t=2$ ms as in \cite{AsuriMukundan2020}

\begin{figure}[h!]
\centering
 	\begin{subfigure}[b]{0.48\textwidth}
     	\centering
         \includegraphics[width=\textwidth]{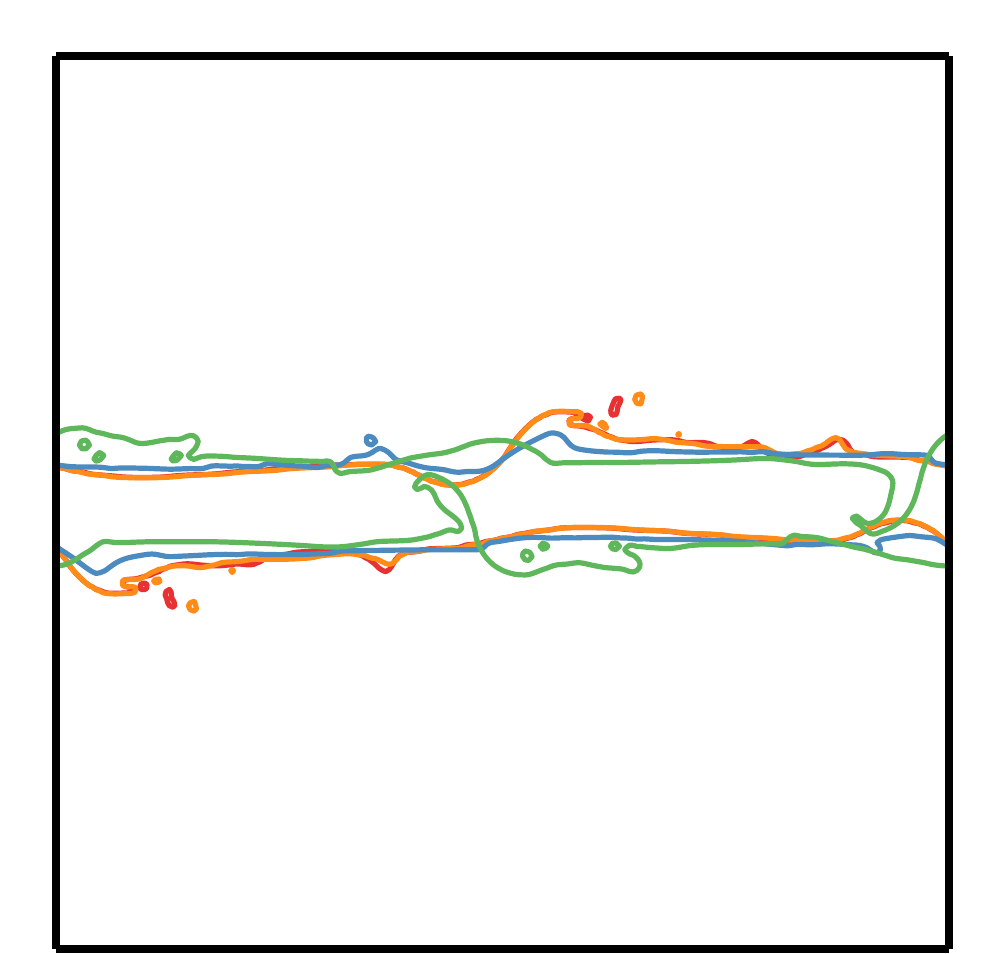}
    		\caption{$t=0.5$ ms}
     \end{subfigure}
     \hfill
     \begin{subfigure}[b]{0.48\textwidth}
     	\centering
         \includegraphics[width=\textwidth]{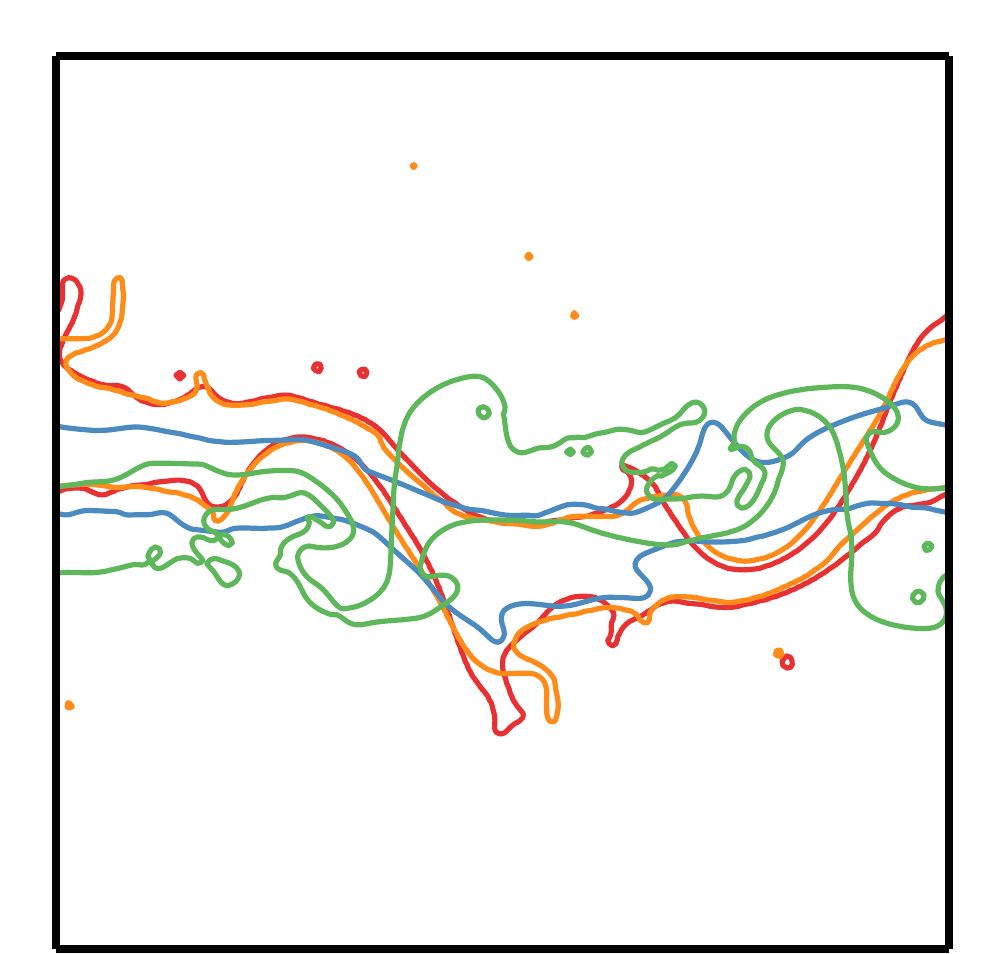}
    		\caption{$t=1$ ms}
     \end{subfigure}
     \hfill
     \begin{subfigure}[b]{0.48\textwidth}
     	\centering
         \includegraphics[width=\textwidth]{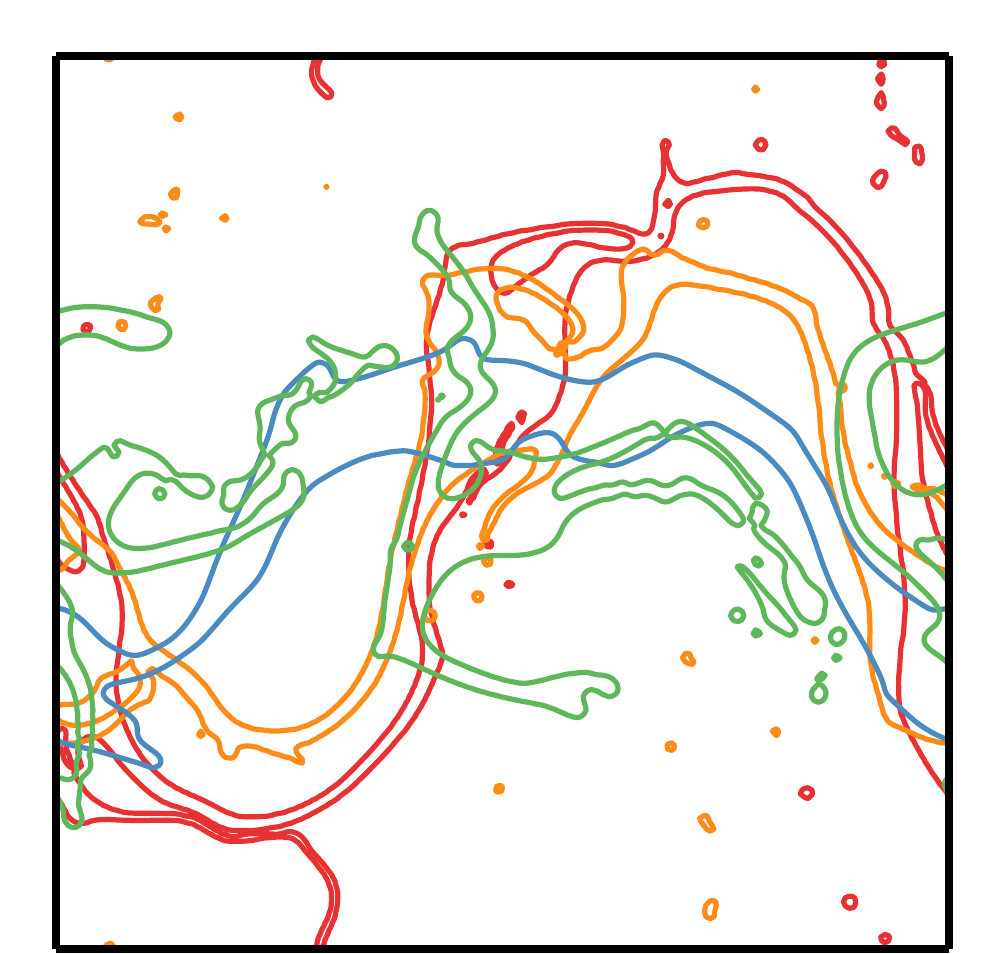}
    		\caption{$t=1.65$ ms}
     \end{subfigure}
     \hfill
     \begin{subfigure}[b]{0.48\textwidth}
     	\centering
         \includegraphics[width=\textwidth]{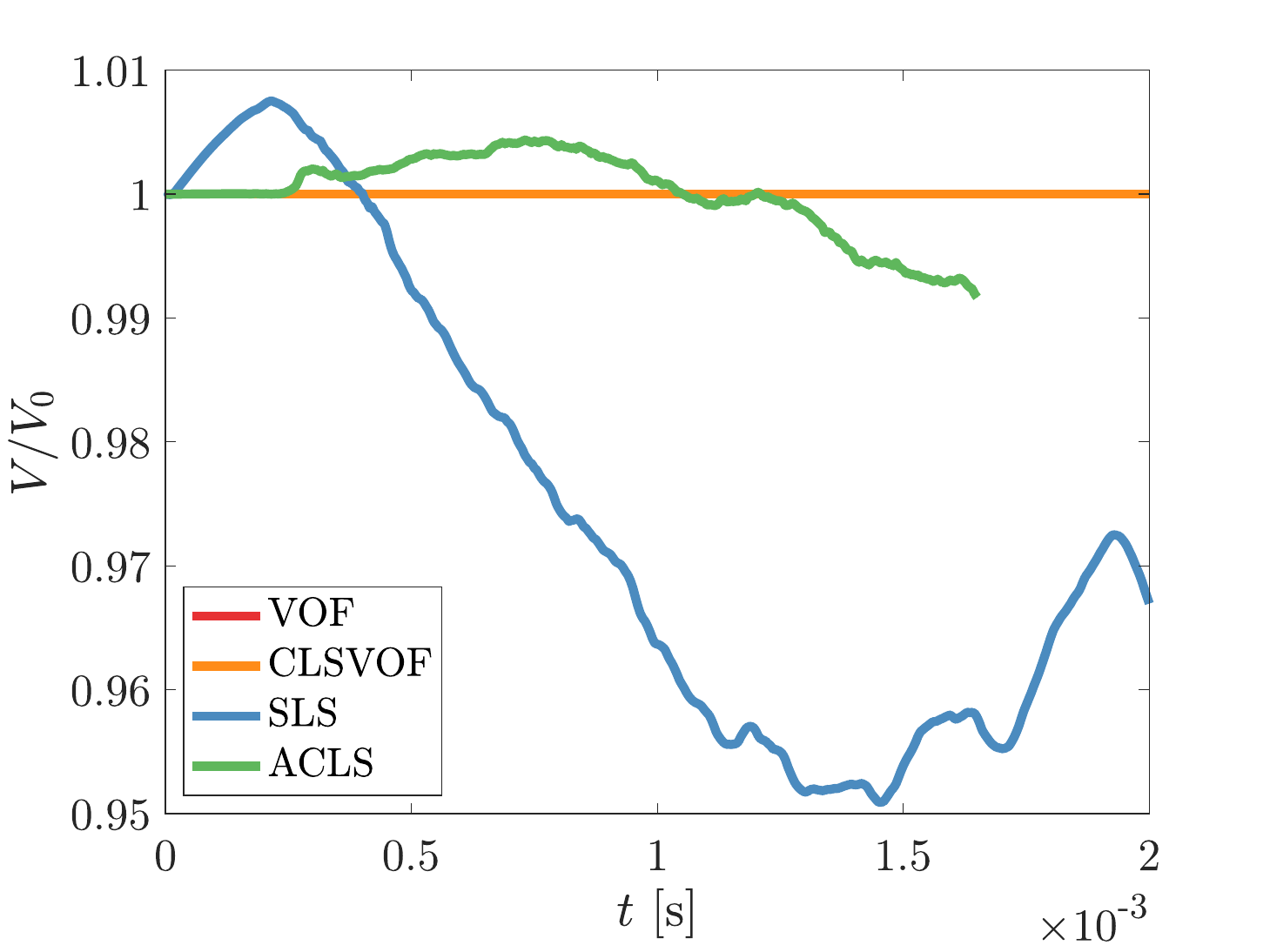}
    		\caption{Temporal $V/V_0$}
    		\label{fig:shear_mass}
     \end{subfigure}
     \caption{Liquid layer isocontour for \textbf{(a)} $t=0.5$ ms, \textbf{(b)} $t=1$ ms, \textbf{(c)} $t=1.65$ ms for VOF (red line), CLSVOF (orange line), SLS (blue line) and ACLS (green line) and \textbf{(d)} the temporal evolution of mass for the shear layer test case}
     \label{fig:shear_field}
\end{figure}

In Fig.~\ref{fig:shear_field} are given isocontour of the four interface capturing methods at different times. While VOF and CLSVOF are behaving similarly at the beginning, the discrepancy is increasing with time. Moreover, SLS and ACLS are showing even more different behaviour compared to VOF-based methods. This shows how interface capturing methods impact the topology of a liquid jet under high convective effects when the structure are not properly captured anymore. In this test case it is difficult to conclude on what method is the best, even if the ACLS exhibits some stability problems. This issue arises when very thin structures are created and density is approximated poorly, leading to very high velocities without any diffusive process to damp them. It is also important to notice that SLS while robust has lost about $4\%$ of mass during the whole simulation time while ACLS managed to maintain a mass error above $1\%$. This last test case enlightens that under-resolved structures behaviour which are present in atomization simulations will be highly impacted by the interface capturing choice. The outcome at a large physical time is thus completely different.

\section{Conclusion}
The comparison of four popular methods of the literature has been presented focusing on mass and momentum conservations, and geometrical accuracy. Overall, the interface capturing methods presented here are able to provide a good physical description of two-phase flows with high density ratio and capillarity effects, with specific strengths and weaknesses. \\
While VOF and CLSVOF are exactly mass conservative and show good momentum conservation, SLS provides a representation of the interface able to compute normal and curvature easily and with high accuracy even in dynamic cases.
ACLS shows improvements in the mass and momentum conservation compared to SLS at the cost of a loss of geometrical accuracy which can be very severe at high resolution.
For coarse to medium resolutions, ACLS accuracy is slightly better than VOF or CLSVOF for dynamic cases. 
The head-on collision has demonstrated the robustness of all methods, showing that they are all able to retrieve the satellite droplet. However, the atomization case has also shown that the mesh resolution has a great impact on the accuracy and the development of thin structures, all methods giving very different results for long times of simulation. This calls for further in-depth statistical analysis in such chaotic test cases.
\\
Overall, Coupling VOF with LS seems to be the most promising choice in a Cartesian finite volume framework as the conservation properties of VOF are preserved while curvature and normal are easier to compute and more accurate. The versatility of such strategy, taking advantage of the forces of each method, will be particularly of interest when additional physics is required, such as phase change. Our unified framework is in that sense the adequate vessel for such developments.\\
It has to be reminded that our conclusions hold for this specific unified framework.
Indeed, the under-resolved problems met in the head-on collision could be better handled by using local mesh refinement \cite{popinet2009accurate,Herrmann2008,Janodet2019}.
Moreover, our VOF and CLSVOF transport rely on dimensional splitting and the geometrical operations are straightforward on a Cartesian mesh. This is no longer true in unstructured meshes, and the ACLS method can be a good alternative to keep good accuracy and conservation. \\

As perspectives, some numerical aspects are still challenging and require further investigations.
As an example, it could be interesting to explore the possibility of higher order computation of distance function in the ACLS method based on PDE \cite{Jiang2000}, or a higher order reconstruction of the interface in the case of VOF and CLSVOF such as parabolas \cite{Renardy2002} and quadratic splines \cite{Diwakar2009} in order to obtain a consistent curvature estimation in dynamic cases.
Recent reinitialization modifications for SLS would also be useful, in order to enhance mass conservation \cite{Solomenko2017}.

\section{Acknowledgements}
The support of the ANR Project MIMETYC (ANR-17-CE22-0003) and SubSuperJet (ANR-14-CE22- 0014) is acknowledged. A part of this work was performed using HPC resources from the mésocentre computing center of Ecole CentraleSupélec and Ecole Normale Supérieure Paris- Saclay supported by CNRS and Région Ile-de-France. The authors would like to thank Romain Janodet, Vincent Moureau, Ghislain Lartigue and Robert Chiodi for fruitful discussions about Interface capturing methods and numerical implementation.

\bibliographystyle{unsrt}
\bibliography{paper}

\end{document}